\documentclass[11pt,a4paper]{article}
\pdfoutput=1

\usepackage{jheppub}
\usepackage{amsmath}
\usepackage{amssymb}
\usepackage{graphics}
\usepackage[active]{srcltx}
\usepackage{pdfsync}
\usepackage{shuffle}
\usepackage{slashed}
\usepackage{hyperref}
\usepackage{subfigure}

\newcommand{\bea}{\begin{eqnarray}}
\newcommand{\eea}{\end{eqnarray}}
\newcommand{\nn}{\nonumber \\}

\def\W #1{\widetilde{#1}}

\def\Tr{\mathop{\rm Tr}}

\def\eref#1{(\ref{#1})}

\def\a{{\alpha}}

\def\b{{\beta}}

\allowdisplaybreaks

\title{Understanding zeros and splittings of ordered tree amplitudes via Feynman diagrams}
 \author[a]{Kang Zhou}

\affiliation[a]{Center for Gravitation and Cosmology, College of Physical Science and Technology, Yangzhou University,\\
No.180, Siwangting Road, Yangzhou, 225009, P.R. China}

\emailAdd{zhoukang@yzu.edu.cn}

\date{\today}
\abstract{In this paper, we propose new understandings for recently discovered hidden zeros and novel splittings, by utilizing Feynman diagrams. The study focus on ordered tree level amplitudes of three theories, which are ${\rm Tr}(\phi^3)$, Yang-Mills, and non-linear sigma model. We find three universal ways of cutting Feynman diagrams, which are valid for any diagram, allowing us to separate a full amplitude into two/three pieces. As will be shown, the first type of cuttings leads to hidden zeros, the second one gives rise to $2$-splits, while the third one corresponds to $3$-splits called smooth splittings. Throughout this work, we frequently use the helpful auxiliary technic of thinking the resulting pieces as in orthogonal spaces. However, final results are independent of this auxiliary picture.
}


\begin{document}

\maketitle \flushbottom

\section{Introduction}
\label{sec-intro}

The factorizations of scattering amplitudes play the crucial role in modern researches on S-matrix. As well known, the residue of a tree level scattering amplitude at any physical pole factorizes into the product of two lower-point ones.
This factorization behavior enables effective recursive computation of amplitudes, without the aid of traditional Lagrangian and Feynman rules. A prime example is the Britto-Cachazo-Feng-Witten (BCFW) recursion, which elegantly utilizes the factorization as physical input \cite{Britto:2004ap,Britto:2005fq}. At the loop level, the powerful method called (generalized) unitarity cut exploits factorizations to determine integrands of Feynman integrals \cite{Bern:1994zx,Bern:1994cg,Britto:2004nc}.

Recently, a class of novel factorizations, without taking residue at any pole, have been discovered. In chronological order, the first one is the so called smooth splitting found in \cite{Cachazo:2021wsz}, where a certain tree level amplitude factorizes into three amputate currents, when setting special Mandelstam variables to be zero (see also \cite{GimenezUmbert:2024jjn} for recent study of $\phi^p$ theory). Such behavior holds for amplitudes of bi-adjoint scalar (BAS) theory, non-linear sigma model (NLSM) and special  Galileon (SG) theory. Subsequently, an amazing property of tree level amplitudes called hidden zeros, as well as factorizations near such zeros, are discovered in \cite{Arkani-Hamed:2023swr}, based on the great progress in generalizing the combinatorial and geometric structure of amplitudes from toy models towards theories for the real world \cite{Arkani-Hamed:2023lbd,Arkani-Hamed:2023mvg,Arkani-Hamed:2023swr,Arkani-Hamed:2023jry,Arkani-Hamed:2024nhp,Arkani-Hamed:2024vna,
Arkani-Hamed:2024yvu,Arkani-Hamed:2024tzl}. The hidden zeros found in \cite{Arkani-Hamed:2023swr} are for ${\rm Tr}(\phi^3)$, non-linear sigma model (NLSM) and Yang-Mills (YM) amplitudes at the tree level. Taking ${\rm Tr}(\phi^3)$ ones as the example, zeros occur when certain Mandelstam variables vanish. By turning on one of vanished Mandelstam variables to be non-zero, each amplitude factorizes into three pieces. Zeros and factorizations near zeros for NLSM and YM amplitudes are similar. Very recently, a more basic type of facgtorization/splitting called $2$-split, holds for a larger class of particle and string amplitudes, has been studied in \cite{Cao:2024gln,Arkani-Hamed:2024fyd,Cao:2024qpp}.
The $2$-split describes the phenomenon that an amplitude factorizes into two amputated currents, on simple loci in the kinematic space. For instance, when
\bea
k_a\cdot k_b=0\,,~~~~{\rm for}~a\in\{j+1,\cdots,k-1\}\,,~b\in\{k+1,\cdots,j-1\}\setminus i\,,~~\label{kine-condi-0}
\eea
the $n$-point ordered ${\rm Tr}(\phi^3)$ amplitude factorizes as
\bea
{\cal A}^{\phi^3}_{n}(1,\cdots,n)\,\rightarrow\,{\cal J}^{\phi^3}_{n_1}(i^*,j,\dots,k)\,\times\,{\cal J}^{\phi^3}_{n+3-n_1}(k,\cdots,i^*,\cdots,j)\,,
\eea
where $i$, $j$ and $k$ are three external legs in $\{1,\cdots,n\}$. In the above, ${\cal J}^{\phi^3}_{n_1}$ and ${\cal J}_{n+3-n_1}^{\phi^3}$ are two currents, each of them carries an off-shell leg $i^*$. The two types of $3$-splits, which are smooth splittings and factorizations near zeros,
follow from further specializations of the $2$-splits.

The above remarkable hidden zeros and splittings are discovered by considering novel formulas of scattering amplitudes, such like CHY formulas \cite{Cachazo:2013hca,Cachazo:2013iea,Cachazo:2014nsa,Cachazo:2014xea}, or stringy curve integrals \cite{Arkani-Hamed:2023lbd,Arkani-Hamed:2023mvg,Arkani-Hamed:2023jry,Arkani-Hamed:2024nhp}. Comparing with traditional factorizations which are guaranteed by fundamental unitarity and locality, the underlying principle which governs novel zeros and splittings, is still lacking. Therefore, it is worth to derive and interpret such zeros and splittings via other different paths. The new derivation and interpretation may hint the new insight, which is beneficial for seeking the underlying principle allowing us to understand these zeros and splittings as conceptually and intuitionally as traditional factorizations.

Motivated by the above expectation, in this paper we study zeros and splittings of tree level ordered amplitudes, including ${\rm Tr}(\phi^3)$, YM and NLSM ones, by utilizing Feynman diagrams. We find three types of universal cuttings which are valid for any diagram, allowing us to separate the full amplitude into two/three pieces. By setting resulting pieces to be in orthogonal spaces, we can figure out that the first type of cuttings leads to zeros of amplitudes, the second one corresponds to $2$-splits, while the third one gives rise to smooth splittings. The final results are independent of the auxiliary picture of orthogonal spaces.

More specifically, for ${\rm Tr}(\phi^3)$ amplitudes, we will give two methods for understanding zeros and splittings. The first one based on the assumption that any scattering process in extra dimensions is insensible for an observer lives in ordinary dimensions, therefore two amplitudes localized in two orthogonal spaces are independent of each other. This method determines the special loci in kinematic space for zeros and splittings, but fail to answer the corresponding theory (theories) of resulting amplitudes in splittings. To interpret the resulting amplitudes,
we also introduce the second method, by utilizing the observation which can be called "factorization of propagators". Via the second method, the resulting amplitudes in $2$-split and smooth splitting can be fixed as currents of ${\rm Tr}(\phi^3)$ theory, while zeros can also be reproduced. Notice that the second method is independent of the assumption in the first one.

The above first method also leads to the correct loci in kinematic space for zeros and splittings of YM and NLSM amplitudes, while exact interpretations for resulting amplitudes should be determined via other approaches. The second method for ${\rm Tr}(\phi^3)$ amplitudes can not be directly applied to the YM case, due to the non-trivial contributions from vertices. We extend the second method by exploiting locality and gauge invariance. Using such generalized second method, one can completely determine the resulting amplitudes in splittings of YM amplitudes, and reproduce zeros. We have not successfully generalized the second method to NLSM amplitudes. Instead, we adopt an alternative method which generates zeros and splittings of NLSM amplitudes from zeros and splittings of YM ones, by applying transmutation operators. This method gives rise to exact interpretations of resulting amplitudes in splittings of NLSM amplitudes.

The rest of this paper is organized as follows. In section \ref{sec-phi3}, we consider zeros and splittings of tree level ${\rm Tr}(\phi^3)$ amplitudes, and explain the main ideas of this paper. In section \ref{sec-YM}, we apply the similar approach to understand zeros and splittings of tree level YM amplitudes. In section \ref{sec-NLSM}, zeros and splittings of tree level NLSM amplitudes are considered. Finally, we end with a brief discussion in section \ref{sec-conclusion}.

\section{${\rm Tr}(\phi^3)$ amplitudes}
\label{sec-phi3}

In this section, we discuss zeros and splittings of tree level amplitudes of ${\rm Tr}(\phi^3)$ theory.
This theory describe the cubic interaction of colored massless scalars, with Lagrangian
\bea
{\cal L}_{{\rm Tr}(\phi^3)}={\rm Tr}(\partial\phi)^2+g\,{\rm Tr}(\phi^3)\,,
\eea
where $\phi$ is an $N\times N$ matrix, with two indices in the fundamental and anti-fundamental representations of
$SU(N)$. The ordered tree level amplitudes in this theory, with coupling constants stripped off, consist solely of propagators for massless scalars.

The structure of this section is as follows. In subsections \ref{subsec-5p0} and \ref{subsec-5pfac}, we use the simple $5$-point examples, to demonstrate our main approach based on Feynman diagrams, for understanding zeros and splittings. In subsection \ref{subsec-idea}, we give a general discussion for the idea of decomposing the amplitude into those in orthogonal complementary subspaces, which is helpful for subsequent discussions in this paper. In subsection \ref{subsec-zero-phi3}, we use this idea to figure out the special loci in kinematic space, for zeros of ${\rm Tr}(\phi^3)$ amplitudes. In subsection \ref{subsec-1to2-phi3},
we use this idea to consider the $2$-split. The picture of orthogonal spaces determines the corresponding constraints on external kinematics, but can not answer the theory (theories) for resulting amplitudes in $2$-splits.
In subsection \ref{subsec-verification-phi3}, we generalize the phenomenon in subsections \ref{subsec-5p0} and \ref{subsec-5pfac}, which can be called "factorization of propagators", to the general case with arbitrary number of external legs, to give the interpretation to the resulting amplitudes in $2$-splits in subsection \ref{subsec-1to2-phi3}, as well as another alternative understanding for zeros in subsection \ref{subsec-zero-phi3}. Finally, in subsection \ref{subsec-fac-zero}, we discuss two types of interesting $3$-splits, which are factorizations near zeros discovered in \cite{Arkani-Hamed:2023swr}, and smooth splittings found in \cite{Cachazo:2021wsz}.

Although focusing on ${\rm Tr}(\phi^3)$ theory, the discussions and conclusions in subsections \ref{subsec-idea}, \ref{subsec-zero-phi3} and \ref{subsec-1to2-phi3} are quiet general, and can be directly extended to YM and NLSM theories in sections \ref{subsec-zero-YM} and \ref{subsec-zero-NLSM}.

\subsection{The $5$-point example of zero}
\label{subsec-5p0}

In this subsection, we use a simple $5$-point example, to demonstrate our main idea for understanding zeros of tree level ${\rm Tr}(\phi^3)$ amplitudes via Feynman diagrams.

\begin{figure}
  \centering
  \includegraphics[width=9cm]{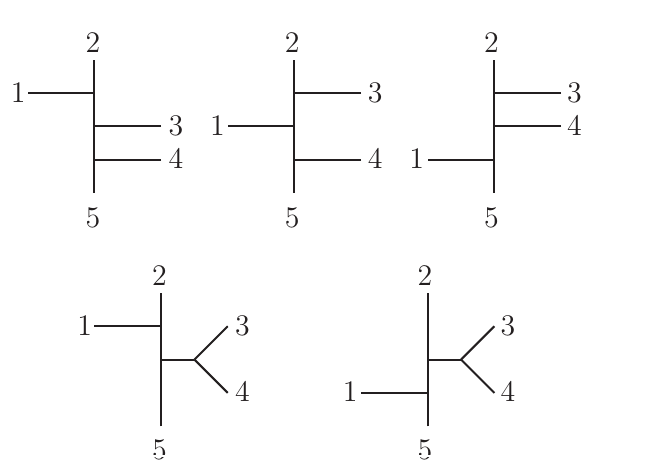} \\
  \caption{Five Feynman diagrams for the $5$-point amplitude ${\cal A}_5^{\phi^3}(1,2,3,4,5)$. These diagrams are arranged for illustrating the zero occurs when $s_{13}=s_{14}=0$.}\label{5p01}
\end{figure}

Consider the $5$-point tree level ${\rm Tr}(\phi^3)$ amplitude ${\cal A}_5^{\phi^3}(1,2,3,4,5)$. Let us arrange five Feynman diagrams as in Fig.\ref{5p01}, the amplitude is expressed as
\bea
{\cal A}_5^{\phi^3}(1,2,3,4,5)&=&{1\over s_{12}}\,{1\over s_{123}}+{1\over s_{23}}\,{1\over s_{123}}+{1\over s_{23}}\,{1\over s_{234}}\nn
& &+{1\over s_{12}}\,{1\over s_{34}}+{1\over s_{34}}\,{1\over s_{234}}\,,~~\label{5p0-1}
\eea
where the Mandelstam variables $s_{a\cdots b}$ is defined as usual, namely, $s_{a\cdots b}=k_{a\cdots b}^2$, with $k_{a\cdots b}=k_a+\cdots+k_b$. In the above, three terms in the first line correspond to three diagrams in the first line of Fig.\ref{5p01},
while two terms in the second line correspond to two diagrams in the second line.
Now we set $s_{13}=s_{14}=0$. Under this constraint, the first line of \eref{5p0-1} behaves as
\bea
& &{1\over s_{12}}\,{1\over s_{123}}+{1\over s_{23}}\,{1\over s_{123}}+{1\over s_{23}}\,{1\over s_{234}}\,\xrightarrow[]{s_{13}=s_{14}=0}\,
{1\over s_{12}}\,{1\over s_{23}}+{1\over s_{23}}\,{1\over s_{234}}=0\,,~~\label{5p0-1st}
\eea
since
\bea
s_{12}+s_{23}\,&\xrightarrow[]{s_{13}=0}&\,s_{12}+s_{23}+s_{13}=s_{123}\,,\nn
s_{12}+s_{234}\,&\xrightarrow[]{s_{13}=s_{14}=0}&\,s_{12}+s_{234}+s_{13}+s_{14}=s_{1234}=0\,.~~\label{alg-pro}
\eea
Meanwhile, the second line of \eref{5p0-1} behaves as
\bea
{1\over s_{12}}\,{1\over s_{34}}+{1\over s_{34}}\,{1\over s_{234}}\,&\xrightarrow[]{s_{13}=s_{14}=0}&\,0\,,~~\label{5p0-2nd}
\eea
again due to
\bea
s_{12}+s_{234}\,&\xrightarrow[]{s_{13}=s_{14}=0}&\,s_{1234}=0\,.
\eea
Combining \eref{5p0-1st} and \eref{5p0-2nd} leads to
\bea
{\cal A}_5^{\phi^3}\,&\xrightarrow[]{s_{13}=s_{14}=0}&\,0\,.~~\label{5p0}
\eea

To understand the zero exhibited in \eref{5p0}, we consider a special path for achieving the constraint $s_{13}=s_{14}=0$. We decompose the full $D$-dimensional spacetime ${\cal S}$ into two orthogonal complementary spaces, namely, ${\cal S}={\cal S}_1\oplus{\cal S}_2$, where ${\cal S}_1$ and ${\cal S}_2$ are expanded by basis vectors in $a^{\rm th}$ and $b^{\rm th}$ dimensions respectively, with $\{a\}\cup\{b\}=\{0,1,\cdots,D-1\}$, $\{a\}\cap\{b\}=\emptyset$. Assuming $k_1$ is purely in the subspace ${\cal S}_1$, while $k_3$ and $k_4$ are purely in ${\cal S}_4$, the kinematic condition $s_{13}=s_{14}=0$ is satisfied automatically.
Notice that the above decomposition is based on the assumption that external momenta are allowed to take complex values, which supports the configuration that two
null momenta lie in orthogonal dimensions, in a special inertial frame. For example, in $4$-dimensional spacetime, we can have $k_1^\mu=(p,p,0,0)$, $k_3^\mu=(0,0,q,iq)$, i.e., $k_1$
lies in $0^{\rm th}$ and $1^{\rm th}$ dimensions, while $k_3$ lies in $2^{\rm th}$ and $3^{\rm th}$.

\begin{figure}
  \centering
  \includegraphics[width=14cm]{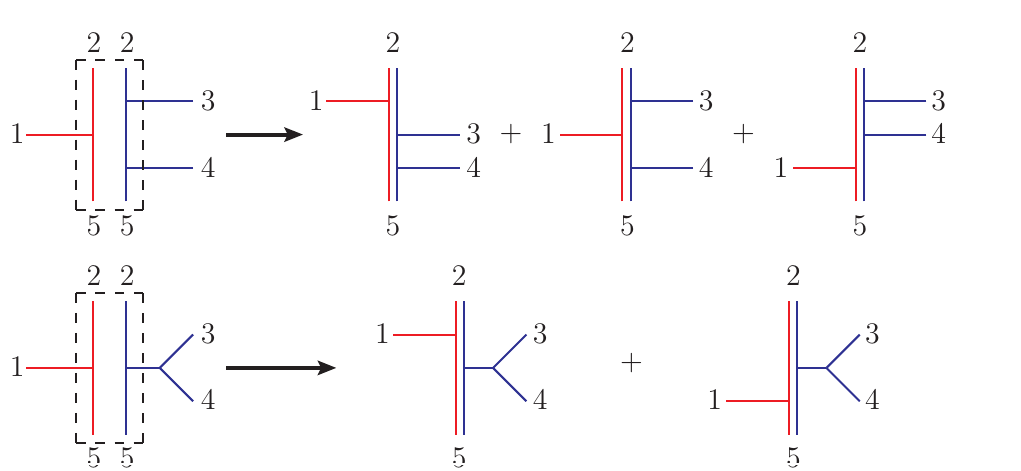} \\
  \caption{Generating fives diagrams in Fig.\ref{5p01}, by merging pairs of diagrams in two orthogonal complementary spaces ${\cal S}_1$ and ${\cal S}_2$. The red lines carry momenta in ${\cal S}_1$, while the blue ones carry momenta in ${\cal S}_2$.}\label{5p02}
\end{figure}

Then we consider the reduction of the $5$-point Feynman diagrams in Fig.\ref{5p01} to subspaces ${\cal S}_1$ and ${\cal S}_2$. For an observer in ${\cal S}_1$, only the $3$-point red Feynman diagram in Fig.\ref{5p02}, with external momenta $k_1$, $k_2^{{\cal S}_1}$ and $k_5^{{\cal S}_1}$, can be observed. Here $k_2^{{\cal S}_1}$ and $k_5^{{\cal S}_1}$ are ${\cal S}_1$ components of $k_2$ and $k_5$, respectively. Similarly, for an observer in ${\cal S}_2$, only the $4$-point blue diagrams ($s$ and $t$ channels), with external momenta $k_2^{{\cal S}_2}$, $k_3$, $k_4$ and $k_5^{{\cal S}_2}$, can be observed. Based on the above observation, one can think the $5$-point amplitude ${\cal A}_5^{\phi^3}(1,2,3,4,5)$ as the resulting object of joining the $3$-point amplitude $A_3^{{\cal S}_1}(1,2,5)$ in ${\cal S}_1$ and the $4$-point $A^{{\cal S}_2}_4(2,3,4,5)$ in ${\cal S}_2$, by gluing lines $L^{{\cal S}_1}_{2,5}$ and $L^{{\cal S}_2}_{2,5}$ which connects external legs $2$ and $5$. In particular, three diagrams in the first line of Fig.\ref{5p01} are obtained by merging the trivalent diagram of $A_3^{{\cal S}_1}(1,2,5)$ and the $s$-channel diagram of $A^{{\cal S}_2}_4(2,3,4,5)$, as shown in the first line of Fig.\ref{5p02}. Analogously, two diagrams in the second line of Fig.\ref{5p01} are obtained by merging the trivalent diagram of $A_3^{{\cal S}_1}(1,2,5)$ and the $t$-channel diagram of $A^{{\cal S}_2}_4(2,3,4,5)$,
as can be seen in the second line of Fig.\ref{5p02}. Two amplitudes $A_3^{{\cal S}_1}(1,2,5)$ and $A^{{\cal S}_2}_4(2,3,4,5)$ are off-shell ones, since the external momenta carried by them are $k_\ell^{{\cal S}_a}$ with $\ell=2,5$ and $a=1,2$, rather than $k_\ell$. Throughout this paper, we use the notation $A$ to denote amplitudes which can be either on-shell or off-shell, to distinguish them from on-shell ones denoted as ${\cal A}$. We have not given the superscript such as $\phi^3$ to $A_3^{{\cal S}_1}(1,2,5)$ and $A^{{\cal S}_2}_4(2,3,4,5)$, since the above diagrammatical analysis does not specify the theory (theories) for them. Notice that in the above merging, the pairs of diagrams in subspaces, and the resulting diagrams generated by gluing lines $L^{{\cal S}_1}_{2,5}$ and $L^{{\cal S}_2}_{2,5}$, are not one-to-one correspondent: the merging in the first line of Fig.\ref{5p02} creates three diagrams, while the merging in the second line creates two.
We can call such phenomenon the summation over all shuffles for lines from ${\cal S}_1$ and ${\cal S}_2$, since in all diagrams in the summation, the ordering among lines from ${\cal S}_1$ (${\cal S}_1$) is kept. For example, in the first line of Fig.\ref{5p02}, we say the summation is over shuffles for the external line $1$ from ${\cal S}_1$ and external lines $3$ and $4$ from ${\cal S}_2$, since the ordering among lines $3$ and $4$ is kept. In the second line of Fig.\ref{5p01}, the line from ${\cal S}_1$ which is connected to $L_{2,5}$ is an internal one, thus we say the summation is over shuffles for the external line $1$ and this internal line.

It is natural to assume that any scattering in extra dimensions is insensible for an observer who lives in ordinary dimensions, therefore, two scattering processes in ${\cal S}_1$ and ${\cal S}_2$ are independent of each other. Based on this assumption, we can understand zeros in \eref{5p0-1st} and \eref{5p0-2nd} in two alternative ways. The first one is as follows. In the first line of Fig.\ref{5p02}, the $3$-point diagram of $A_3^{{\cal S}_1}(1,2,5)$, and the $4$-point $s$ channel diagram of $A^{{\cal S}_2}_4(2,3,4,5)$, are two independent diagrams, since they are localized in ${\cal S}_1$ and ${\cal S}_2$ respectively. These two independent diagrams do not share any common vertex in the merging, thus the merging does not create any $5$-point connected diagram. Since amplitudes are defined for connected diagrams, the contribution to the $5$-point amplitude, arises from the above merging, must vanish. This is the reason for the zero in \eref{5p0-1st}. The zero in \eref{5p0-2nd} can be understood analogously.

\begin{figure}
  \centering
  \includegraphics[width=11cm]{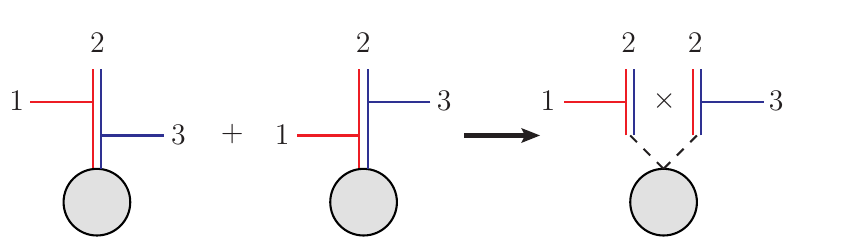} \\
  \caption{The summation of two diagrams at the l.h.s of $\to$ can be understood as the merging of two propagators in ${\cal S}_1$ and ${\cal S}_2$ respectively. At the r.h.s of $\to$, the left propagator is in ${\cal S}_1$ while the right one is in ${\cal S}_2$. But we still use the double-lines to draw them. For the left one, the blue line represents the effective mass. Similarly, for the right one, the red line represents the effective mass.}\label{fac-2pro}
\end{figure}
\begin{figure}
  \centering
  \includegraphics[width=14cm]{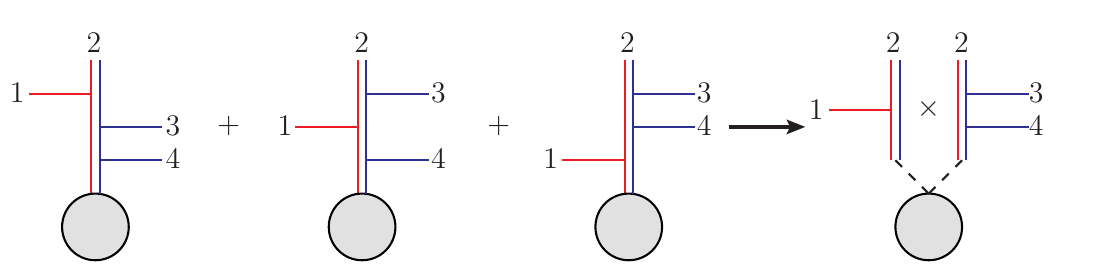} \\
  \caption{Interpreting the summation of three diagrams, which correspond to terms at the l.h.s in the second line of \eref{5p-facpropa}, as the merging of propagators in ${\cal S}_1$ and ${\cal S}_2$.}\label{fac-3pro}
\end{figure}

The second understanding focus on explicit expressions. We first observe that, if the external leg $5$ in first lines of Fig.\ref{5p01} and Fig.\ref{5p02} are replaced by the internal virtual massless particle which carries the off-shell momentum $k_{1234}$, then algebraic relations in \eref{5p0-1st} read
\bea
{1\over s_{12}}\,{1\over s_{123}}+{1\over s_{23}}\,{1\over s_{123}}\,&\xrightarrow[]{k_1\in{\cal S}_1\,,~k_3\in{\cal S}_2}&\,
{1\over s_{12}}\,\times\,{1\over s_{23}}\,,\nn
{1\over s_{12}}\,{1\over s_{123}}\,{1\over s_{1234}}+{1\over s_{23}}\,{1\over s_{123}}\,{1\over s_{1234}}+{1\over s_{23}}\,{1\over s_{234}}\,{1\over s_{1234}}\,&\xrightarrow[]{k_1\in{\cal S}_1\,,~k_{3,4}\in{\cal S}_2}&\,
{1\over s_{12}}\,\times\,\Big({1\over s_{23}}\,{1\over s_{234}}\Big)\,,~~\label{5p-facpropa}
\eea
as diagrammatically represented in Fig.\ref{fac-2pro} and Fig.\ref{fac-3pro}. The above factorization of propagators in the first line of \eref{5p-facpropa}, can be understood as, the summation of two diagrams at the l.h.s of $\to$ in Fig.\ref{fac-2pro} can be regarded as the merging of one propagator in ${\cal S}_1$
and another one in ${\cal S}_2$, as shown at the r.h.s of $\to$, due to the independence of scattering processes in ${\cal S}_1$ and ${\cal S}_2$. More explicitly, the propagator $1/s_{12}$ is reinterpreted as the one in ${\cal S}_1$, while $1/s_{23}$ is reinterpreted as another one in ${\cal S}_2$
\bea
{1\over \big(k_1+k_2^{{\cal S}_1}\big)^2+\big(k_2^{{\cal S}_2}\big)^2}&=&{1\over s_{12}}\,,\nn
{1\over \big(k_3+k_2^{{\cal S}_2}\big)^2+\big(k_2^{{\cal S}_1}\big)^2}&=&{1\over s_{23}}\,.~~\label{pro-S12}
\eea
In the above formula, we see that from the ${\cal S}_1$ point of view, the virtual particle has the effective mass
\bea
\big(m^{{\cal S}_1}\big)^2=-\big(k_2^{{\cal S}_2}\big)^2\,,~~\label{5p-mass1}
\eea
arises from the component $k_2^{{\cal S}_2}$ in extra dimensions. Similarly, for an observer in ${\cal S}_2$, the virtual particle acquires the effective mass
\bea
\big(m^{{\cal S}_2}\big)^2=-\big(k_2^{{\cal S}_1}\big)^2\,,~~\label{5p-mass2}
\eea
from the component $k_2^{{\cal S}_1}$ in extra dimensions. This is the reason why we use the double-line (one in red and another one in blue)
to denote the propagator in ${\cal S}_1$ or ${\cal S}_2$ at the r.h.s of $\to$, i.e., one of two lines is used to represent the effective mass.
The factorization of propagators in the second line of \eref{5p-facpropa} and Fig.\ref{fac-3pro}, can be understood analogously.
We can interpret the zero in \eref{5p0-1st} as the consequence of the factorization of propagators in Fig.\ref{fac-3pro}, since it indicates
\bea
{1\over s_{12}}\,{1\over s_{123}}+{1\over s_{23}}\,{1\over s_{123}}+{1\over s_{23}}\,{1\over s_{234}}\,&\xrightarrow[]{k_1\in{\cal S}_1\,,~k_{3,4}\in{\cal S}_2}&\,
s_{1234}\,\Big({1\over s_{12}}\,{1\over s_{23}}\,{1\over s_{234}}\Big)\,,~~\label{inter-5p0-1}
\eea
vanishing when $s_{1234}=0$.

\begin{figure}
  \centering
  \includegraphics[width=11cm]{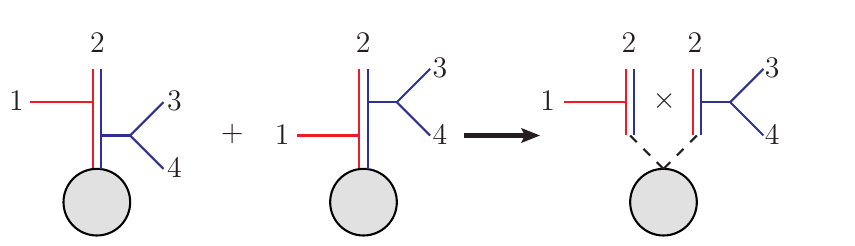} \\
  \caption{Interpreting the summation of two diagrams, which correspond to terms at the l.h.s of \eref{5p-fac-2pros}, as the merging of propagators in ${\cal S}_1$ and ${\cal S}_2$.}\label{fac-2pros}
\end{figure}

Clearly, the zero in \eref{5p0-2nd} can be understood in the similar way, as illustrated in Fig.\ref{fac-2pros} which implies
\bea
{1\over s_{12}}+{1\over s_{234}}\,\xrightarrow[]{k_1\in{\cal S}_1\,,~k_{3,4}\in{\cal S}_2}\,s_{1234}\,\Big({1\over s_{12}}\,{1\over s_{234}}\Big)\,&\xrightarrow[]{s_{1234}=0}&\,0\,.~~\label{5p-fac-2pros}
\eea

The factorizations of propagators, such as those in Fig.\ref{fac-2pro}, Fig.\ref{fac-3pro} and Fig.\ref{fac-2pros}, also play the crucial role when considering the splitting, as will be seen immediately. In subsection \ref{subsec-verification-phi3}, we will extend such factorizations to the more general case, with the arbitrary number of external legs.

Until now, we have explained that the amplitude ${\cal A}^{\phi^3}_5(1,2,3,4,5)$ tends zero if $k_1$ lies in ${\cal S}_1$, while $k_3$ and $k_4$ lie in ${\cal S}_2$. The amplitude solely depends on Mandelstam variables, thus can never distinguish if the locus $s_{13}=s_{14}=0$ in kinematic space is caused by the orthogonality of spaces or other reasons. Therefore, the amplitude vanishes under the constraint $s_{13}=s_{14}=0$, without respecting the picture of orthogonal spaces.

\begin{figure}
  \centering
  \includegraphics[width=9cm]{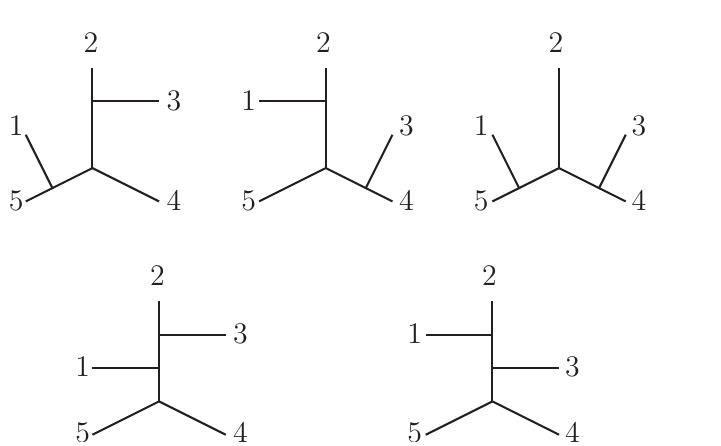} \\
  \caption{Five Feynman diagrams for \eref{5pfac-1}.}\label{5pfac1}
\end{figure}
%

\subsection{The $5$-point example of $2$-split}
\label{subsec-5pfac}

In this subsection, we use a simple example of $2$-split of the same $5$-point amplitude ${\cal A}_5^{\phi^3}(1,2,3,4,5)$, to demonstrate the idea for understanding splittings.

This time we arrange Feynman diagrams as in Fig.\ref{5pfac1}, and order five terms of the amplitude as
\bea
{\cal A}_5^{\phi^3}(1,2,3,4,5)&=&{1\over s_{51}}\,{1\over s_{23}}+{1\over s_{12}}\,{1\over s_{34}}+{1\over s_{51}}\,{1\over s_{34}}\nn
& &+{1\over s_{23}}\,{1\over s_{123}}+{1\over s_{12}}\,{1\over s_{123}}\,.~~\label{5pfac-1}
\eea
Three terms in the first line of \eref{5pfac-1} correspond to three diagrams in the first line of Fig.\ref{5pfac1}, while two terms in the second line of \eref{5pfac-1} correspond to two diagrams in the second line. Setting $s_{13}=0$, the second line of \eref{5pfac-1}
becomes
\bea
{1\over s_{23}}\,{1\over s_{123}}+{1\over s_{12}}\,{1\over s_{123}}\,&\xrightarrow[]{s_{13}=0}&\,{1\over s_{12}}\,{1\over s_{23}}\,,~~\label{2nd-turnto}
\eea
therefore leads to the following splitting of \eref{5pfac-1},
\bea
{\cal A}_5^{\phi^3}(1,2,3,4,5)\,&\xrightarrow[]{s_{13}=0}&\,{\cal J}^{\phi^3}_4(5,1,2,4^*)\,\times\,{\cal J}^{\phi^3}_4(5^*,2,3,4)\,,~~\label{5pfac-2}
\eea
where
\bea
{\cal J}^{\phi^3}_4(5,1,2,4^*)={1\over s_{51}}+{1\over s_{12}}\,,~~\label{5p-JL}
\eea
and
\bea
{\cal J}^{\phi^3}_4(5^*,2,3,4)={1\over s_{23}}+{1\over s_{34}}\,.~~\label{5p-JR}
\eea
In the above, ${\cal J}^{\phi^3}_4(5,1,2,4^*)$ is a current of the ${\rm Tr}(\phi^3)$ theory, in which the external leg $4^*$
carries the off-shell momentum, where the superscript $*$ is used to denote the off-shell leg. Similarly, ${\cal J}^{\phi^3}_4(5^*,2,3,4)$ is also a current of the ${\rm Tr}(\phi^3)$ theory,
with an off-shell external leg $5^*$. One can also understand these currents by choosing other external legs to be off-shell. For example,
we can interpret the current in \eref{5p-JR} as ${\cal J}^{\phi^3}_4(5,2,3,4^*)$, where the off-shell external leg is $4^*$. The momentum conservation
forces $k_4^*$ to be $k_4^*=k_4+k_1$, thus the propagator $1/s_{34}$ should be understood as
\bea
{1\over s_{34}}={1\over \big(k_3+k_4^*\big)^2-\big(k^*_4\big)^2}\,,~~\label{additonalmass}
\eea
where the assumption $s_{13}=0$ has been used. In other words, the virtual particle along the propagator $1/s_{34}$ acquires the effective mass satisfying $m^2=\big(k^*_4\big)^2$.

\begin{figure}
  \centering
  \includegraphics[width=13cm]{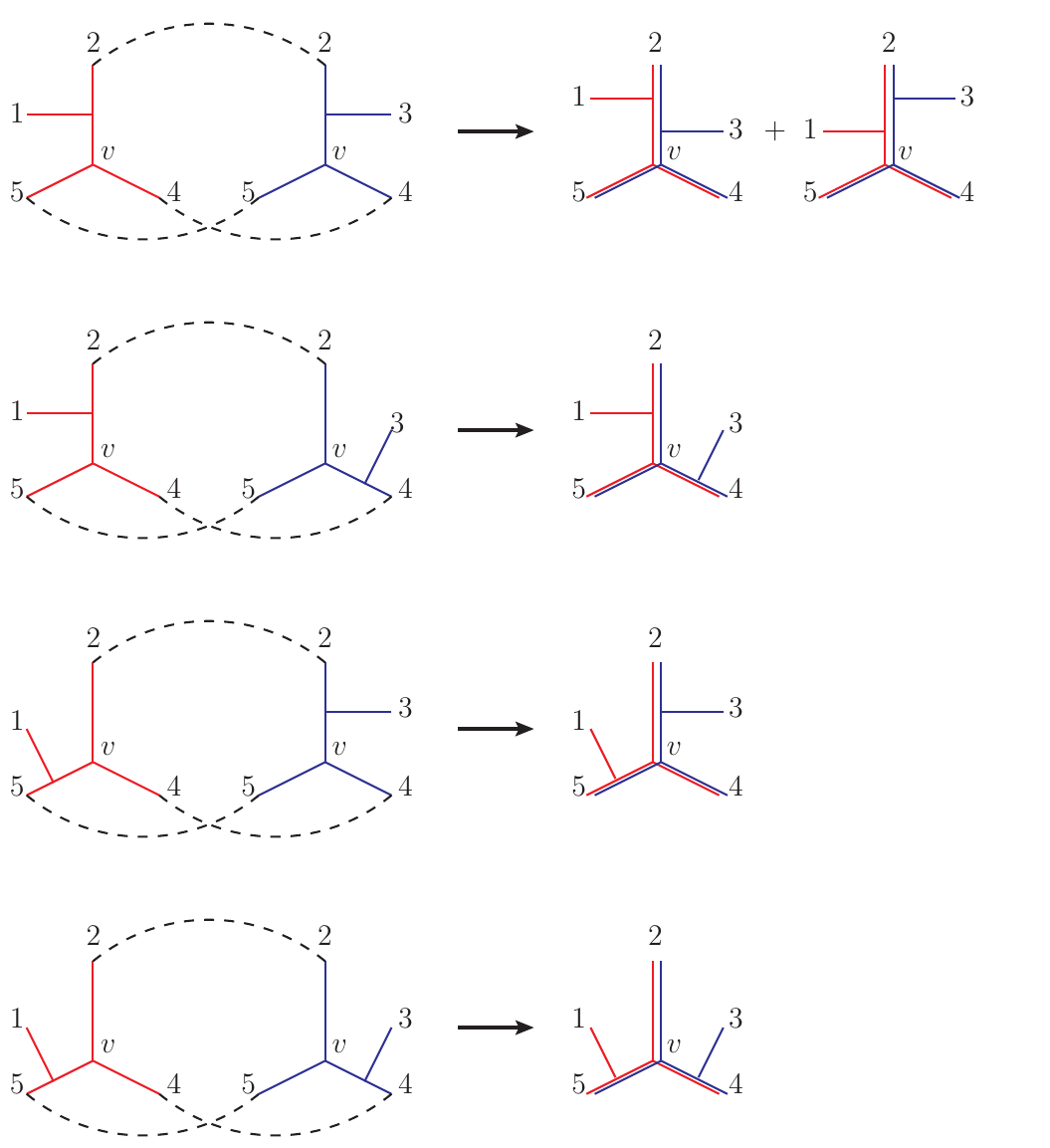} \\
  \caption{Generating five diagrams in Fig.\ref{5pfac1}, by merging $4$-point cubic diagrams of $A_4(5,1,2,4)$ and $A_4(5,2,3,4)$.}\label{5pfac2}
\end{figure}

The splitting in \eref{5pfac-2} can also be understood by using the idea of orthogonal spaces. Let us localize $k_1$ and $k_3$ as $k_1\in{\cal S}_1$, $k_3\in{\cal S}_2$, which serves as a special case of the condition $s_{13}=0$. Then five diagrams in Fig.\ref{5pfac1} can be thought as the merging of $4$-point cubic diagrams of $A^{{\cal S}_1}_4(1,2,4,5)$ and $A^{{\cal S}_2}_4(2,3,4,5)$, as illustrated in Fig.\ref{5pfac2}. Here $A^{{\cal S}_1}_4(1,2,4,5)$ is the off-shell amplitude in ${\cal S}_1$, carries external momenta $k_1$, $k_2^{{\cal S}_1}$, $k_4^{{\cal S}_1}$ and $k_5^{{\cal S}_1}$. Analogously, $A^{{\cal S}_2}_4(2,3,4,5)$ is the off-shell amplitude in ${\cal S}_2$, with external momenta $k_2^{{\cal S}_2}$, $k_3$, $k_4^{{\cal S}_2}$, $k_5^{{\cal S}_2}$. More explicitly, the first line of Fig.\ref{5pfac2} is the merging of two $s$-channel diagrams of $A^{{\cal S}_1}_4(1,2,4,5)$ and $A^{{\cal S}_2}_4(2,3,4,5)$, creates two diagrams correspond to those in the second line of Fig.\ref{5pfac1}. The remaining three lines of Fig.\ref{5pfac2} create three diagrams in the first line of Fig.\ref{5pfac1}, respectively. Therefore, the amplitude ${\cal A}_5^{\phi^3}(1,2,3,4,5)$ can be regarded as the merging of $A^{{\cal S}_1}_4(1,2,4,5)$ and $A^{{\cal S}_2}_4(2,3,4,5)$ by gluing three pairs of lines $(L^{{\cal S}_1}_{\ell,v},L^{{\cal S}_2}_{\ell,v})$ with $\ell=2,4,5$. In this merging, $A^{{\cal S}_1}_4(1,2,4,5)$ and $A^{{\cal S}_2}_4(2,3,4,5)$ share a common vertex $v$, thus the $5$-point amplitude does not vanish since the resulting diagrams are connected ones.
Based on the assumption that scattering processes in orthogonal spaces are independent, one can conclude that the $5$-point amplitude factorizes as
\bea
{\cal A}_5^{\phi^3}(1,2,3,4,5)\,&\xrightarrow[]{k_1\in{\cal S}_1\,,\,k_3\in{\cal S}_2}&\,A^{{\cal S}_1}_4(1,2,4,5)\,\times\,A^{{\cal S}_2}_4(2,3,4,5)\,.~~\label{5p-split1}
\eea

The above diagrammatical discussion cannot specify the corresponding theory (theories) of resulting amplitudes $A^{{\cal S}_1}_4(1,2,4,5)$ and $A^{{\cal S}_2}_4(2,3,4,5)$.
However, it predicts that the $2$-split occurs when $s_{13}=0$, since the amplitude only depends on Mandelstam variables without distinguishing the mechanism for causing $s_{13}=0$.

To determine the theory (theories) for amplitudes $A^{{\cal S}_1}_4(1,2,4,5)$ and $A^{{\cal S}_2}_4(2,3,4,5)$, one need to use the explicit formulas in \eref{5pfac-2}, \eref{5p-JL} and \eref{5p-JR}, which shows that $A^{{\cal S}_1}_4(1,2,4,5)$ and $A^{{\cal S}_2}_4(2,3,4,5)$ in orthogonal subspaces are equivalent to ${\rm Tr}(\phi^3)$ currents in the full spacetime ${\cal S}$, namely,
\bea
& &A^{{\cal S}_1}_4(1,2,4,5)\,\sim\,{\cal J}^{\phi^3}_4(5,1,2,4^*)\,,\nn
& &A^{{\cal S}_2}_4(2,3,4,5)\,\sim\,{\cal J}^{\phi^3}_4(5^*,2,3,4)\,.~~\label{equivalence}
\eea
The above equivalences can be understood as follows.
In the previous derivation of the $2$-split \eref{5pfac-2}, the key step is the relation \eref{2nd-turnto}, which can be recognized as the factorization of propagators in Fig.\ref{fac-2pro} in the orthogonal spaces context.
Thus propagators $1/s_{12}$ and $1/s_{23}$ can be reinterpreted as in \eref{5p-facpropa}, which means the observer in ${\cal S}_1$ (${\cal S}_2$) will regard the virtual particle along $L_{2,v}^{{\cal S}_1}$ ($L_{2,v}^{{\cal S}_2}$) as a massive one, with the mass $m_2^{{\cal S}_1}$ ($m_2^{{\cal S}_2}$)
from the momentum $k_2^{{\cal S}_2}$ ($k_2^{{\cal S}_1}$) in extra dimensions.
Similarly, for an observer in ${\cal S}_1$, the virtual particle propagates in $L_{5,v}^{{\cal S}_1}$ acquires the mass satisfying
\bea
\big(m_5^{{\cal S}_1}\big)^2=-\big(k_5^{{\cal S}_2}\big)^2\,,~~\label{5p-mass5}
\eea
while for an observer in ${\cal S}_2$, the virtual particle along $L_{4,v}^{{\cal S}_2}$ has the mass satisfying
\bea
\big(m_4^{{\cal S}_2}\big)^2=-\big(k_4^{{\cal S}_1}\big)^2\,.~~\label{5p-mass4}
\eea
The above masses $m_2^{{\cal S}_1}$, $m_2^{{\cal S}_2}$, $m_4^{{\cal S}_2}$ and $m_5^{{\cal S}_1}$, ensure the equivalences in \eref{equivalence} between off-shell amplitudes in subspaces and ${\rm Tr}(\phi^3)$ currents in the full spacetime, since they indicates the equivalence between propagators in subspaces and propagators in the full spacetime, as can be seen in \eref{pro-S12}.

Before ending this subsection, we emphasize that each merging considered in this subsection is universal for all pairs of Feynman diagrams: in Fig.\ref{5p02}, all pairs of diagrams are merged by gluing lines $L_{2,5}^{{\cal S}_1}$ and $L_{2,5}^{{\cal S}_2}$, and in Fig.\ref{5pfac2}, all pairs of diagrams are merged by gluing pairs of lines $(L^{{\cal S}_1}_{\ell,v},L^{{\cal S}_2}_{\ell,v})$ with $\ell=2,4,5$, and the vertex $v$. The importance of such universality will be explained in the next subsection.

\subsection{Merging amplitudes from orthogonal complementary spaces}
\label{subsec-idea}

In previous subsections, we have seen that the merging of two amplitudes from two orthogonal complementary spaces is helpful for understanding zeros and the splittings. We now give a general discussion about such merging. The main part of this subsection focus on splittings of amplitudes. Zeros arise as a special case, when the corresponding mergings can be called "spurious" ones.

Suppose an $n$-point amplitude $A_n$ has the following specific behavior. For an observer in ${\cal S}_1$ who can not sense the scattering in ${\cal S}_2$, $A_n$ is reduced to an $n_1$-point amplitude $A^{{\cal S}_1}_{n_1}$,
with $n_1\leq n$; for an observer in ${\cal S}_2$, it is reduced to an $n_2$-point one $A^{{\cal S}_2}_{n_2}$, with $n_2\leq n$. Notice that all of above amplitudes are allowed to be off-shell, thus we use the notation $A$. As discussed in previous subsections \ref{subsec-5p0} and \ref{subsec-5pfac}, we assume that two scattering processes in two orthogonal spaces are independent of each other. Such independence guarantees the factorization
\bea
A_n\to A^{{\cal S}_1}_{n_1}\,\times\,A^{{\cal S}_2}_{n_2}\,.~~\label{split-general}
\eea
Thus, we are interested in the splitting that the full amplitude is factorized into two sub-amplitudes in orthogonal complementary spaces.
To find such splitting, it is convenient to study the opposite direction: we consider whether two amplitudes $A^{{\cal S}_1}_{n_1}$ and $A^{{\cal S}_2}_{n_2}$ in orthogonal spaces can be merged together without adding any new line or new vertex, to obtain a bigger $D$-dimensional amplitude. If such merging can be realized, then the reversing of merging accesses the splitting of the bigger amplitude into $A^{{\cal S}_1}_{n_1}$ and $A^{{\cal S}_2}_{n_2}$.

\begin{figure}
  \centering
  \includegraphics[width=8cm]{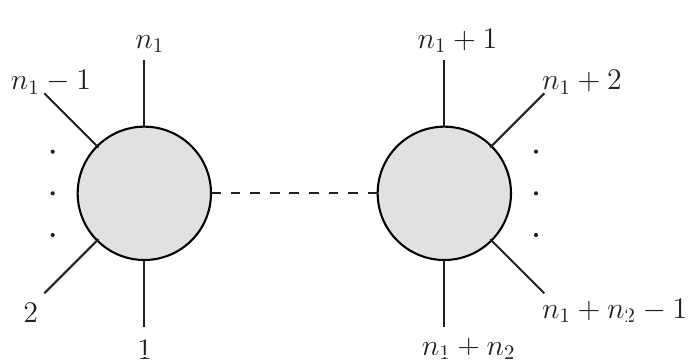} \\
  \caption{In order to combine an $n_1$-point diagram and an $n_2$-point diagram to an $(n_1+n_2)$-point one, an additional propagator labeled by the dashed line must be added.}\label{n1+n2}
\end{figure}

Without the additional line, two amplitudes $A^{{\cal S}_1}_{n_1}$ and $A^{{\cal S}_2}_{n_2}$ can never be combined to a full $(n_1+n_2)$-point
amplitude $A_{n_1+n_2}$, as diagrammatically illustrated in Fig.\ref{n1+n2}. Thus, if the desired merging exist, the resulting amplitude $A_n$ should satisfy $n<n_1+n_2$.
It means $A^{{\cal S}_1}_{n_1}$ and $A^{{\cal S}_2}_{n_2}$ should share some external legs. Thus, one need to choose $m$ pairs of external legs.
In each pair, one leg is from $A^{{\cal S}_1}_{n_1}$, and another one is from $A^{{\cal S}_2}_{n_2}$. Then each pair of lines is glued to form a new external line, doing this for all pairs gives rise to $n=n_1+n_2-m$. A new external leg, arises from merging a pair of legs,
carries momentum in the full $D$-dimensional spacetime, where the component in ${\cal S}_1$ is contributed by the external momentum of $A^{{\cal S}_1}_{n_1}$,
while component in ${\cal S}_2$ is contributed by the external momentum of $A^{{\cal S}_2}_{n_2}$. For convenience, we will give the same label for legs in each pair, as well as the associated new leg created by gluing. For instance, consider an amplitude $A^{{\cal S}_1}_4(1,2,3,4)$ in ${\cal S}_1$,
and another one $A^{{\cal S}_2}_3(5,6,1)$ in ${\cal S}_2$. The labels of external legs imply that we have chosen only one pair $(1,1)$. Gluing two legs in this pair together leads to the new external line $1$, with $k_1=k_1^{{\cal S}_1}+k_1^{{\cal S}_2}$, where $k_1^{{\cal S}_1}$ is the external momentum carried by
$A_4(1,2,3,4)$, and $k_1^{{\cal S}_2}$ is the external momentum carried by
$A_3(5,6,1)$. Notice that in the above example we have not required that the merging of $A^{{\cal S}_1}_4(1,2,3,4)$
and $A^{{\cal S}_2}_3(5,6,1)$ gives rise to a physically acceptable $A_6(1,2,3,4,5,6)$.

The effective merging which indicates the splitting of resulting amplitude, should also create the interaction between two pieces in ${\cal S}_1$ and ${\cal S}_2$. Otherwise the resulting object can not be understood as a bigger amplitude, since each amplitude is defined for connected diagrams. It means $A^{{\cal S}_1}_{n_1}$ and $A^{{\cal S}_2}_{n_2}$ should also share at least one vertex in the effective merging. Therefore, besides gluing external legs, one need to glue at least one pair of vertices, one vertex is from $A^{{\cal S}_1}_{n_1}$, and another one is from $A^{{\cal S}_2}_{n_2}$. A "spurious" merging without sharing any vertex, corresponds to the zero of the amplitude in ${\cal S}$, as can be seen in the example in subsection \ref{subsec-5p0}, and will be further discussed in subsection \ref{subsec-zero-phi3}.

\begin{figure}
  \centering
  \includegraphics[width=4cm]{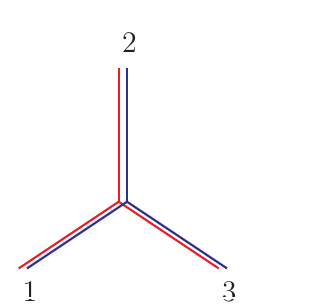}
   \includegraphics[width=5cm]{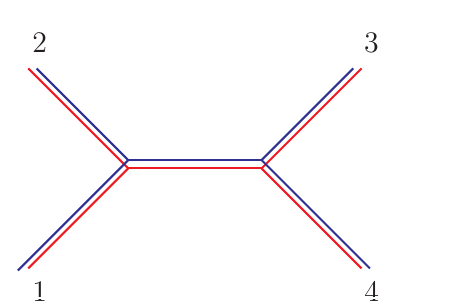}  \\
  \caption{Gluing diagrams from ${\cal S}_1$ and ${\cal S}_2$. The red lines carry momenta in ${\cal S}_1$, while blue lines carry momenta in ${\cal S}_2$.}\label{gluing}
\end{figure}

\begin{figure}
  \centering
  \includegraphics[width=9cm]{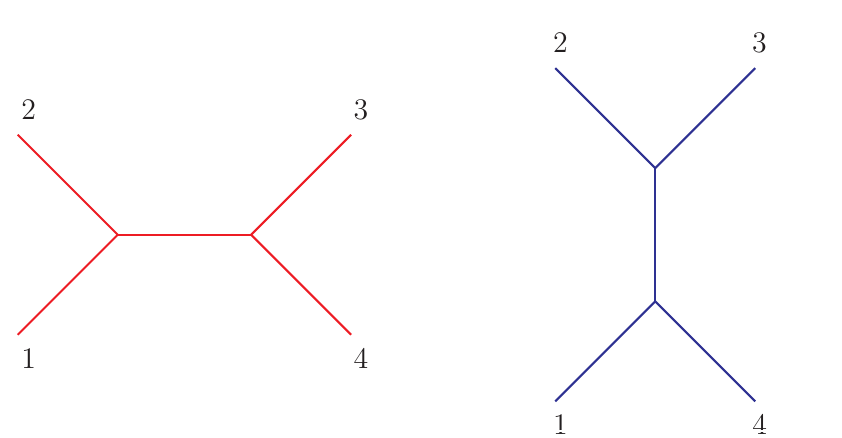} \\
  \caption{Two diagrams from ${\cal S}_1$ and ${\cal S}_2$ which can not be glued together. The red lines carry momenta in ${\cal S}_1$, while blue lines carry momenta in ${\cal S}_2$.}\label{1234b}
\end{figure}

Before seeking systematic mergings in subsequent subsections, let us consider two very simple intuitive examples, to gain some general insights about the allowed mergings.
The first one is $A^{{\cal S}_1}_3(1,2,3)=1$ in ${\cal S}_1$ and $A^{{\cal S}_2}_3(1,2,3)=1$ in ${\cal S}_2$, where three pairs of external legs are chosen as $(1,1)$, $(2,2)$, $(3,3)$.
Gluing external legs and the trivalent vertices gives
\bea
A_3(1,2,3)=A^{{\cal S}_1}_3(1,2,3)\times A^{{\cal S}_2}_3(1,2,3)=1
\eea
in the full space ${\cal S}$. The corresponding Feynman diagram is given in the left graph of
Fig.\ref{gluing}. The second example is $A^{{\cal S}_1}_4(1,2,3,4)$ and $A^{{\cal S}_2}_4(1,2,3,4)$, one can never merge the corresponding diagrams together via an universal scheme, due to
the difference between two channels $s$ and $t$. As can be seen in the right graph in Fig.\ref{gluing}, when both two processes in ${\cal S}_1$ and ${\cal S}_2$ are $s$-channel ones, one can glue the corresponding Feynman diagrams together to get the $s$-channel diagram in ${\cal S}$. However, when diagrams in ${\cal S}_1$ and ${\cal S}_2$ correspond to different channels, as shown in Fig.\ref{1234b}, gluing lines in four pairs $(1,1)$, $(2,2)$, $(3,3)$ and $(4,4)$ does not give rise to any Feynman diagram for the scattering in ${\cal S}$. In other words, the scheme in the right graph of Fig.\ref{gluing} does not make sense for
the situation in Fig.\ref{1234b}. The above two examples tell us that, an allowed merging should be an universal one, which is valid for all Feynman diagrams under consideration. In subsequent subsections, we will follow this requirement to seek appropriate mergings.

From the above discussion, we know that if the scheme of merging is universal, then the merging/splitting of individual diagrams can be extended to the merging/spitting of the whole amplitudes. Here a subtle point is, when merging two diagrams from ${\cal S}_1$ and ${\cal S}_2$ by gluing lines, one should sum over all shuffles for lines from ${\cal S}_1$ and ${\cal S}_2$. An example is the summations in Fig.\ref{5p02}, as explained in subsection \ref{subsec-5p0}.
After summing over shuffles, the summation over Feynman diagrams for $A^{{\cal S}_1}_{n_1}$ and $A^{{\cal S}_2}_{n_2}$ automatically yields the summation over Feynman diagrams for $A_n$, and vice versa. Notice that since the merging under consideration does not add any new vertex or propagator, it does not affect the scattering process in ${\cal S}_1$ or ${\cal S}_2$, thus the amplitude in ${\cal S}_1$ or ${\cal S}_2$ will not be altered. The above discussion indicates that the splitting in \eref{split-general} can be ensured by the universal merging for individual diagrams. On the other hand, since the merging is totally diagrammatical, without exploiting explicit Feynman rules, it fails to answer the corresponding theory (theories) of resulting $A^{{\cal S}_1}_{n_1}$ and $A^{{\cal S}_2}_{n_2}$ in the splitting.
For instance, the merging in Fig.\ref{5pfac2} predicts the splitting in \eref{5p-split1}, but can not give the interpretation of resulting currents.

There are various natural generalizations of the above idea. Firstly, one can decompose the full spacetime into more orthogonal subspaces, and split the full amplitude into more pieces. We will consider the decomposition ${\cal S}={\cal S}_1\oplus{\cal S}_2\oplus{\cal S}_3$ in subsection \ref{subsec-fac-zero}, which leads to the smooth splitting that the full amplitude factorizes into $3$ currents, as discovered in \cite{Cachazo:2021wsz}. Secondly, we will also extend this idea to YM and NLSM cases in sections \ref{sec-YM} and \ref{sec-NLSM}.

\subsection{Zeros of tree ${\rm Tr}(\phi^3)$ amplitudes}
\label{subsec-zero-phi3}

Based on the requirement found in section \ref{subsec-idea}, in this subsection we will find the first type of universal (but spurious) mergings, which leads to zeros of amplitudes. In subsection \ref{subsec-5p0}, we gave two alternative interpretation for the zero, one is based on the assumption that scattering processes in orthogonal spaces are independent of each other, while another one utilizes the factorizations of propagators. In this subsection, the first interpretation will be generalized to the general case with the arbitrary number of external legs. The second one will be extended to the general case in subsection \ref{subsec-verification-phi3}.

\begin{figure}
  \centering
  \includegraphics[width=9cm]{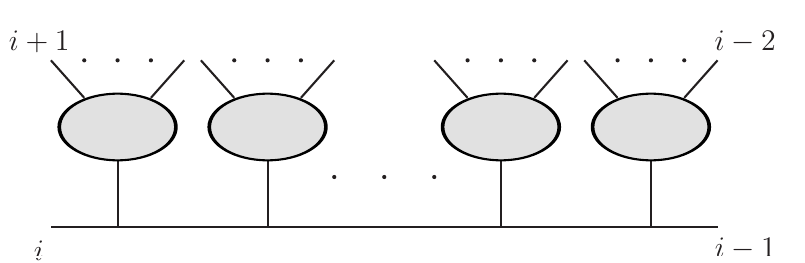} \\
  \caption{First common structure of tree level ordered Feynman diagrams with cubic interaction.}\label{line}
\end{figure}

To find an universal merging which is valid for any Feynman diagram, it is natural to use the common feature of diagrams. In this subsection, we utilize the observation that any ordered diagram at the tree level, with cubic vertices, can be thought as planting trees to the line from one external leg $i-1$ to another one $i$, as shown in Fig.\ref{line}. As in subsection \ref{subsec-5p0}, we denote the line connects external legs $i$ and $j$ as $L_{i,j}$.

\begin{figure}
  \centering
  \includegraphics[width=11cm]{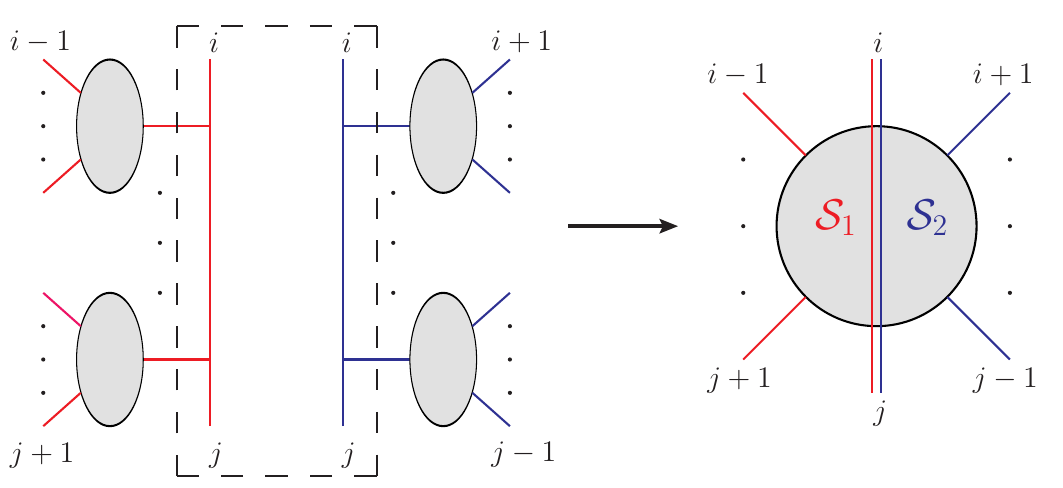} \\
  \caption{Merging two sub-amplitudes by gluing $L^{{\cal S}_1}_{i,j}$ and $L^{{\cal S}_2}_{i,j}$. The red lines carry momenta in ${\cal S}_1$, while blue lines carry momenta in ${\cal S}_2$.}\label{share2}
\end{figure}

Now we consider two ordered tree level amplitudes $A^{{\cal S}_1}_{n_1}(i,j,j+1,\cdots,i-1)$ and $A^{{\cal S}_2}_{n_2}(j,i,i+1,\cdots,j-1)$, with only cubic interactions, localized in ${\cal S}_1$ and ${\cal S}_2$ respectively. Due to the common feature described above, gluing the line $L^{{\cal S}_1}_{i,j}$ for $A^{{\cal S}_1}_{n_1}(i,j,j+1,\cdots,i-1)$ and the line $L^{{\cal S}_2}_{i,j}$ for $A^{{\cal S}_2}_{n_2}(j,i,i+1,\cdots,j-1)$ together makes sense for any pair of diagrams. Thus, one can use this scheme to merge two amplitudes together without ambiguity, as illustrated in Fig.\ref{share2}. To verify if the resulting object includes all topologically possible diagrams in ${\cal S}$, which also involves only cubic vertices, it is sufficient to observe that, in each diagram for the ordered $A_{n_1+n_2-2}(i,i+1,\cdots,j,j+1,\cdots,i-1)$, one can always find a line $L_{i,j}$ connects legs $i$ and $j$, and think this diagram as a planner one obtained by planting trees onto $L_{i,j}$ according to the color ordering. It is worth to emphasize that if one choose to glue lines $L^{{\cal S}_1}_{i',j'}$ and $L^{{\cal S}_2}_{i',j'}$, such that $i'$ and $j'$ are not adjacent to each other in $A^{{\cal S}_1}_{n_1}$ or $A^{{\cal S}_2}_{n_2}$, then the resulting object can not have the well-defined ordering among all external legs.

However, the resulting bigger amplitude $A_{n_1+n_2-2}$ must vanish, since the above merging is the spurious one discussed in subsection \ref{subsec-idea}, which does not include any interaction between two pieces in ${\cal S}_1$ and ${\cal S}_2$. The gluing only gives a new way of thinking two independent amplitudes in two orthogonal spaces. The bigger amplitude among external particles in $\{1,\cdots,n_1+n_2-2\}$, which describe the scattering in the full spacetime ${\cal S}$, has not been actually produced, since amplitudes are defined for connected diagrams. Therefore, such gluing leads to the zero of $A_{n_1+n_2-2}$. It is straightforward to figure out the constraint on Mandelstam variables, corresponding to this zero. Suppose the amplitude carries $n$ external legs in $\{1,\cdots,n\}$, the above merging indicates that external momenta $k_a$ with $a\in\{j+1,\cdots,i-1\}$ lie in ${\cal S}_1$, while $k_b$ with $b\in\{i+1,\cdots,j-1\}$ lie in ${\cal S}_2$, therefore
\bea
k_a\cdot k_b=0\,,~~~~{\rm for}~a\in\{j+1,\cdots,i-1\}\,,~b\in\{i+1,\cdots,j-1\}\,.~~~~\label{kinematic-condi-zero}
\eea
Since a scalar amplitude only depends on Mandelstam variables, this amplitude can not distinguish if the locus \eref{kinematic-condi-zero} in kinematic space is caused by the orthogonality of spaces or other reasons. Thus, we arrive at the zeros discovered in \cite{Arkani-Hamed:2023swr}: an $n$-point tree ${\rm Tr}(\phi^3)$ amplitude ${\cal A}^{\phi^3}_n(1\cdots,n)$ vanish
if the external kinematic variables satisfy \eref{kinematic-condi-zero}, with $i,j\in\{1,\cdots,n\}$ and $i<j-1$.

\begin{figure}
  \centering
  \includegraphics[width=8cm]{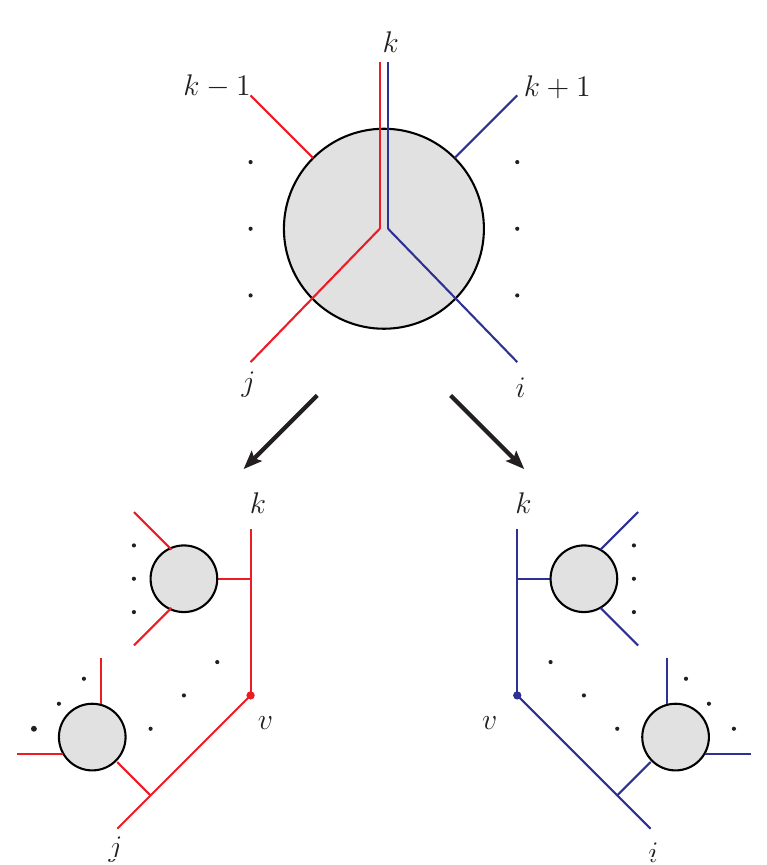} \\
  \caption{Gluing only one pair of legs $(k,k)$. Here $i$ and $j$ are adjacent in the full ordered ${\cal A}_n$. The red lines carry momenta in ${\cal S}_1$, while blue lines carry momenta in ${\cal S}_2$. Thus, $k_a$ with $a\in\{j,\cdots,k-1\}$ are in ${\cal S}_1$, while $k_b$ with $b\in\{k+1,\cdots,i\}$ are in ${\cal S}_2$. Split a Feynman diagram for ${\cal A}_n$ in this way creates two bivalent vertices $v_L$ and $v_R$.}\label{share1}
\end{figure}

The mergings which lead to zeros of $A_n$ glue two pairs of external legs. In the next subsection, we will show that
the mergings which correspond to the $2$-split of $A_n$ glue three pairs. Before ending this subsection, we point out that the merging which glues
only one pair of legs does not make sense. As can be seen in the top graph in Fig.\ref{share1},
such merging need to add a new vertex which connects lines from external legs $i$, $j$ and $k$, thus violates our requirement. If one naively split a Feynman diagram for $A_n$ in this manner, each obtained sub-diagram contains an bivalent vertex which is physically un-acceptable, as can be seen in two bottom graphs in Fig.\ref{share1}.

\subsection{Loci in kinematic space for $2$-splits}
\label{subsec-1to2-phi3}

Now we study a new type of mergings by exploiting another common structure of ordered Feynman diagrams. The new merging leads to the
splitting of an amplitude in ${\cal S}$, which can be called the $2$-split, since the amplitude factorizes into two sub-pieces. In subsection \ref{subsec-5pfac}, we have seen that although the picture of merging can not provide the interpretation of resulting amplitudes, it completely fix the particular loci in kinematic space for splittings. We will generalize this idea to amplitudes with arbitrary numbers of external legs, and determine the loci in kinematic space for $2$-splits. The interpretation of the resulting amplitudes will be given in subsection \ref{subsec-verification-phi3}, by using the factorizations of propagators like in \eref{5p-facpropa}.

\begin{figure}
  \centering
  \includegraphics[width=9cm]{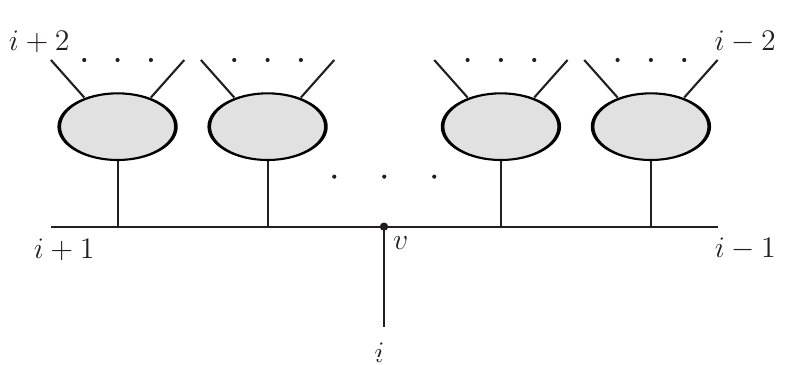} \\
  \caption{Second common structure of tree level ordered Feynman diagrams with cubic interaction.}\label{vertex}
\end{figure}
\begin{figure}
  \centering
  \includegraphics[width=10cm]{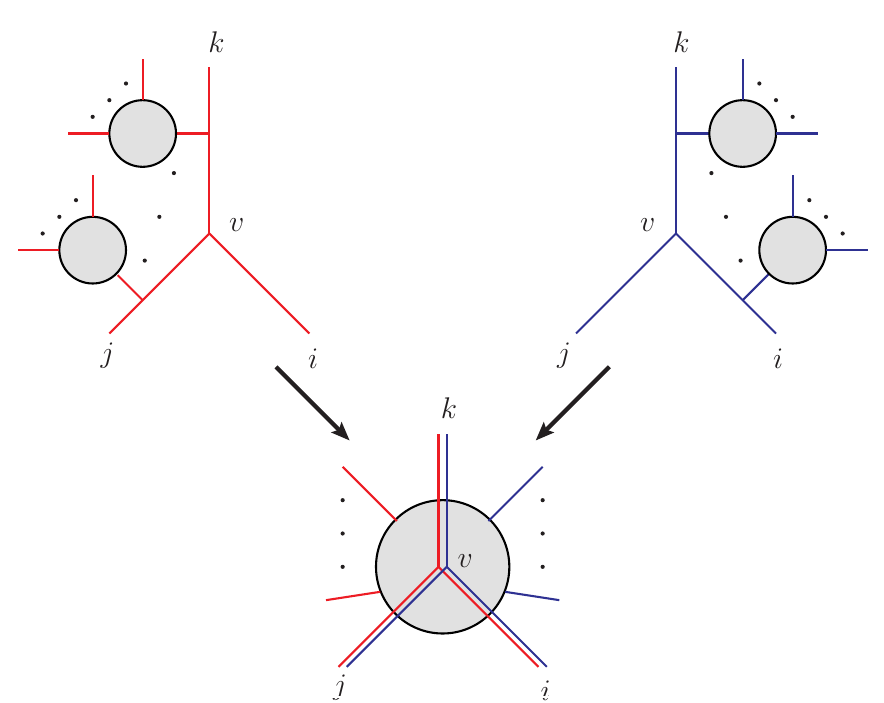} \\
  \caption{Merging two sub-amplitudes by gluing $\big(L^{{\cal S}_1}_{i,v},L^{{\cal S}_2}_{i,v}\big)$, $\big(L^{{\cal S}_1}_{j,v},L^{{\cal S}_2}_{j,v}\big)$ and
$\big(L^{{\cal S}_1}_{k,v},L^{{\cal S}_2}_{k,v}\big)$. The red lines carry momenta in ${\cal S}_1$, while blue lines carry momenta in ${\cal S}_2$.}\label{share3-onshell}
\end{figure}

As illustrated in Fig.\ref{vertex}, any ordered tree level Feynman diagram with cubic vertices can also be thought as that three external legs $i-1$, $i$ and $i+1$ are connected by the trivalent vertex $v$, and other trees are planted onto lines $L_{i-1,v}$ and $L_{i+1,v}$. Using this observation, we
can merge two amplitudes $A^{{\cal S}_1}_{n_1}$ and $A^{{\cal S}_2}_{n_2}$ with only cubic vertices to get the bigger one $A_{n_1+n_2-3}$ also with only cubic interactions, by gluing three pairs of lines $\big(L^{{\cal S}_1}_{i,v},L^{{\cal S}_2}_{i,v}\big)$, $\big(L^{{\cal S}_1}_{j,v},L^{{\cal S}_2}_{j,v}\big)$,
$\big(L^{{\cal S}_1}_{k,v},L^{{\cal S}_2}_{k,v}\big)$, as shown in Fig.\ref{share3-onshell}.
This is an effective merging which indicates the splitting of $A_{n_1+n_2-3}$, rather than a spurious one, since a pair of vertices $(v,v)$ is shared. We first restricted our selves to the case that $i$ and $j$ are two adjacent external legs in the resulting ordered amplitude $A_{n_1+n_2-3}$, namely $j=i+1$. The more general configurations with $i<j$ will be discussed soon.
Then, for the $n$-point ordered tree amplitude $A_n$, with $n=n_1+n_2-3$,
the independence of scattering processes in ${\cal S}_1$ and ${\cal S}_2$ forces the splitting
\bea
A_n(1,\cdots,n)\,&\xrightarrow[a\in\{i+2,\cdots,k-1\}\,,\,b\in\{k+1,\cdots,i-1\}]{k_a\in{\cal S}_1\,,\,k_b\in{\cal S}_2}&\,A^{{\cal S}_1}_{n_1}(i,\cdots,k)\,\times\,A^{{\cal S}_2}_{n+3-n_1}(k,\cdots,i+1)\,,
~~\label{fac-adj-sub}
\eea
which can be understood as the reversing of merging.
External momenta $k_a$ with $a\in\{i+2,\cdots,k-1\}$ lie in ${\cal S}_1$, while $k_b$ with $b\in\{k+1,\cdots,i-1\}$ lie in ${\cal S}_2$, thus the constraints on external kinematics read
\bea
k_a\cdot k_b=0\,,~~~~{\rm for}~a\in\{i+2,\cdots,k-1\}\,,~b\in\{k+1,\cdots,i-1\}\,.~~~~\label{kinematic-condi-1to2}
\eea
Again, this is the condition which leads to the splitting of $A_n$ into $A^{{\cal S}_1}_{n_1}$ and $A^{{\cal S}_2}_{n+3-n_1}$, without distinguishing whether it is caused by the orthogonality of two spaces, since scalar amplitudes solely depend on Mandelstam variables.

\begin{figure}
  \centering
  \includegraphics[width=10cm]{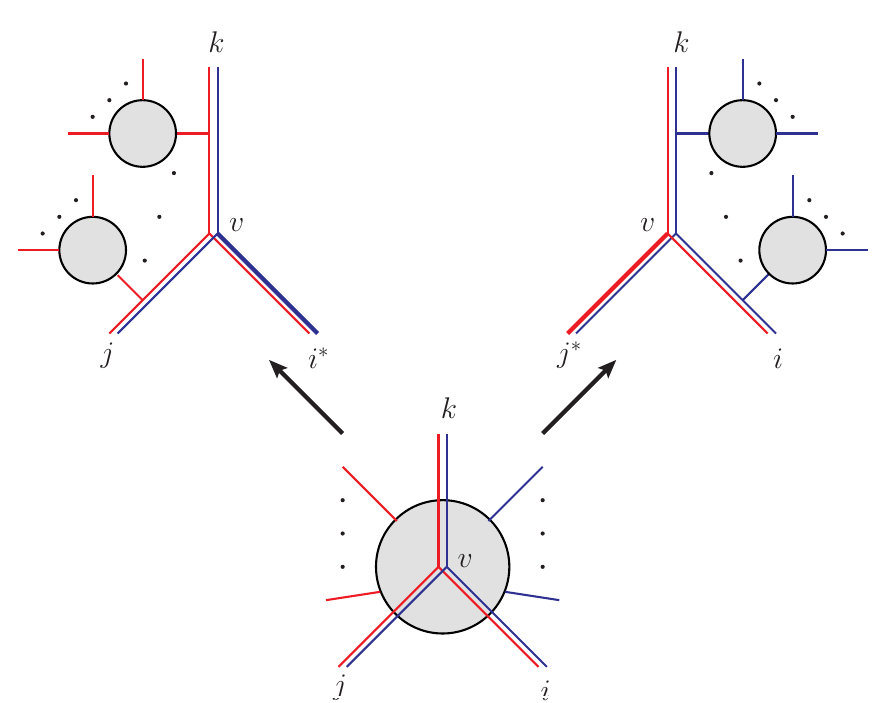} \\
  \caption{Split ${\cal A}_n$ into two currents, when $i$ and $j$ are two adjacent legs. We put the graph for ${\cal A}_n$ at the bottom and put two resulting amplitudes at the top, to highlight that such splitting is determined by inverse the merging in Fig.\ref{share3-onshell}. In the left top diagram, the external momentum $k_i^*=k_i+\sum_{b=k+1}^{i-1}k_b$ is off shell, thus we labeled the leg $i$ as $i^*$. Furthermore, since the ${\cal S}_2$ component receives additional contribution $\sum_{b=k+1}^{i-1}k_b$, we use the bold blue line to represent $(k^*_i)^{{\cal S}_2}$. Similar representation holds for the right diagram.}\label{share3-flow}
\end{figure}

Replacing $A_n$ in \eref{fac-adj-sub} by the ${\rm Tr}(\phi^3)$ amplitude ${\cal A}^{\phi^3}_n$, we arrive at the $2$-split
\bea
{\cal A}^{\phi^3}_n(1,\cdots,n)\,&\xrightarrow[]{\eref{kinematic-condi-1to2}}&\,A^{{\cal S}_1}_{n_1}(i,\cdots,k)\,\times\,A^{{\cal S}_1}_{n+3-n_1}(k,\cdots,i+1)\,,
~~\label{facphi-adj-sub}
\eea
as illustrated in Fig.\ref{share3-flow}. The Fig.\ref{share3-flow} also involves the information for interpreting amplitudes $A^{{\cal S}_1}_{n_1}$ and
$A^{{\cal S}_2}_{n+3-n_1}$, which will be discussed in subsection \ref{subsec-verification-phi3}.
In the $2$-split \eref{facphi-adj-sub}, the external momenta $k_k$, $k_i$ and $k_{i+1}$ carried by the amplitude ${\cal A}^{\phi^3}_n$ are regarded as
\bea
k_\ell=k_\ell^{{\cal S}_1}+k_\ell^{{\cal S}_2}\,,~~~~{\rm with}~\ell\in\{i,i+1,k\}\,,
\eea
where $k_\ell^{{\cal S}_1}$ are external momenta of $A^{{\cal S}_1}_{n_1}$, while $k_\ell^{{\cal S}_2}$ are external momenta of $A^{{\cal S}_2}_{n+3-n_1}$, satisfying
\bea
\Big(k_\ell^{{\cal S}_1}\Big)^2+\Big(k_\ell^{{\cal S}_2}\Big)^2=k_\ell^2=0\,,
\eea
due to the orthogonality of ${\cal S}_1$ and ${\cal S}_2$.

\begin{figure}
  \centering
  \includegraphics[width=10cm]{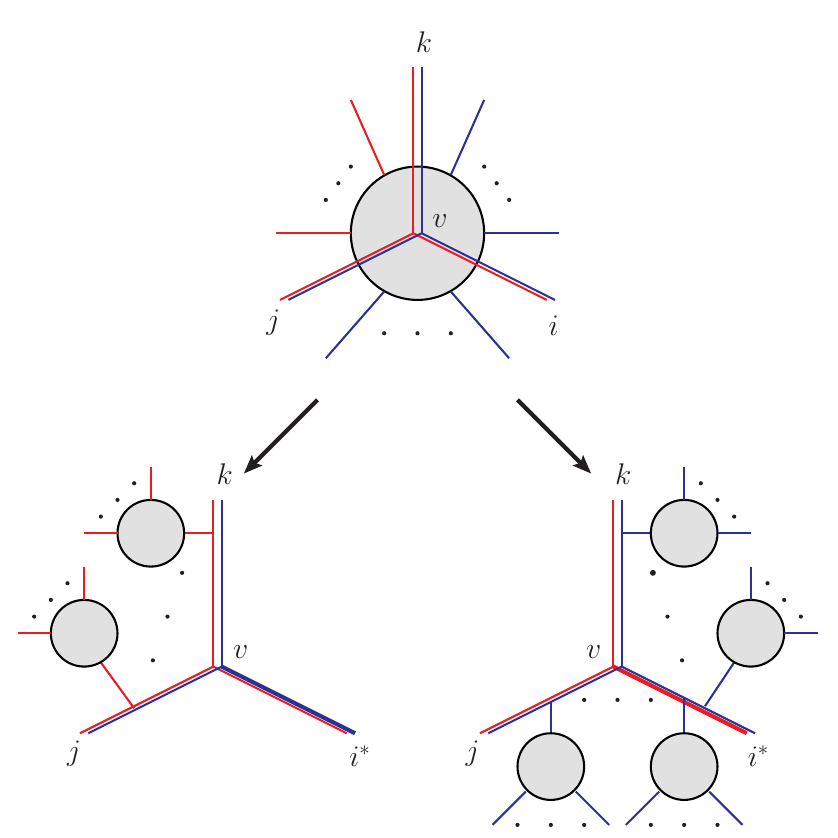} \\
  \caption{Split ${\cal A}_n$ into two currents, for the case that $i$ and $j$ are not adjacent legs. The representation for gluons $i^*$ in two down diagrams are similar as that in Fig.\ref{share3-flow}: each off-shell external particle carries a superscript $*$, and the components of momenta in subspaces which are different from those in the original amplitude, are represented by bold lines.}\label{share3-gen}
\end{figure}

Now we consider the more general case that $i$ and $j$ are not adjacent external legs in the ordered amplitude $A_n$, as shown in the top graph of Fig.\ref{share3-gen}. We assume that external momenta $k_a$ with $a\in\{j+1,\cdots,k-1\}$ lie in ${\cal S}_1$, while $k_b$ with $b\in\{k+1,\cdots,j-1\}\setminus i$ lie in ${\cal S}_2$, then the reversing of merging gives rise to the more general $2$-split
\bea
A_n(1,\cdots,n)\,&\xrightarrow[a\in\{j+1,\cdots,k-1\}\,,\,b\in\{k+1,\cdots,j-1\}\setminus i]{k_a\in{\cal S}_1\,,\,k_b\in{\cal S}_2}&\,A^{{\cal S}_1}_{n_1}(i,j,\cdots,k)\,\times\,A^{{\cal S}_2}_{n+3-n_1}(k,\cdots,j)\,,
~~\label{fac-nadj-sub}
\eea
The constraint on external kinematics reads
\bea
& &k_a\cdot k_b=0\,,~~~~{\rm for}~a\in\{j+1,\cdots,k-1\}\,,~b\in\{k+1,\cdots,j-1\}\setminus i\,,~~~~\label{kinematic-condi2-1to2}
\eea
coincide with that found in \cite{Cao:2024gln,Arkani-Hamed:2024fyd,Cao:2024qpp}. Replacing $A_n$ in \eref{fac-nadj-sub} by the ${\rm Tr}(\phi^3)$
amplitude ${\cal A}^{\phi^3}_n$, we get the $2$-split of the ${\rm Tr}(\phi^3)$ amplitude
\bea
{\cal A}^{\phi^3}_n(1,\cdots,n)\,&\xrightarrow[]{\eref{kinematic-condi2-1to2}}&\,A^{{\cal S}_1}_{n_1}(i,j,\cdots,k)\,\times\,A^{{\cal S}_2}_{n+3-n_1}(k,\cdots,j)\,,
~~\label{facphi-nadj-sub}
\eea
as diagrammatically illustrated in Fig.\ref{share3-gen}.

As will be explained in the next subsection, for the splitting in \eref{facphi-adj-sub},
one can interpret $A^{{\cal S}_1}_{n_1}$ and $A^{{\cal S}_2}_{n+3-n_1}$ as ${\rm Tr}(\phi^3)$ currents ${\cal J}^{\phi^3}_{n_1}(i^*,j,\cdots,k)$ and ${\cal J}^{\phi^3}_{n+3-n_1}(k,\cdots,j^*)$, similar as in \eref{5pfac-2}, where $i^*$ and $j^*$ are off-shell massless external particles. For the splitting in \eref{facphi-nadj-sub}, the interpretation for $A^{{\cal S}_1}_{n_1}$ is the same, but it is more natural to understand $A^{{\cal S}_2}_{n+3-n_1}$ as ${\cal J}^{\phi^3}_{n+3-n_1}(k,\cdots,i^*,\cdots,j)$, similar as discussed around \eref{5p-JR}, where $i^*$ is the off-shell massless external particle.

At the end of this subsection, let us discuss a subtle point when generalizing the splittings \eref{fac-adj-sub} and \eref{fac-nadj-sub} to other theories beyond ${\rm Tr}(\phi^3)$. Since the vertex $v$ is shared by $A^{{\cal S}_1}_{n_1}$ and $A^{{\cal S}_2}_{n+3-n_1}$, splittings in \eref{fac-adj-sub} and \eref{fac-nadj-sub} require that the contribution from $v$ is separated into two parts, one is absorbed by $A^{{\cal S}_1}_{n_1}$, while another one is absorbed by
$A^{{\cal S}_2}_{n+3-n_1}$. For the ${\rm Tr}(\phi^3)$ theory, such separation is the trivial $1\to1\times1$, due to the simplicity of vertices. However, for a new theory with complicated vertices, it is possible that the vertex $v$ contributes $m$ terms where $m\geq 2$ is an integer. Clearly, the contribution from each term can be split into two pieces, which can be naturally allocated to $A^{{\cal S}_1}_{n_1}$ and $A^{{\cal S}_2}_{n+3-n_1}$ respectively, since $v^{{\cal S}_1}\cdot v^{{\cal S}_2}=0$ for any Lorentz vectors $v^{{\cal S}_1}$ in ${\cal S}_1$ and $v^{{\cal S}_2}$ in ${\cal S}_2$. Hence, splittings in \eref{fac-adj-sub} and \eref{fac-nadj-sub} should be generalized to
\bea
A_n(1,\cdots,n)\,&\xrightarrow[a\in\{i+2,\cdots,k-1\}\,,\,b\in\{k+1,\cdots,i-1\}]{k_a\in{\cal S}_1\,,\,k_b\in{\cal S}_2}&\,\sum_{s=1}^m\,A^{{\cal S}_1}_{s;n_1}(i,\cdots,k)\,\times\,A^{{\cal S}_2}_{s;n+3-n_1}(k,\cdots,i+1)\,,
~~\label{fac-adj-sub-gen}
\eea
and
\bea
A_n(1,\cdots,n)\,&\xrightarrow[a\in\{j+1,\cdots,k-1\}\,,\,b\in\{k+1,\cdots,j-1\}\setminus i]{k_a\in{\cal S}_1\,,\,k_b\in{\cal S}_2}&\,\sum_{s=1}^m\,A^{{\cal S}_1}_{s;n_1}(i,j,\cdots,k)\,\times\,A^{{\cal S}_2}_{s;n+3-n_1}(k,\cdots,j)\,,
~~\label{fac-nadj-sub-gen}
\eea
where each pair of amplitudes $A^{{\cal S}_1}_{s;n_1}$ and $A^{{\cal S}_2}_{s;n+3-n_1}$ with the given $s$ absorbs one term from the vertex $v$.
The above generalized splittings will not occur in YM and NLSM cases under consideration in this paper, since in these two cases the vertex $v$ carried by one of $A^{{\cal S}_1}_{n_1}$ and $A^{{\cal S}_2}_{n+3-n_1}$ is the ${\rm Tr}(\phi^3)$ one, as will be explained in sections \ref{sec-YM} and \ref{sec-NLSM}.

\subsection{Factorization of propagators}
\label{subsec-verification-phi3}

In this subsection, we show that from the ${\cal S}$ point of view, two amplitudes $A^{{\cal S}_1}_{n_1}$ and $A^{{\cal S}_2}_{n+3-n_1}$ in the splittings \eref{facphi-adj-sub} and \eref{facphi-nadj-sub} are currents of the ${\rm Tr}(\phi^3)$
theory, by generalizing the factorization of propagators in \eref{5p-facpropa} to the general case. Similar as in subsection \ref{subsec-5p0}, the general factorization of propagators also leads to the alternative understanding for zeros discussed in subsection \ref{subsec-zero-phi3}. It is worth to emphasize that, if the assumption that scattering processes in orthogonal spaces are independent of each other is not correct, the derivations in this subsection still hold.

Let us first focus on the interpretation of $A^{{\cal S}_1}_{n_1}$ and $A^{{\cal S}_2}_{n+3-n_1}$ in \eref{facphi-nadj-sub}. As can be seen in Fig.\ref{share3-gen}, most of propagators and vertices in the full amplitude ${\cal A}^{\phi^3}_n$ purely lie in ${\cal S}_1$ or ${\cal S}_2$, except those along the lines $L_{\ell,v}$, with $\ell=i,j,k$. The ingredients in ${\cal S}_1$ or ${\cal S}_2$ automatically satisfy Feynman rules of ${\rm Tr}(\phi^3)$ theory, thus we only need to understand the behavior of vertices and propagators along three lines after splitting. For the ${\rm Tr}(\phi^3)$ theory, each vertex is a trivial constant. Therefore, it is sufficient to deal with propagators along these lines.
The key observation is, for any $A$ and $B$, where $A$ is a subset of
$\{j+1,\cdots,k-1\}$ while $B$ is a subset of $\{k+1,\cdots,j-1\}\setminus i$, we have
\bea
\Big(k_k+k_A+k_B\Big)^2=\Big(k_k+k_A\Big)^2+\Big(k_k+k_B\Big)^2\,,~~\label{key1}
\eea
due to the on-shell condition $k_k^2=0$ and the condition \eref{kinematic-condi2-1to2}.
Based on the above observation, we can show that propagators along $L_{k,v}$ and $L_{j,v}$ factorize into two sectors, which naturally contribute to ${\cal J}^{{\cal S}_1}_{n_1}$ and ${\cal J}^{{\cal S}_2}_{n+3-n_1}$ respectively.

\begin{figure}
  \centering
  \includegraphics[width=11cm]{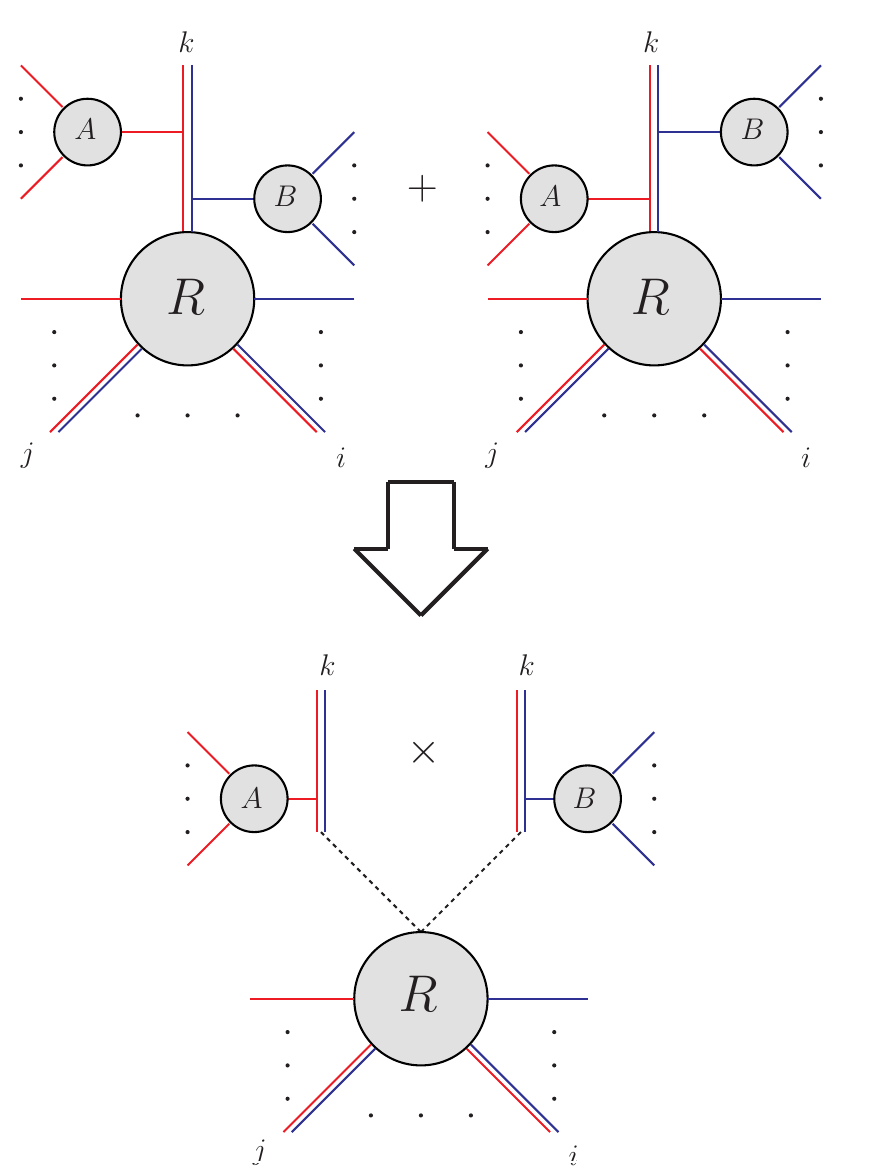} \\
  \caption{The first example of the factorization of propagators in \eref{fac-mkb}.}\label{verify1}
\end{figure}

We begin with a simple case in Fig.\ref{verify1}, where two blocks labeled as $A$ and $B$ are planted onto $L_{k,v}$. As shown in Fig.\ref{verify1}, the block $A$ lies in ${\cal S}_1$, while $B$ lies in ${\cal S}_2$. The left top diagram
contributes
\bea
P_1={\cal J}_A\,{\cal J}_B\,{1\over s_{kA}}\,{1\over s_{kAB}}\,g_R\,,
\eea
while the right one contributes
\bea
P_2={\cal J}_A\,{\cal J}_B\,{1\over s_{kB}}\,{1\over s_{kAB}}\,g_R\,,
\eea
where $s_{kA}=(k_k+k_A)^2$, and the analogous definitions hold for $s_{kB}$ and $s_{kAB}$. For simplicity, here the capital letter $A$ denotes
either the block, or the set of external legs belonging to this block, and similar does $B$.
Two rational functions ${\cal J}_A$ and ${\cal J}_B$ stand for Berends-Giele currents from two blocks $A$ and $B$. The rational function $g_R$ encodes the contribution from the rest part $R$ of diagram, and is related to ${\cal J}_R$ as ${\cal J}_R=g_R/ s_{kAB}$. Summing over the left and right diagrams,
and substituting the observation \eref{key1},
we immediately get
\bea
P_1+P_2=\Big({\cal J}_A\,{1\over s_{kA}}\Big)\,\Big({\cal J}_B\,{1\over s_{kB}}\Big)\,g_R\,,~~\label{fac-2b}
\eea
where the coefficient of $R$ factorizes into two pieces. The first block is independent of $B$, while the second is independent of $A$.
Since two blocks $A$ and $B$ are purely in ${\cal S}_1$ and ${\cal S}_2$ respectively, the non-trivial part of \eref{fac-2b} is the phenomenon
\bea
{1\over s_{kA}}\,{1\over s_{kAB}}+{1\over s_{kB}}\,{1\over s_{kAB}}={1\over s_{kA}}\,\times\,{1\over s_{kB}}\,,~~\label{fac-2b-pro}
\eea
which serves as the generalization of the first line in \eref{5p-facpropa}

\begin{figure}
  \centering
  \includegraphics[width=13cm]{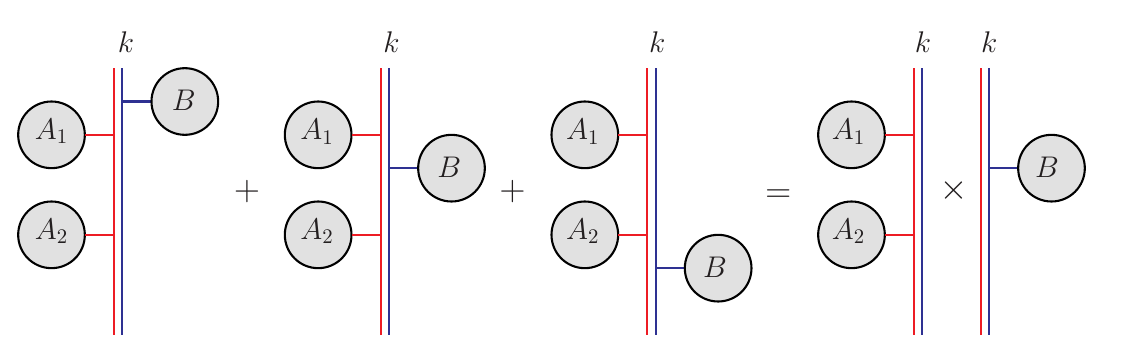} \\
  \caption{The second example of the factorization of propagators in \eref{fac-mkb}. Comparing with Fig.\ref{verify1}, here we omitted external lines enter three blocks, and the remaining block $R$, to shrink the area of graph.}\label{verify2}
\end{figure}

Then we consider the case in Fig.\ref{verify2}, three blocks $A_1$, $A_2$ and $B$, with $A_1$, $A_2$ lie in ${\cal S}_1$ and $B$ lies in ${\cal S}_2$, are planted to $L_{k,v}$.
Contributions from three diagrams are
\bea
P_1&=&{\cal J}_{A_1}\,{\cal J}_{A_2}\,{\cal J}_B\,{1\over s_{kB}}\,{1\over s_{kA_1B}}\,{1\over s_{kA_1A_2B}}\,g_R\,,\nn
P_2&=&{\cal J}_{A_1}\,{\cal J}_{A_2}\,{\cal J}_B\,{1\over s_{kA_1}}\,{1\over s_{kA_1B}}\,{1\over s_{kA_1A_2B}}\,g_R\,,\nn
P_3&=&{\cal J}_{A_1}\,{\cal J}_{A_2}\,{\cal J}_B\,{1\over s_{kA_1}}\,{1\over s_{kA_1A_2}}\,{1\over s_{kA_1A_2B}}\,g_R\,.
\eea
We can use the previous result \eref{fac-2b} to sum over contributions from first two diagrams,
\bea
P_1+P_2
&=&{\cal J}_{A_2}\,\Big({\cal J}_{A_1}\,{1\over s_{kA_1}}\Big)\,\Big({\cal J}_B\,{1\over s_{kB}}\Big)\,{1\over s_{kA_1A_2B}}\,g_R\,.
\eea
Then, using the observation \eref{key1}, we obtain
\bea
P_1+P_2+P_3
&=&\Big({\cal J}_{A_1}\,{\cal J}_{A_2}\,{1\over s_{kA_1}}\,{1\over s_{kA_1A_2}}\Big)\,\Big({\cal J}_B\,{1\over s_{kB}}\Big)\,g_R\,,~~\label{fac-3b}
\eea
Again, the coefficient of $g_R$ factorizes into two pieces. The first one is independent of $B$, while the second one is independent of $A_1$,
$A_2$. The non-trivial part in \eref{fac-3b} is
\bea
{1\over s_{kB}}\,{1\over s_{kA_1B}}\,{1\over s_{kA_1A_2B}}+{1\over s_{kA_1}}\,{1\over s_{kA_1B}}\,{1\over s_{kA_1A_2B}}
+{1\over s_{kA_1}}\,{1\over s_{kA_1A_2}}\,{1\over s_{kA_1A_2B}}
&=&\Big({1\over s_{kA_1}}\,{1\over s_{kA_1A_2}}\Big)\,\times\,{1\over s_{kB}}\,,~~\label{fac-3b-pro}
\eea
which is the generalization of the second line in \eref{5p-facpropa}.

\begin{figure}
  \centering
  \includegraphics[width=13cm]{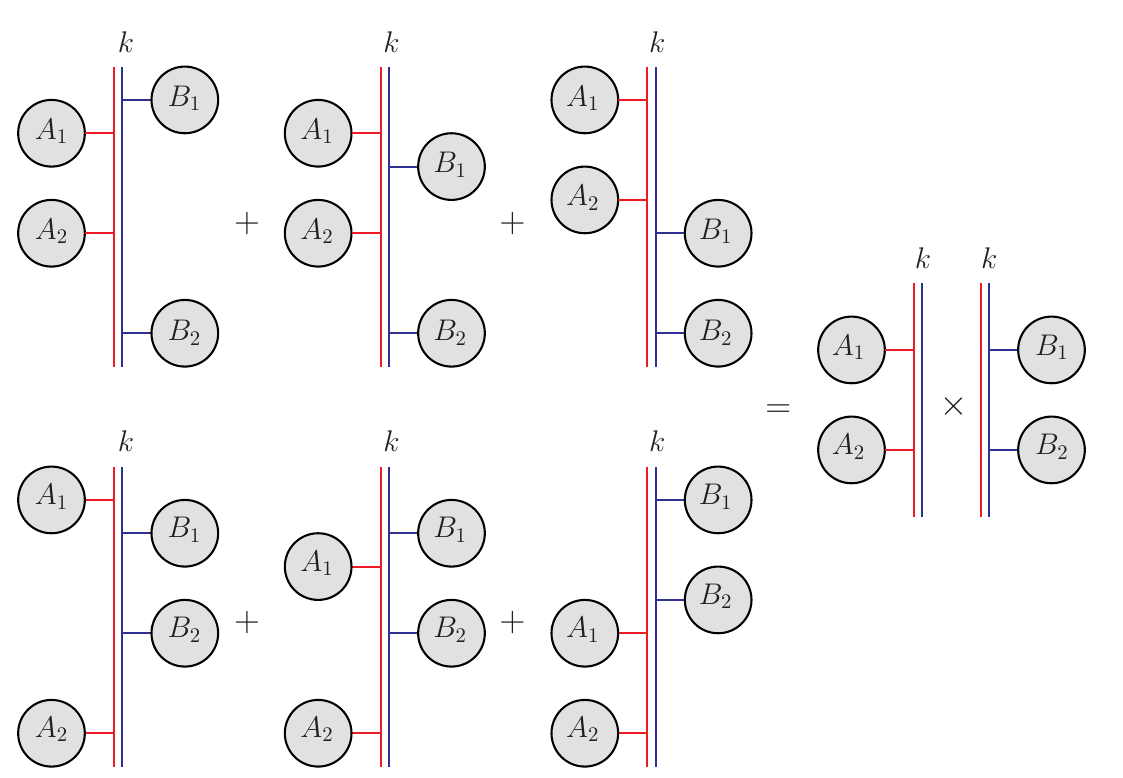} \\
  \caption{The third example of the factorization of propagators in \eref{fac-mkb}. Again, we omitted external lines enter four blocks, and the remaining block $R$.}\label{verify3}
\end{figure}

The next example is the configuration in Fig.\ref{verify3}, with four blocks $A_1$, $A_2$, $B_1$ and $B_2$ planted onto $L_{k,v}$.
One can apply the factorization in \eref{fac-3b} to three diagrams in the first line of Fig.\ref{verify3}, to obtain
\bea
P_1+P_2+P_3={\cal J}_{B_2}\,\Big({\cal J}_{A_1}\,{\cal J}_{A_2}\,{1\over s_{kA_1}}\,{1\over s_{kA_1A_2}}\Big)\,\Big({\cal J}_{B_1}\,{1\over s_{kB_1}}\Big)\,{1\over s_{kA_1A_2B_1B_2}}\,g_R\,.
\eea
Analogously, summing over three diagrams in the second line leads to
\bea
P_4+P_5+P_6={\cal J}_{A_2}\,\Big({\cal J}_{A_1}\,{1\over s_{kA_1}}\Big)\,\Big({\cal J}_{B_1}\,{\cal J}_{B_2}\,{1\over s_{kB_1}}\,{1\over s_{kB_1B_2}}\Big)\,{1\over s_{kA_1A_2B_1B_2}}\,g_R\,.
\eea
Therefore, summing over all six diagrams in Fig.\ref{verify3} gives rise to the factorization
\bea
\sum_{j=1}^6\,P_j=\Big({\cal J}_{A_1}\,{\cal J}_{A_2}\,{1\over s_{kA_1}}\,{1\over s_{kA_1A_2}}\Big)\,\Big({\cal J}_{B_1}\,{\cal J}_{B_2}\,{1\over s_{kB_1}}\,{1\over s_{kB_1B_2}}\Big)\,g_R\,,~~\label{fac-4b}
\eea
where the observation \eref{key1} has been used again. The non-trivial part in \eref{fac-4b} reads
\bea
\sum_{j=1}^6\,H_j=\Big({1\over s_{kA_1}}\,{1\over s_{kA_1A_2}}\Big)\,\times\,\Big({1\over s_{kB_1}}\,{1\over s_{kB_1B_2}}\Big)\,,~~\label{fac-4b-pro}
\eea
where
\bea
H_j={P_j\over {\cal J}_{A_1}\,{\cal J}_{A_2}\,{\cal J}_{B_1}\,{\cal J}_{B_2}\,g_R}\,.
\eea

Now the recursive pattern arises, allowing us to extend the factorizations in \eref{fac-2b-pro}, \eref{fac-3b-pro} and \eref{fac-4b-pro} to the general situation that blocks in $\{A_1,\cdots,A_m\}\cup\{B_1,\cdots,B_h\}$, with each $A_a$ lies in ${\cal S}_1$ and each $B_b$ lies in ${\cal S}_2$, are planted to $L_{k,v}$.
Suppose we have the following factorizations
\bea
\sum_{\Gamma^{k,v}_{m-1,h}}\,\prod_{t=1}^{m+h-1}{1\over{\cal D}^{k,v}_t}=\Big(\prod_{a=1}^{m-1}\,{1\over s_{kA_1\cdots A_a}}\Big)\,\times\,\Big(\prod_{b=1}^h\,{1\over s_{kB_1\cdots B_b}}\Big)\,,~~\label{fac-(m-1,l)}
\eea
and
\bea
\sum_{\Gamma^{k,v}_{m,h-1}}\,\prod_{t=1}^{m+h-1}{1\over{\cal D}^{k,v}_t}=\Big(\prod_{a=1}^{m}\,{1\over s_{kA_1\cdots A_a}}\Big)\,\times\,\Big(\prod_{b=1}^{h-1}\,{1\over s_{kB_1\cdots B_b}}\Big)\,,~~\label{fac-(m,l-1)}
\eea
hold for $\{A_1,\cdots,A_{m-1}\}\cup\{B_1,\cdots,B_h\}$ and $\{A_1,\cdots,A_m\}\cup\{B_1,\cdots,B_{h-1}\}$. In \eref{fac-(m-1,l)},
${1/{\cal D}^{(k,v)}_t}$ are propagators along the line $L_{k,v}$, and the summation is among diagrams $\Gamma^{k,v}_{m-1,h}$
those all blocks in $\{A_1,\cdots,A_{m-1}\}\cup\{B_1,\cdots,B_h\}$ are planted onto $L_{k,v}$. Notice that these blocks are ordered when planting to $L_{k,v}$, with the ordering $(A_{m-1},\cdots,A_1,B_1,\cdots,B_h)$. Thus, the summation over $\Gamma^{k,v}_{m-1,h}$, is equivalent to the summation over all shuffles for lines from ${\cal S}_1$ and ${\cal S}_1$ which are connected to $L_{k,v}$. Analogous interpretation holds for \eref{fac-(m,l-1)}.
Then, for blocks in $\{A_1,\cdots,A_m\}\cup\{B_1,\cdots,B_h\}$, we have
\bea
\sum_{\Gamma^{k,v}_{m,h}}\,\prod_{t=1}^{m+h}{1\over{\cal D}^{k,v}_t}
&=&\Big(\sum_{\Gamma^{k,v}_{m-1,h}}\,\prod_{j=1}^{m+h-1}{1\over{\cal D}^{k,v}_j}\,+\,\sum_{\Gamma^{k,v}_{m,h-1}}\,\prod_{j=1}^{m+h-1}{1\over{\cal D}^{k,v}_j}\Big)\,{1\over s_{iA_1\cdots A_mB_1\cdots B_h}}\nn
&=&\Big(\prod_{a=1}^{m-1}\,{1\over s_{kA_1\cdots A_a}}\Big)\,\Big(\prod_{b=1}^{h-1}\,{1\over s_{kB_1\cdots B_b}}\Big)\,
\Big({1\over s_{kA_1\cdots A_m}}+{1\over s_{kB_1\cdots B_h}}\Big)\,{1\over s_{iA_1\cdots A_mB_1\cdots B_h}}\nn
&=&\Big(\prod_{a=1}^{m}\,{1\over s_{kA_1\cdots A_a}}\Big)\,\times\,\Big(\prod_{b=1}^{h}\,{1\over s_{kB_1\cdots B_b}}\Big)\,,~~\label{fac-mkb}
\eea
where the second equality uses factorizations in \eref{fac-(m-1,l)} and \eref{fac-(m,l-1)}, the third uses the observation \eref{key1}.
Our previous results in \eref{fac-2b-pro}, \eref{fac-3b-pro} and \eref{fac-4b-pro} obviously satisfy \eref{fac-(m-1,l)} and \eref{fac-(m,l-1)}, thus one can
determine the general factorization \eref{fac-mkb} recursively. In the last line of \eref{fac-mkb}, the first piece can be recognized as propagators
along the line $L^{{\cal S}_1}_{k,v}$ in the left down diagram in Fig.\ref{share3-gen}, satisfying Feynman rules of ${\rm Tr}(\phi^3)$ theory. Meanwhile, the second piece in the last line of \eref{fac-mkb} are propagators along $L^{{\cal S}_2}_{k,v}$ in the right down one in Fig.\ref{share3-gen}, and also obey Feynman rules of ${\rm Tr}(\phi^3)$ theory. The factorization formula in \eref{fac-mkb}, is the generalization of factorization in \eref{5p-facpropa}.

The similar manipulation can be performed to propagators along the line $L_{j,v}$. Hence, when blocks in $\{A_1,\cdots,A_m\}\cup\{B_1,\cdots,B_h\}$ are planted to $L_{j,v}$, with each $A_a$ lies in ${\cal S}_1$ and each $B_b$ lies in ${\cal S}_2$, summing over diagrams gives the factorization of propagators
\bea
\sum_{\Gamma^{j,v}_{m,h}}\,\prod_{t=1}^{m+h}{1\over{\cal D}^{j,v}_t}
&=&\Big(\prod_{a=1}^{m}\,{1\over s_{jA_1\cdots A_a}}\Big)\,\times\,\Big(\prod_{b=1}^{h}\,{1\over s_{jB_1\cdots B_b}}\Big)\,.~~\label{fac-pqb}
\eea
Two pieces respectively represent propagators along two lines $L^{{\cal S}_1}_{j,v}$ and $L^{{\cal S}_2}_{j,v}$ in two bottom graphs in Fig.\ref{share3-gen},
again satisfying Feynman rules of ${\rm Tr}(\phi^3)$ theory. On the other hand, propagators along $L_{i,v}$ only contribute to the right graph
at the bottom of Fig.\ref{share3-gen}, since there is no propagator along $L^{{\cal S}_1}_{i,v}$ in the first graph.

With factorizations of propagators obtained above, we are ready to figure out the theory (theories) corresponds to resulting amplitudes in \eref{facphi-nadj-sub}. Clearly, any Feynman diagram of ${\cal A}_n^{\phi^3}$ can be thought as planting a variety of blocks to lines $L_{\ell,v}$, with $\ell=i,j,k$. We label each block as follows
\bea
& &A^\ell_r~~~~{\rm if}~{\rm it}~{\rm is}~{\rm planted}~{\rm to}~L_{\ell,v}\,,~{\rm and}~{\rm lies}~{\rm in}~{\cal S}_1\,,\nn
& &B^\ell_r~~~~{\rm if}~{\rm it}~{\rm is}~{\rm planted}~{\rm to}~L_{\ell,v}\,,~{\rm and}~{\rm lies}~{\rm in}~{\cal S}_2\,.
\eea
Notice that all of blocks planted to $L_{i,v}$ are in ${\cal S}_2$, due to the color ordering. In the above, $r$ is the integer introduced for labeling different blocks in the same sector. For a given set of blocks, suppose any external leg in $\{1,\cdots,n\}\setminus\{i,j,k\}$
belongs to one of these blocks, then we call such set a complete set of blocks.
For a complete set of blocks,
we denote the numbers of $A^\ell_r$ and $B^\ell_r$ as $m_\ell$ and $h_\ell$, respectively.
Then, the summation over Feynman diagrams can be formulated as
\bea
{\cal A}_n^{\phi^3}&=&\sum_{\rm c.s.b}\,\Big(\sum_{\Gamma^{k,v}_{m_k,h_k}}\,\prod_{t=1}^{m_k+h_k}{1\over{\cal D}^{k,v}_t}\Big)\,\Big(\sum_{\Gamma^{j,v}_{m_j,h_j}}\,\prod_{t=1}^{m_j+h_j}{1\over{\cal D}^{j,v}_t}\Big)\,\Big(\sum_{\Gamma^{i,v}_{\ell_i}}\,\prod_{t=1}^{h_i}{1\over{\cal D}^{i,v}_t}\Big)\nn
& &~~~~~~\Big(\prod_{\ell=j,k}\,\prod_{r=1}^{m_\ell}\,{\cal J}_{A^\ell_r}\Big)\,
\Big(\prod_{\ell=i,j,k}\,\prod_{r=1}^{h_\ell}\,{\cal J}_{B^\ell_r}\Big)\,,~~\label{sum-diagrams}
\eea
where the first summation is for all possible complete sets of blocks (c.s.b). Substituting factorizations in \eref{fac-mkb}
and \eref{fac-pqb} into the above \eref{sum-diagrams}, we obtain
\bea
{\cal A}_n^{\phi^3}&=&\sum_{\rm c.s.b}\,\Big[\,\Big(\prod_{a=1}^{m_k}\,{1\over s_{kA^k_1\cdots A^k_a}}\Big)\,\Big(\prod_{a=1}^{m_j}\,{1\over s_{jA^j_1\cdots A^j_a}}\Big)\,\Big(\prod_{\ell=j,k}\,\prod_{r=1}^{m_\ell}\,{\cal J}_{A^\ell_r}\Big)\,\Big]\nn
& &~~~~~\Big[\,\Big(\prod_{b=1}^{h_k}\,{1\over s_{kB^k_1\cdots B^k_b}}\Big)\,\Big(\prod_{b=1}^{h_j}\,{1\over s_{jB^j_1\cdots B^j_b}}\Big)\,\Big(\sum_{\Gamma^{i,v}_{h_i}}\,\prod_{t=1}^{h_i}{1\over{\cal D}^{i,v}_t}\Big)\,\Big(\prod_{\ell=i,j,k}\,\prod_{r=1}^{h_\ell}\,{\cal J}_{B^\ell_r}\Big)\,\Big]\,.
\eea
Therefore, two resulting amplitudes in \eref{facphi-nadj-sub} read
\bea
A^{{\cal S}_1}_{n_1}(i,j,\cdots,k)&=&{\cal J}_{n_1}(i^*,j,\dots,k)\nn
&=&\sum_{\rm c.s.b}\,\Big(\prod_{a=1}^{m_k}\,{1\over s_{kA^k_1\cdots A^k_a}}\Big)\,\Big(\prod_{a=1}^{m_j}\,{1\over s_{jA^j_1\cdots A^j_a}}\Big)\,\Big(\prod_{\ell=j,k}\,\prod_{r=1}^{m_\ell}\,{\cal J}_{A^\ell_r}\Big)\,,~~\label{JL}
\eea
and
\bea
A^{{\cal S}_2}_{n+3-n_1}(k,\cdots,j)&=&{\cal J}_{n+3-n_1}(k,\cdots,i^*,\cdots,j)\nn
&=&\sum_{\rm c.s.b}\,\Big(\prod_{b=1}^{h_k}\,{1\over s_{kB^k_1\cdots B^k_b}}\Big)\,\Big(\prod_{b=1}^{h_j}\,{1\over s_{jB^j_1\cdots B^j_b}}\Big)\,\Big(\sum_{\Gamma^{i,v}_{h_i}}\,\prod_{t=1}^{\ell_i}{1\over{\cal D}^{i,v}_t}\Big)\,\Big(\prod_{\ell=i,j,k}\,\prod_{r=1}^{h_\ell}\,{\cal J}_{B^\ell_r}\Big)\,,~~\label{JR}
\eea
where the complete sets of blocks (c.s.b) in \eref{JL} are defined for external legs $\{i,j,\cdots,k\}\setminus\{i,j,k\}$,
while those in \eref{JR} are defined for external legs $\{k,\cdots,j\}\setminus\{i,j,k\}$.
Similar as in \eref{pro-S12}, from the ${\cal S}_1$ point of view, propagators in \eref{JL} along lines $L^{{\cal S}_1}_{k,v}$ and $L^{{\cal S}_1}_{j,v}$ should be understood
as
\bea
{1\over s_{kA^k_1\cdots A^k_a}}&=&{1\over\big(k_k^{{\cal S}_1}+k_{A^k_1}+\cdots +k_{A^k_a}\big)^2+\big(k_k^{{\cal S}_2}\big)^2}\,,\nn
{1\over s_{jA^j_1\cdots A^j_a}}&=&{1\over\big(k_j^{{\cal S}_1}+k_{A^j_1}+\cdots +k_{A^j_a}\big)^2+\big(k_j^{{\cal S}_2}\big)^2}\,,
\eea
namely, virtual particles propagate along $L^{{\cal S}_1}_{k,v}$ acquire the effective mass $m^{{\cal S}_1}_k$, satisfying
\bea
\big(m^{{\cal S}_1}_k\big)^2=-\big(k_k^{{\cal S}_2}\big)^2\,,
\eea
while virtual particles propagate along $L^{{\cal S}_1}_{j,v}$ acquire the effective mass $m^{{\cal S}_1}_j$, satisfying
\bea
\big(m^{{\cal S}_1}_j\big)^2=-\big(k_j^{{\cal S}_2}\big)^2\,.
\eea
From the ${\cal S}_2$ point of view, the analogous interpretation holds for propagators in \eref{JR}, along lines $L^{{\cal S}_2}_{k,v}$ and $L^{{\cal S}_2}_{j,v}$.

From the ${\cal S}$ point of view, two currents in \eref{JL} and \eref{JR} satisfy Feynman rules of ${\rm Tr}(\phi^3)$ theory, thus the splitting in \eref{facphi-nadj-sub} can be understood as
\bea
{\cal A}^{\phi^3}_{n}(1,\cdots,n)\,&\xrightarrow[]{\eref{kinematic-condi2-1to2}}&\,{\cal J}^{\phi^3}_{n_1}(i^*,j,\dots,k)\,\times\,{\cal J}^{\phi^3}_{n+3-n_1}(k,\cdots,i^*,\cdots,j)\,,~~~~\label{1to2express-2-fix}
\eea
where two currents carry the superscripts $\phi^3$, since the corresponding theory is ${\rm Tr}(\phi^3)$.
In the above, the interpretation is from the $S$ point of view, then the momentum conservation requires
\bea
k_i^*=k_i+\sum_{b\in\{k+1,\cdots,j-1\}\setminus i}\,k_b\,,~~\label{offi-L}
\eea
carried by the off-shell external leg $i^*$ in ${\cal J}^{\phi^3}_{n_1}$, and
\bea
k_i^*=k_i+\sum_{a\in\{j+1,\cdots,k-1\}}\,k_a\,,~~\label{offi-R}
\eea
for the off-shell external leg $i^*$ in ${\cal J}^{\phi^3}_{n+3-n_1}$. In general, one can choose any one of external particles in $\{i,j,k\}$
to be off-shell. However, from two bottom graphs in Fig.\ref{share3-gen}, and from the formulas \eref{JL} and \eref{JR}, one can see that the line $L_{i,v}$ is more special for both ${\cal J}^{\phi^3}_{n_1}$ and ${\cal J}^{\phi^3}_{n+3-n_1}$, thus it is natural to choose $i^*$ as the off-shell one, as diagrammatically illustrated in Fig.\ref{share3-gen}. The shift of $k_i$ in \eref{offi-L}
does not affect any propagator in \eref{JL}. On the other hand, for an observer in ${\cal S}$, virtual particles in ${\cal J}^{\phi^3}_{n+3-n_1}$ along the line $L_{i,v}$ acquire an additional effective mass. In \eref{JR}, each propagator $1/{\cal D}^{i,v}_t$ has the formula
\bea
{1\over{\cal D}^{i,v}_t}={1\over \big(k_i+K_t\big)^2}\,,
\eea
where $K_t$ stands for the combinatorial momentum from blocks planted to $L_{i,v}$. When the external momentum $k_i$ is shifted as in \eref{offi-R},
such propagator should be reinterpreted as
\bea
{1\over{\cal D}^{i,v}_t}={1\over \big(k_i^*+K_t\big)^2-\big(k^*_i\big)^2}\,,~~\label{additionM}
\eea
similar as in \eref{additonalmass}. In the above, we have used the observation $k_A\cdot K_t=0$, since all blocks planted onto $L_{i,v}$ are in ${\cal S}_2$.
Therefore, the virtual particles along $L_{i,v}$ have the effective mass satisfying $m^2=(k_i^*)^2$.

Applying the factorization of propagators in \eref{fac-mkb}, one can also interpret the splitting in \eref{facphi-adj-sub} as
\bea
{\cal A}^{\phi^3}_n(1,\cdots,n)\,&\xrightarrow[]{\eref{kinematic-condi-1to2}}&\, {\cal J}^{\phi^3}_{n_1}(i^*,\dots,k)\,\times\,{\cal J}^{\phi^3}_{n+3-n_1}(k,\cdots,i+1^*)\,,~~~~\label{1to2express-fix}
\eea
where
\bea
{\cal J}^{\phi^3}_{n_1}(i^*,\dots,k)
&=&\sum_{\rm c.s.b}\,\Big(\prod_{a=1}^{m_k}\,{1\over s_{kA^k_1\cdots A^k_a}}\Big)\,\Big(\prod_{a=1}^{m_{i+1}}\,{1\over s_{(i+1)A^{i+1}_1\cdots A^{i+1}_a}}\Big)\,\Big(\prod_{\ell=i+1,k}\,\prod_{r=1}^{m_\ell}\,{\cal J}_{A^\ell_r}\Big)\,,~~\label{JL-adj}
\eea
and
\bea
{\cal J}^{\phi^3}_{n+3-n_1}(k,\cdots,i+1^*)
&=&\sum_{\rm c.s.b}\,\Big(\prod_{b=1}^{h_k}\,{1\over s_{kB^k_1\cdots B^k_b}}\Big)\,\Big(\prod_{b=1}^{h_i}\,{1\over s_{iB^i_1\cdots B^i_b}}\Big)\,\Big(\prod_{\ell=i,k}\,\prod_{r=1}^{h_\ell}\,{\cal J}_{B^\ell_r}\Big)\,.~~\label{JR-adj}
\eea
For the current case, it is more convenient to choose the off-shell external leg of ${\cal J}^{\phi^3}_{n+3-n_1}$ as $i+1^*$, as diagrammatically illustrated in Fig.\ref{share3-flow}, since shifting $k_{i+1}$ as
\bea
k^*_{i+1}=k_{i+1}+\sum_{a\in\{i+2,\cdots,k-1\}}\,k_a
\eea
will not affect any propagator in \eref{JR-adj}.

The factorization of propagators in \eref{fac-mkb} also generalizes the second interpretation for zeros in subsection \ref{subsec-5p0},
which is discussed around \eref{inter-5p0-1} and \eref{5p-fac-2pros}, to the general case with the arbitrary number of external legs.
According to the merging in Fig.\ref{share2}, we can formulate the amplitude as
\bea
{\cal A}^{\phi^3}_n(1,\cdots,n)&=&\sum_{\rm c.s.b}\,\Big(\sum_{\Gamma^{i,j}_{m,h}}\,\prod_{t=1}^{m+h-1}{1\over{\cal D}^{i,j}_t}\Big)\,\Big(\prod_{r=1}^m\,{\cal J}_{A_r}\Big)\,
\Big(\prod_{r=1}^{h}\,{\cal J}_{B_r}\Big)\,,~~\label{np-for0}
\eea
where the the blocks planted onto $L_{i,j}$ are labeled as $\{A_1,\cdots,A_m\}\cup\{B_1,\cdots,B_h\}$. Notice that the number of propagators $1/{\cal D}^{i,j}_t$ along the line $L_{i,j}$ is $m+h-1$ rather than $m+h$, since both $i$ and $j$ are external legs. If we add the $(m+h)^{\rm th}$ propagator
\bea
{1\over {\cal D}^{i,j}_{m+h}}={1\over s_{iA_1\cdots A_m B_1\cdot B_h}}~~\label{fro0-step0}
\eea
into \eref{np-for0}, then the factorization in \eref{fac-mkb} reads
\bea
\sum_{\Gamma^{i,j}_{m,h}}\,\prod_{t=1}^{m+h}{1\over{\cal D}^{i,j}_t}
&=&\Big(\prod_{a=1}^{m}\,{1\over s_{iA_1\cdots A_a}}\Big)\,\times\,\Big(\prod_{b=1}^{h}\,{1\over s_{iB_1\cdots B_b}}\Big)\,.~~\label{for0-step1}
\eea
The above formula indicates
\bea
\sum_{\Gamma^{i,j}_{m,h}}\,\prod_{t=1}^{m+h-1}{1\over{\cal D}^{i,j}_t}
&=&s_{iA_1\cdots A_m B_1\cdot B_h}\,\Big(\prod_{a=1}^{m}\,{1\over s_{iA_1\cdots A_a}}\Big)\,\Big(\prod_{b=1}^{h}\,{1\over s_{iB_1\cdots B_b}}\Big)\,,~~\label{for0-step2}
\eea
which vanishes when $k_j$ is an on-shell massless momentum, since the momentum conservation requires $s_{iA_1\cdots A_m B_1\cdot B_h}=k_j^2$.

\subsection{Factorization near zero, and smooth splitting}
\label{subsec-fac-zero}

In this subsection, we consider two interesting $3$-splits, in which the amplitude factorizes into three pieces. The first one is the factorization near zero discovered in \cite{Arkani-Hamed:2023swr}, the second one is the so called smooth splitting found in \cite{Cachazo:2021wsz}.
As explained in \cite{Cao:2024gln,Arkani-Hamed:2024fyd,Cao:2024qpp}, these $3$-splits can be achieved by performing the $2$-split twice. For reader's convenience, we will give a brief review for this method. Then, we will show that the smooth splitting can also be reproduced by applying the idea of orthogonal spaces.

To do the $2$-split twice, we first notice that all arguments in subsection \ref{subsec-verification-phi3} still hold if one of $k_a$ and $k_b$ in \eref{kinematic-condi-1to2} or \eref{kinematic-condi2-1to2} is taken to be off-shell (with $s_{ab}$ replaced by $k_a\cdot k_b$). Consequently, the $2$-splits in \eref{1to2express-2-fix} and \eref{1to2express-fix} are valid for the current with one off-shell external particle. This observation allows us to split one of ${\cal J}^{\phi^3}_{n_1}$ and ${\cal J}^{\phi^3}_{n_2}$ in \eref{1to2express-2-fix} or \eref{1to2express-fix} further, via totally the same procedure.

Let us split ${\cal J}^{\phi^3}_{n_1}$ in \eref{1to2express-2-fix} by cutting along three lines $L_{k,v'}$, $L_{j,v'}$ and $L_{h,v'}$, where $h\in\{j+1,\cdots,k-1\}$, and $v'$ denotes the cubic vertex which connects three lines. This splitting gives rise to
\bea
{\cal J}^{\phi^3}_{n_1}(i^*,j,\cdots,k)\,&\xrightarrow[]{\eref{constraint-add}}&\, {\cal J}^{\phi^3}_4(i^*,j,h^*,k)\,\times \,{\cal J}^{\phi^3}_{n_1-1}(j,\cdots,h^*,\cdots,k)\,,
\eea
in the manner of \eref{1to2express-2-fix}, therefore the original amplitude ${\cal A}^{\phi^3}_n$ factorizes as
\bea
{\cal A}^{\phi^3}_n(1,\cdots,n)\,&\xrightarrow[]{\eref{condi-fac-near0}}&\,{\cal J}^{\phi^3}_4(i^*,j,h^*,k)\,\times \,{\cal J}^{\phi^3}_{n_1-1}(j,\cdots,h^*,\cdots,k)\,\times \,{\cal J}^{\phi^3}_{n+3-n_1}(k,\cdots,i^*,\cdots,j)\,.
~~\label{fac-near0}
\eea
The picture of orthogonal spaces leads to the following new constraint on kinematics for ${\cal J}^{\phi^3}_{n_1}$,
\bea
k_i^*\cdot k_c=0\,,~~~~{\rm for}~c\in\{j+1,\cdots,k-1\}\setminus h\,.~~\label{constraint-add}
\eea
Combining the constraint in \eref{kinematic-condi-1to2}, and the off-shell momentum $k_i^*$ determined in \eref{offi-R},
we see that the new condition \eref{constraint-add} yields
\bea
k_i\cdot k_c=0\,,~~~~{\rm for}~c\in\{j+1,\cdots,k-1\}\setminus h\,.~~\label{condi-factori-near0}
\eea
Putting conditions \eref{kinematic-condi-1to2} and \eref{condi-factori-near0} together gives the full constraint on kinematics,
\bea
k_{a'}\cdot k_{b'}=0\,,~~~~{\rm for }~a'\in\{k+1,\cdots,j-1\}\,,~b'\in\{j+1,\cdots,k-1\}\,,~~~~{\rm except}~a'=i,\,b'=h\,,~~\label{condi-fac-near0}
\eea
which can be obtained by turning on $s_{ih}$ in the condition \eref{kinematic-condi-zero} for zeros (with $i$ in \eref{kinematic-condi-zero} replaced by $k$). The splitting in \eref{fac-near0} is the factorization near zero
found in \cite{Arkani-Hamed:2023swr}.

Along the line of split twice, it is straightforward to arrive at the smooth splitting found in \cite{Cachazo:2021wsz}.
One can separate ${\cal J}^{\phi^3}_{n+3-n_1}$ in \eref{1to2express-2-fix} as
\bea
{\cal J}^{\phi^3}_{n+3-n_1}(k,\cdots,i^*,\cdots,j)\,&\xrightarrow[]{\eref{condi-add}}&\,{\cal J}^{\phi^3}_{n_2}(j^*,k,\cdots,i)\,\times\,{\cal J}^{\phi^3}_{n+6-n_1-n_2}(k^*,i,\cdots,j)\,,~~\label{verify-1to3}
\eea
in the manner of \eref{1to2express-fix}, with the additional constraint on kinematics
\bea
k_b\cdot k_c=0\,,~~~~~~{\rm for}~b\in\{k+1,\cdots,i-1\}\,,~~c\in\{i+1,\cdots,j-1\}\,.~~\label{condi-add}
\eea
Here a subtle point is, the external leg $i^*$ at the l.h.s is an off-shell one, while legs $i$ at the r.h.s are two on-shell ones. The reason is, although $i^*$ at the l.h.s carries the off-shell momentum $k_i^*$, the effective mass $m^2=(k_i^*)^2$ of virtual particles along the line $L_{i,v}$, forces the expression of ${\cal J}^{\phi^3}_{n+3-n_1}$ to be exactly the same as the on-shell amplitude ${\cal A}^{\phi^3}_{n+3-n_1}$.
Thus, when the momentum conservation can be ensured by shifting momenta carried by new off-shell legs $j^*$ and $k^*$ in two currents at the r.h.s,
it is natural to interpret the external particle $i$ as an on-shell one. Substituting \eref{verify-1to3} into \eref{1to2express-2-fix}, we get the smooth splitting
\bea
{\cal A}^{\phi^3}_n(1,\cdots,n)\,&\xrightarrow[]{\eref{condi-1to3split}}&\, {\cal J}^{\phi^3}_{n_1}(i^*,j,\cdots,k)\,\times\,{\cal J}^{\phi^3}_{n_2}(j^*,k,\cdots,i)\,\times\,{\cal J}^{\phi^3}_{n+6-n_1-n_2}(k^*,i,\cdots,j)\,,~~\label{1to3express-fix}
\eea
\begin{figure}
  \centering
  \includegraphics[width=13cm]{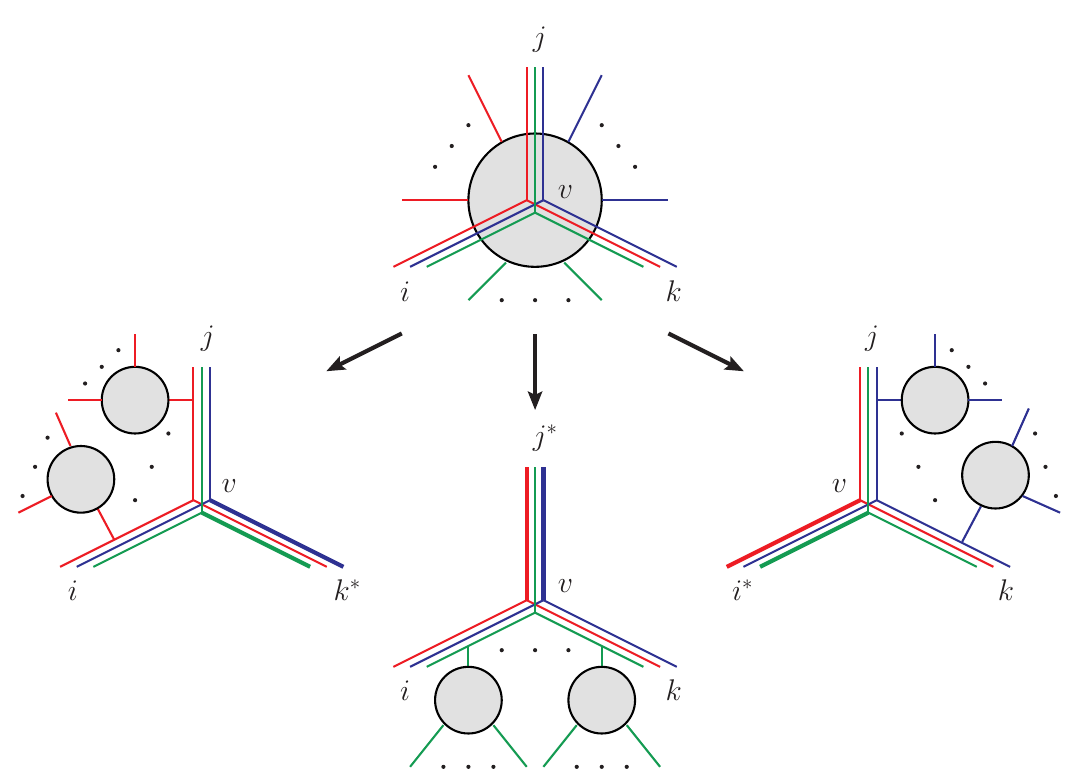} \\
  \caption{The smooth spitting. Again, each off-shell external particle carries a superscript $*$, and the components of momenta in subspaces which are different from those in the original amplitude, are represented by bold lines. The red lines carry momenta in ${\cal S}_1$, the green lines carry momenta in ${\cal S}_2$, while blue lines carry momenta in ${\cal S}_3$.}\label{1to3split}
\end{figure}

Using the methods in this paper, we can give another alternative understanding of the smooth splitting. We can decompose the full spacetime into three orthogonal subspaces as ${\cal S}={\cal S}_1\oplus{\cal S}_2\oplus{\cal S}_3$. Consider the merging of three amplitudes $A^{{\cal S}_1}(i,j,\cdots,k)$, $A^{{\cal S}_2}(j,k,\cdots,i)$ and $A^{{\cal S}_3}(k,i,\cdots,j)$, by gluing lines in each triplet $\big(L^{{\cal S}_1}_{\ell,v},L^{{\cal S}_2}_{\ell,v},L^{{\cal S}_3}_{\ell,v}\big)$, with $\ell=i,j,k$, as shown in
the top graph of Fig.\ref{1to3split}. The resulting amplitude automatically factorizes into the original three ones.
Then the argument similar as that for obtaining \eref{fac-adj-sub} and \eref{fac-nadj-sub} leads to the $3$-split
\bea
& &A_{n}(1,\cdots,n)\,\xrightarrow[a\in\{j+1,\cdots,k-1\}\,,\,b\in\{k+1,\cdots,i-1\}\,,\,c\in\{i+1,\cdots,j-1\}]{k_a\in{\cal S}_1\,,\,k_b\in{\cal S}_2\,,\,k_c\in{\cal S}_3}\,\nn
& &~~~~~~~~~~~~~~~~~~~~~~~~A^{{\cal S}_1}_{n_1}(i,j,\dots,k)\,\times\,A^{{\cal S}_2}_{n_2}(j,k,\cdots,i)\,\times\,A^{{\cal S}_3}_{n+6-n_1-n_2}(k,i,\cdots,j)\,,~~\label{1to3express}
\eea
as illustrated in Fig.\ref{1to3split}.
The constraint on kinematics arises from the arrangement that $k_a$, $k_b$ and $k_c$ lie in ${\cal S}_1$, ${\cal S}_2$ and ${\cal S}_3$ respectively, thus can be given as
\bea
& &k_a\cdot k_b=0\,,~~~~k_b\cdot k_c=0\,,~~~~k_c\cdot k_a=0\,,\nn
& &{\rm for}~a\in\{j+1,\cdots,k-1\}\,,~~b\in\{k+1,\cdots,i-1\}\,,~~c\in\{i+1,\cdots,j-1\}\,,~~\label{condi-1to3split}
\eea
which is the same as that in \cite{Cachazo:2021wsz}.
Again, we can forget the picture of orthogonal spaces, and take \eref{condi-1to3split} as the constraint for $3$-split, since the amplitude solely depends on Lorentz invariants, without distinguishing the ways for achieving \eref{condi-1to3split}.
The final formula of such splitting in \eref{1to3express-fix}, with three resulting amplitudes interpreted as ${\rm Tr}(\phi^3)$ currents, can be obtained by applying the factorization of propagators in \eref{fac-mkb} to propagators along lines $L_{\ell,v}$ with $\ell=i,j,k$.

\section{YM amplitudes}
\label{sec-YM}

In this section, we discuss zeros and splittings of tree level YM amplitudes. In subsection \ref{subsec-zero-YM}, we directly use the method in sections \ref{subsec-zero-phi3}, \ref{subsec-1to2-phi3} and \ref{subsec-fac-zero}, to determine the special loci in kinematic space for zeros and splittings of YM amplitudes. Then, in subsection \ref{subsec-flow-YM}, we generalize the method in section \ref{subsec-verification-phi3} by utilizing locality and gauge invariance, to determine the exact interpretation of resulting amplitudes in splittings.

\begin{figure}
  \centering
  \includegraphics[width=10cm]{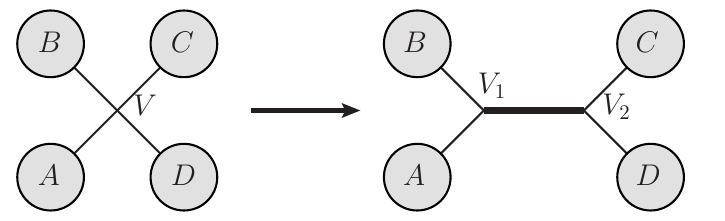} \\
  \caption{Split a quadrivalent vertex $V$ to two trivalent ones $V_1$ and $V_2$, by inserting $s_{AB}/s_{AB}$, where the inserted propagator $1/s_{AB}$ is labeled by the bold line.}\label{split4pv}
\end{figure}

The YM theory contains both trivalent and quadrivalent vertices.
The well known technic of decomposing quadrivalent YM vertices into trivalent ones is useful for discussions in this section.
For later convenience, we briefly discuss this technic as follows.
One can transmutes any $4$-point vertex to $3$-point ones, by inserting $1={{\cal D}\over {\cal D}}$, where
each $1/{\cal D}$ is the propagator of ${\rm Tr}(\phi^3)$ theory, determined by the corresponding channel.
Such decomposition is diagrammatically illustrated in Fig.\ref{split4pv}.
Then an $n$-point tree YM amplitude can be represented as the summation over trivalent diagrams,
\bea
{\cal A}^{\rm YM}_n=\sum_{\Gamma^3}\,{N_{\Gamma^3}\over\prod_i\,{\cal D}_i}\,.~~\label{sumdia-YM}
\eea
In the above, factors ${\cal D}$ from
splitting $4$-point interactions, are absorbed by the numerators $N_{\Gamma^3}$. After the above treatment, each resulting trivalent vertex has mass dimension $1$.

In the original Feynman rules of YM theory, momenta carried by each trivalent vertex are only those along the lines which are connected to this vertex, we can call this feature the locality of the vertex. Then one can ask if such locality is kept by trivalent vertices arise from splitting quadrivalent ones. For a quadrivalent vertex, with mass dimension $0$, the Berends-Giele currents from four corners in Fig.\ref{split4pv} interact at this vertex as
\bea
\a\,({\cal J}_A\cdot{\cal J}_B)\,({\cal J}_C\cdot{\cal J}_D)+\a\,({\cal J}_D\cdot{\cal J}_A)\,({\cal J}_B\cdot{\cal J}_C)
+\b\,({\cal J}_A\cdot{\cal J}_C)\,({\cal J}_B\cdot{\cal J}_D)\,,~~\label{3term}
\eea
where ${\cal J}^\mu_A$, ${\cal J}^\nu_B$, ${\cal J}^\rho_C$ and ${\cal J}^\tau_D$ denote the Berends-Giele currents from corners $A$, $B$, $C$ and $D$ respectively.
For the first term in \eref{3term}, one can split it as in Fig.\ref{split4pv}, namely,
\bea
({\cal J}_A\cdot{\cal J}_B)\,({\cal J}_C\cdot{\cal J}_D)\to ({\cal J}_A\cdot{\cal J}_B)\,{s_{AB}\over s_{AB}}\,({\cal J}_C\cdot{\cal J}_D)\,,~~\label{sp-4p1}
\eea
which means the tensor $g^{\mu\nu}g^{\rho\tau}$ carried by the vertex $V$ is split as
\bea
g^{\mu\nu}\,g^{\rho\tau}\to V_1^{\mu\nu\lambda}\,{g_{\lambda\sigma}\over s_{AB}}\,V_2^{\sigma\rho\tau}\,,
\eea
where
\bea
V_1^{\mu\nu\lambda}=g^{\mu\nu}\,k_{AB}^\lambda\,,~~~~V_2^{\sigma\rho\tau}=k_{AB}^\sigma\,g^{\rho\tau}\,,
\eea
and $g_{\lambda\sigma}/ s_{AB}$ is the propagator of gluon in the Feynman gauge (up to a $-i$). Tensors $V_1^{\mu\nu\lambda}$
and $V_2^{\sigma\rho\tau}$ carried by two new vertices $V_1$ and $V_2$ satisfy the locality discussed above, since the momentum $k_{AB}$ flows along the line $L_{V_1,V_2}$ which is connected to these two vertices. For the second term in \eref{3term}, we change the r.h.s graph in Fig.\ref{split4pv} from the "$s$-channel" to the "$t$-channel",
namely, we split this term as
\bea
({\cal J}_D\cdot{\cal J}_A)\,({\cal J}_B\cdot{\cal J}_C)\to ({\cal J}_D\cdot{\cal J}_A)\,{s_{DA}\over s_{DA}}\,({\cal J}_B\cdot{\cal J}_C)\,,~~\label{sp-4p2}
\eea
which implies
\bea
g^{\tau\mu}\,g^{\nu\rho}\to V_1^{\tau\mu\lambda}\,{g_{\lambda\sigma}\over s_{AB}}\,V_2^{\sigma\nu\rho}\,,
\eea
where
\bea
V_1^{\tau\mu\lambda}=g^{\tau\mu}\,k_{AB}^\lambda\,,~~~~V_2^{\sigma\nu\rho}=k_{AB}^\sigma\,g^{\nu\rho}\,.
\eea
The above $V_1^{\tau\mu\lambda}$ and $V_2^{\sigma\nu\rho}$ also satisfy the locality. However, for the third term in \eref{3term}, the locality is broken when the quadrivalent vertex is separated into trivalent ones. Thus, we conclude that the locality can be kept, if
\bea
{\cal J}_A\cdot{\cal J}_C=0\,,~~~~{\rm or}~~~~{\cal J}_B\cdot{\cal J}_D\,.~~\label{condi-loca}
\eea
The above conclusion will be used frequently in the rest of this section.

\subsection{Kinematic conditions for zeros and splittings}
\label{subsec-zero-YM}

%
\begin{figure}
  \centering
   \includegraphics[width=10cm]{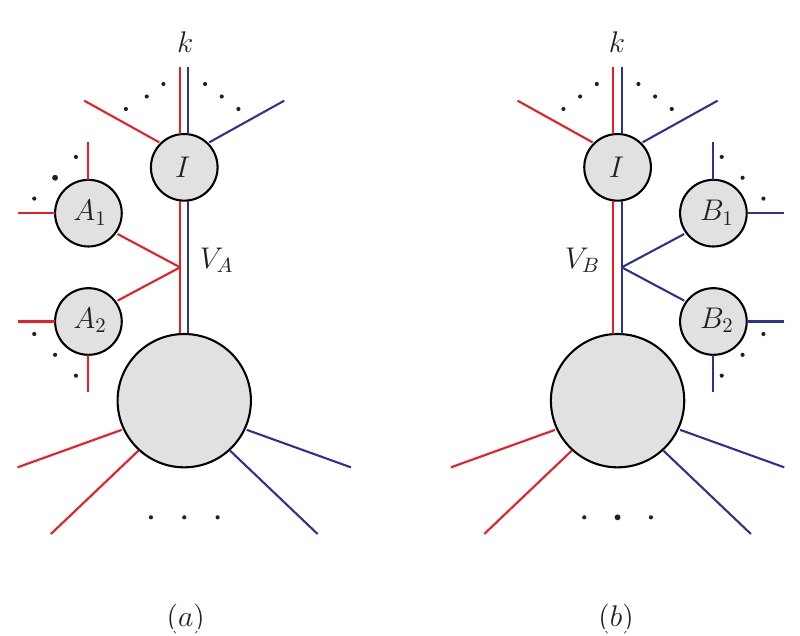}
   ~~~~
   \includegraphics[width=4.6cm]{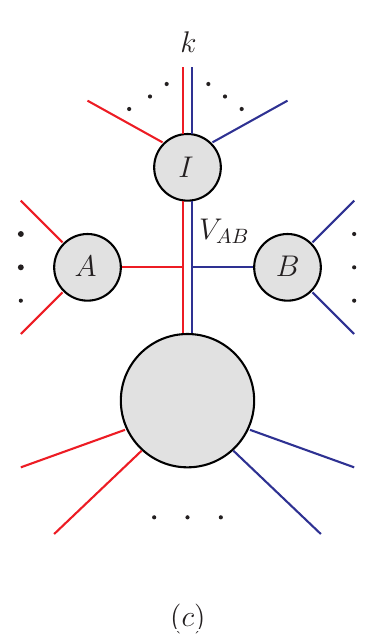} \\
  \caption{Three types of quadrivalent vertices along the line $L_{k,\bullet}$, where $\bullet$ can be either another external leg or a vertex $v$. The former corresponds to zeros, while the latter corresponds to splittings.}\label{4p-vert}
\end{figure}
\begin{figure}
  \centering
  \includegraphics[width=14cm]{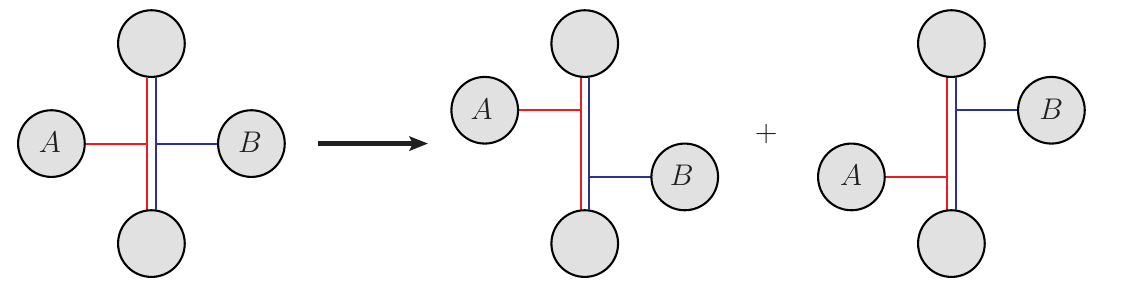} \\
  \caption{Split a quadrivalent vertex along the glued line into two trivalent ones. Two diagrams at the r.h.s are one to one correpondent to splittings \eref{sp-4p1} and \eref{sp-4p2}.}\label{split4to3}
\end{figure}

The loci in kinematic space, for zeros and splittings of YM amplitudes, can also be figured out by applying the idea of orthogonal spaces.
As discussed in section \ref{subsec-zero-phi3}, merging two amplitudes in ${\cal S}_1$ and ${\cal S}_2$ by gluing only two lines, as diagrammatically illustrated in Fig.\ref{share2}, does not create any scattering process in the full space time ${\cal S}$. When applying this idea to the current YM case, the obstacle is, there are three types of quadrivalent vertices along the glued line, as can be seen in Fig.\ref{4p-vert}, the type (c) vertex in the third graph receives blocks from ${\cal S}_1$ and ${\cal S}_2$ simultaneously, therefore threatens our statement that the merging does not introduce any interaction between two scattering processes in orthogonal spaces.
Fortunately, as discussed previously, one can use the technic of inserting ${\cal D}/{\cal D}$, as illustrated in Fig.\ref{split4pv}, to split such type (c) vertices to trivalent ones as in Fig.\ref{split4to3}. For a type (c) vertex in Fig.\ref{4p-vert} or Fig.\ref{split4to3}, two blocks $A$ and $B$ belong to ${\cal S}_1$ and ${\cal S}_2$ respectively, therefore ${\cal J}_A\cdot{\cal J}_B=0$, satisfying the condition \eref{condi-loca} (with relabeling of $A$, $B$, $C$ and $D$). It means,
each new trivalent vertex created by the splitting in Fig.\ref{split4pv} satisfies the locality, thus can never sense two blocks from two orthogonal spaces simultaneously. Such locality ensures that the idea in section \ref{subsec-zero-phi3} can be applied to the YM case, if all type (c) quadrivalent vertices along the glued line are separated to trivalent ones discussed above.
Furthermore, to guarantee a gluon to be purely in ${\cal S}_1$ (${\cal S}_2$), we also need to require its polarization vector to be in ${\cal S}_1$ (${\cal S}_2$).
Therefore, the merging in Fig.\ref{share2} indicates the following condition for the zero of $n$-point YM amplitude with external gluons in $\{1,\cdots,n\}$,
\bea
& &k_a\cdot k_b=k_a\cdot\epsilon_b=\epsilon_a\cdot k_b=\epsilon_a\cdot \epsilon_b=0\,,\nn
& &{\rm for}~a\in\{j+1,\cdots,i-1\}\,,~~~~b\in\{i+1,\cdots,j-1\}\,,~~~~\label{kinematic-condi-zero-YM}
\eea
the same as that found in \cite{Arkani-Hamed:2023swr}.

\begin{figure}
  \centering
  \includegraphics[width=11cm]{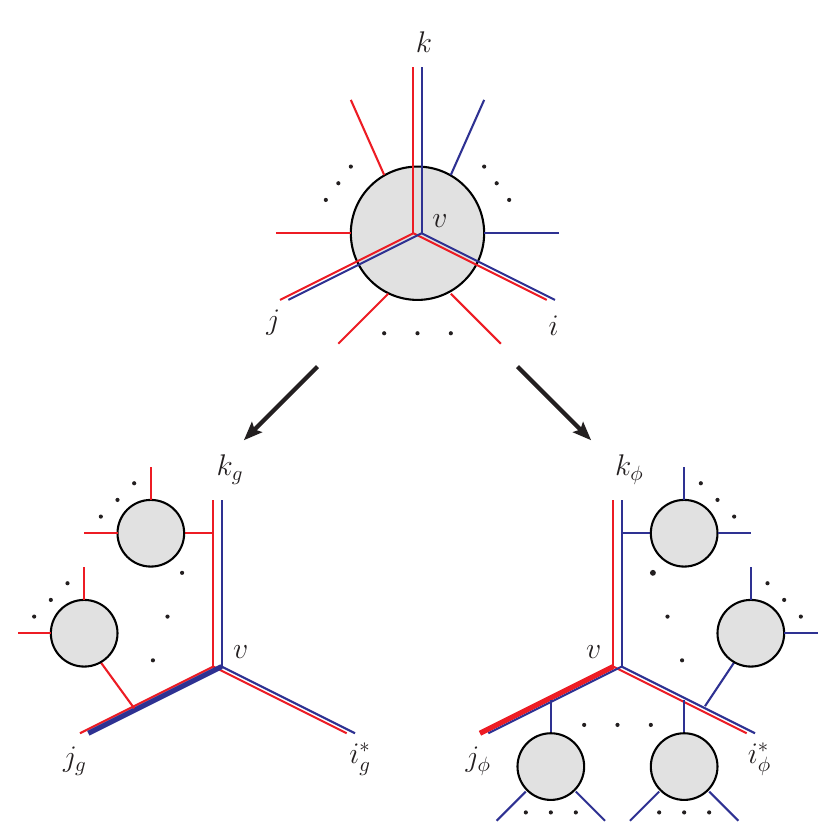} \\
  \caption{$2$-split of YM amplitudes. In the left down graph, all external particles are gluons. In the right one, $i^*_\phi$, $j_\phi$ and $k_\phi$ are ${\rm Tr}(\phi^3)$ scalars, while the remaining external particles are gluons. For simplicity, here we only draw trivalent vertices.}\label{splitYM}
\end{figure}

The consideration for splittings of YM amplitudes is also similar as in the previous section. One can pick up three external legs $i$, $j$ and $k$ in $A_n^{\rm YM}$, and setting gluons in $\{j+1,\cdots,k-1\}$ to be in ${\cal S}_1$, while those in $\{k+1,\cdots,j-1\}\setminus i$ to be in ${\cal S}_2$. These gluons satisfy
\bea
& &k_a\cdot k_b=k_a\cdot\epsilon_b=\epsilon_a\cdot k_b=\epsilon_a\cdot \epsilon_b=0\,,\nn
& &{\rm for}~a\in\{j+1,\cdots,k-1\}\,,~~~~b\in\{k+1,\cdots,j-1\}\setminus i\,,~~~~\label{kinematic-condi-fac-YM}
\eea
due to the orthogonality of two subspaces. We again split the type (c) quadrivalent vertices along the glued lines $L_{\ell,v}$ with $\ell=i,j,k$ into trivalent ones, as shown in Fig.\ref{split4to3}. The orthogonality of ${\cal S}_1$ and ${\cal S}_2$ ensures the condition \eref{condi-loca}, thus the resulting trivalent vertices satisfy the locality. This feature supports our assumption that the scattering processes in ${\cal S}_1$ and ${\cal S}_2$ are independent of each other, since momenta and polarizations in ${\cal S}_2$ (${\cal S}_1$) do not affect vertices in ${\cal S}_1$ (${\cal S}_2$).
However, the condition \eref{kinematic-condi-fac-YM} is not sufficient to govern the splitting of the YM amplitude. Since a YM amplitude is linear in each $\epsilon_\ell$ with $\ell=i,j,k$, external legs $i^*$, $j$, $k$ of two resulting amplitudes in the splitting
can not inherit these polarizations simultaneously, otherwise the linearity will be violated. To handle this, the simplest way is to restrict ourselves to the special case that polarizations $\epsilon_\ell$ also lie in one subspace. Without lose of generality, we set $\epsilon_\ell$ to be in ${\cal S}_1$.
Therefore, we need to impose the additional condition
\bea
& &\epsilon_\ell\cdot\epsilon_b=\epsilon_\ell\cdot k_b=0\,,~~~~{\rm for}~\ell\in\{i,j,k\}\,.~~\label{kinematic-condi-fac-YM-add}
\eea
Constraints in \eref{kinematic-condi-fac-YM} and \eref{kinematic-condi-fac-YM-add} reproduce the loci in kinematic space for the $2$-split in \cite{Cao:2024gln,Cao:2024qpp}.
Then the independence of the scattering processes in ${\cal S}_1$ and ${\cal S}_2$ implies the splitting
\bea
{\cal A}^{\rm YM}_n(1,\cdots,n)\,\xrightarrow[]{\eref{kinematic-condi-fac-YM}\,,\,\eref{kinematic-condi-fac-YM-add}}\,A^{{\cal S}_1}_{n_1}(i,j,\cdots,k)\,\times\,A^{{\cal S}_2}_{n+3-n_1}(k,\cdots,j)\,,
~~\label{facYM-nadj-sub}
\eea
which is diagrammatically illustrated in Fig.\ref{splitYM}. It is worth to point out that, unlike in the ${\rm Tr}(\phi^3)$ case, the vertex
$v$ in Fig.\ref{splitYM} contributes non-trivial terms rather than a constant, to the amplitude ${\cal A}^{\rm YM}_n$. Thus, in general we should expect the splitting to be the generalized one like in \eref{fac-nadj-sub-gen}. However, the vertex $v$ in the resulting sub-amplitude $A^{{\cal S}_2}_{n+3-n_1}$ is a ${\rm Tr}(\phi^3)$ one which contributes only a constant, since external particles
$i$, $j$ and $k$ without any polarization vector should be interpreted as ${\rm Tr}(\phi^3)$ scalars, and three scalar lines are connected by $v$, as will be explained in the next subsection. This phenomenon avoids the generalized formula like in \eref{fac-nadj-sub-gen}, reduces the splitting to the simple one in \eref{facYM-nadj-sub}.

When $i$ and $j$ are two adjacent gluons in ${\cal A}_n^{\rm YM}$, one get the splitting
\bea
{\cal A}^{\rm YM}_n(1,\cdots,n)\,\xrightarrow[]{\eref{kinematic-condi-fac-YM-adj}\,,\,\eref{kinematic-condi-fac-YM-add-adj}}\,A^{{\cal S}_1}_{n_1}(i,\cdots,k)\,\times\,A^{{\cal S}_2}_{n+3-n_1}(k,\cdots,i+1)\,,
~~\label{facYM-adj-sub}
\eea
under the constraints
\bea
& &k_a\cdot k_b=k_a\cdot\epsilon_b=\epsilon_a\cdot k_b=\epsilon_a\cdot \epsilon_b=0\,,\nn
& &{\rm for}~a\in\{i+2,\cdots,k-1\}\,,~~~~b\in\{k+1,\cdots,i-1\}\,,~~~~\label{kinematic-condi-fac-YM-adj}
\eea
and
\bea
& &\epsilon_\ell\cdot\epsilon_b=\epsilon_\ell\cdot k_b=0\,,~~~~{\rm for}~\ell\in\{i,j,k\}\,.~~\label{kinematic-condi-fac-YM-add-adj}
\eea

Decomposing the spacetime into three orthogonal subspaces as ${\cal S}={\cal S}_1\oplus{\cal S}_2\oplus{\cal S}_3$, one can also predict the $3$-split,
\bea
& &{\cal A}^{\rm YM}_{n}(1,\cdots,n)\,\xrightarrow[]{\eref{condi-1to3-YM}\,,\,\eref{condi-1to3-YM-2}}\,A^{{\cal S}_1}_{n_1}(i,j,\dots,k)\,\times\,A^{{\cal S}_2}_{n_2}(j,k,\cdots,i)\,\times\,A^{{\cal S}_3}_{n+6-n_1-n_2}(k,i,\cdots,j)\,,~~\label{1to3-YM}
\eea
with the conditions for kinematics
\bea
& &s_{ab}=k_a\cdot\epsilon_b=\epsilon_a\cdot k_b=\epsilon_a\cdot \epsilon_b=0\,,\nn
& &s_{ac}=k_a\cdot\epsilon_c=\epsilon_a\cdot k_c=\epsilon_a\cdot \epsilon_c=0\,,\nn
& &s_{bc}=k_b\cdot\epsilon_c=\epsilon_b\cdot k_c=\epsilon_b\cdot \epsilon_c=0\,,\nn
& &{\rm for}~a\in\{j+1,\cdots,k-1\}\,,~~b\in\{k+1,\cdots,i-1\}\,,~~c\in\{i+1,\cdots,j-1\}\,,~~\label{condi-1to3-YM}
\eea
as well as
\bea
& &\epsilon_\ell\cdot\epsilon_a=\epsilon_\ell\cdot k_a=
\epsilon_\ell\cdot\epsilon_c=\epsilon_\ell\cdot k_c=0\,,\nn
& &{\rm for}~\ell\in\{i,j,k\}\,,~~~~a\in\{j+1,\cdots,k-1\}\,,~~~~c\in\{i+1,\cdots,j-1\}\,.~~\label{condi-1to3-YM-2}
\eea
Similar as for the $2$-split, we assume that polarizations $\epsilon_\ell$ are localized in ${\cal S}_2$. Therefore, we need to add the condition
in \eref{condi-1to3-YM-2}.

%

\subsection{Determine resulting amplitudes via locality and gauge invariance}
\label{subsec-flow-YM}

Now we need to find the theory (theories) corresponds to resulting amplitudes in \eref{facYM-nadj-sub}, \eref{facYM-adj-sub} and \eref{1to3-YM}. We will do this by using the factorization of propagators found in section \ref{subsec-verification-phi3}, as well as the fundamental principles including locality and gauge invariance. As will be shown at the end of this subsection, the above generalization of the method in section \ref{subsec-verification-phi3} also interprets zeros of YM amplitudes.

Notice that although we will analyse the effects of vertices on lines $L_{\ell,v}$ with $\ell=i,j,k$, our method is independent on explicit formulas of Feynman rules for vertices. Furthermore, similar as in section \ref{subsec-verification-phi3}, the correctness of derivation in this subsection will not be affected, if the assumption that scattering processes in orthogonal spaces are independent of each other does not hold.

\begin{figure}
  \centering
  \includegraphics[width=10cm]{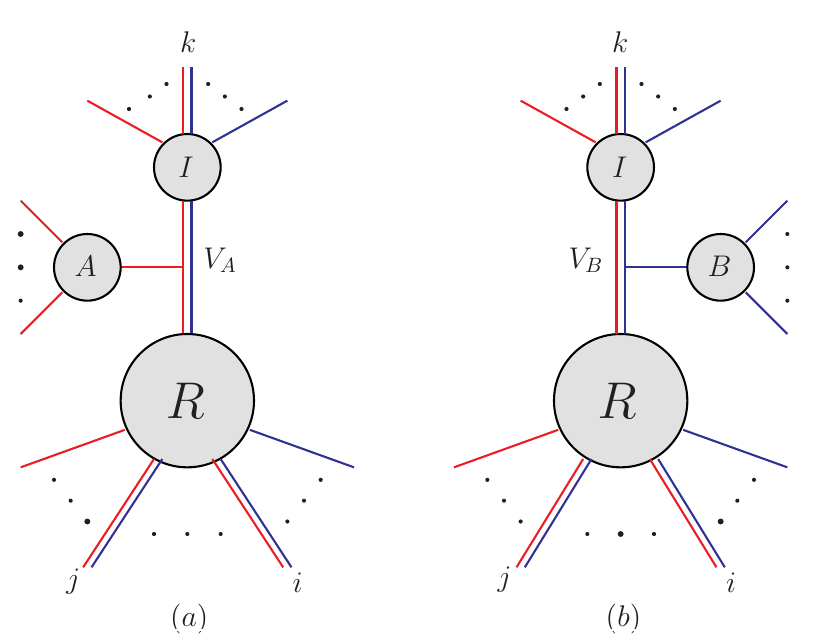} \\
  \caption{Two types of trivalent vertices along the line $L_{k,v}$.}\label{VertexYM}
\end{figure}

We begin by considering the $2$-split \eref{facYM-nadj-sub}. The treatment can be outlined as follows. We first analyse the effects of vertices on lines $L_{i,v}$, $L_{j,v}$ and $L_{k,v}$. The conclusion of such analysis, together with the factorization of propagators in \eref{fac-mkb}, allow us to separate the summation over Feynman diagrams into two pieces corresponding to two resulting amplitudes.
Then two amplitudes can be interpreted via propagators and vertices in the separated formula.

Again, most of vertices and propagators are purely in ${\cal S}_1$ or ${\cal S}_2$, thus we only need to deal with exceptional ones along lines $L_{i,v}$, $L_{j,v}$ and $L_{k,v}$. As before, we split the type (c) quadrivalent vertices along these lines to trivalent ones, as in Fig.\ref{split4to3}. We emphasize again the resulting trivalent vertices satisfy the locality, since the condition \eref{condi-loca} is ensured by the orthogonality of ${\cal S}_1$ and ${\cal S}_2$. After such splitting, let us consider vertices along $L_{k,v}$, except
the vertex $v$ at the end. We first focus on trivalent vertices, which can be clarified into two types, as shown in Fig.\ref{VertexYM}. The first type, a block $A$ from ${\cal S}_1$ is planted to the line, while the second type is that a block $B$ from ${\cal S}_2$ is planted to it. Although explicit Feynman rules for trivalent vertices created by splitting quadrivalent ones are different from those for original trivalent vertices, since our method only utilizes the locality and gauge invariance, it is not necessary to distinguish the source of any trivalent vertex.

\begin{figure}
  \centering
  \includegraphics[width=6.3cm]{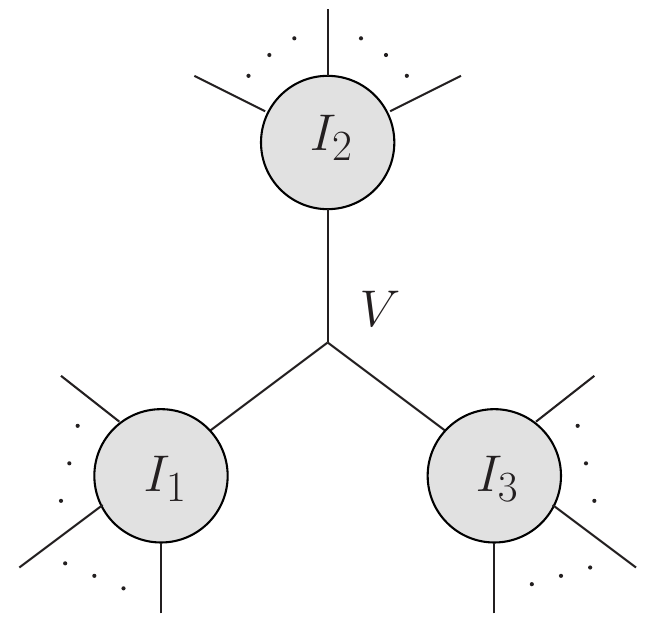}
   \includegraphics[width=7.2cm]{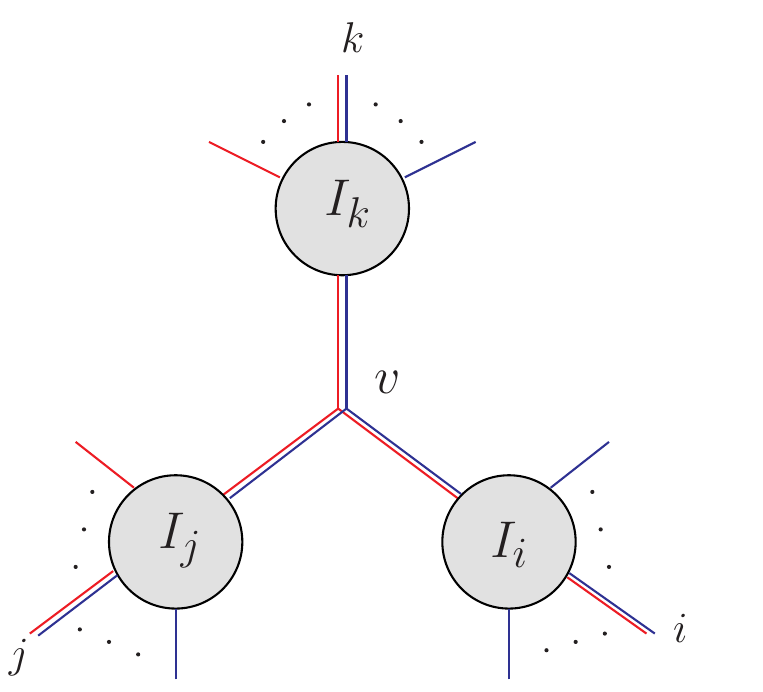}  \\
  \caption{Interactions of YM theory, at trivalent vertices.}\label{vertex-Y}
\end{figure}

To analyse the effect of a trivalent vertex, we discuss two ways of understanding the interaction between gluons at this vertex.
Since each gluon is a vector particle, we can think the interaction at this vertex $V$ as, three vector Berends-Giele currents ${\cal J}^\mu_{I_1}$, ${\cal J}^\mu_{I_2}$ and ${\cal J}^\mu_{I_3}$ meet at $V$, where ${\cal J}_{I_i}$ with $i=1,2,3$ are contributed by three corners $I_i$ in the left diagram of Fig.\ref{vertex-Y}. The mass dimension of the vertex is $1$, thus the interaction between three currents gives rise to the Lorentz invariant
\bea
T_{I_1|I_2|I_3}=({\cal J}^\mu_{I_1}\cdot{\cal J}^\mu_{I_2})\,({\cal J}^\mu_{I_3}\cdot K_3)+{\rm cyclic}\,,~~\label{T-V}
\eea
which serves as one term of the full amplitude.
In the above, $K_i$ with $i=1,2,3$ are three combinatorial momenta,
the locality indicates that each $K_i$ only depends on $k_{I_1}$, $k_{I_2}$ and $k_{I_3}$, carried by propagators from blocks $I_1$, $I_2$ and $I_3$, respectively. Furthermore, the gauge invariance implies
\bea
{\cal J}_{I_i}\cdot k_{I_i}=0\,,~~~~~~{\rm for}~i=1,2,3\,,~~\label{gauge-condi}
\eea
as can be seen in \cite{Berends:1987me,Wu:2021exa}.
This is the additional constraint on ${\cal J}_{I_i}\cdot K_i$ at $V$.

We can alternatively understand the effect of this trivalent vertex $V$
as in the well known Berends-Giele recursion. Two currents ${\cal J}^\mu_{I_1}$ and ${\cal J}^\mu_{I_2}$ enters $V$, create a Lorentz vector ${\cal V}^\mu_{I_1|I_2}$. Then ${\cal J}^\mu_{I_1|I_2}={\cal V}^\mu_{I_1|I_2}/s_{I_1I_2}$ enters the next vertex, interacts with other currents at this new vertex. Obviously, ${\cal J}^\mu_{I_1|I_2}$ serves as a component of the current ${\cal J}^\mu_{I_1\cup I_2}$ for the bigger block $I_1\cup I_2$, where $I_1\cup I_2$ is the block involves all external lines belong to $I_1$ or $I_2$. As an example, consider the term
\bea
{1\over s_{12}}(\epsilon_1\cdot k_2)\,(\epsilon_2\cdot \epsilon_3)\,(k_3\cdot\epsilon_4)\,,~~\label{example}
\eea
which contributes to the $4$-point amplitude. One can understand the interactions in this example as follows. Two currents ${\cal J}_1^\mu=\epsilon_1^\mu$ and ${\cal J}_2^\mu=\epsilon_2^\mu$ meet at the first vertex, create ${\cal V}^\mu_{1|2}$ which includes a component $(\epsilon_1\cdot k_2)\epsilon_2^\mu$. Then, ${\cal J}^\mu_{1|2}={\cal V}^\mu_{1|2}/s_{12}$ and ${\cal J}^\mu_3=\epsilon^\mu_3$ interact at the second vertex, generates ${\cal V}^\mu_{(1|2)|3}$
which contains a component $(\epsilon_1\cdot k_2)(\epsilon_2\cdot \epsilon_3)k^\mu_3/s_{12}$. This component is contracted with $\epsilon_4^\mu$, finally yields the term \eref{example}.

With understandings discussed above, let us focus on the effect of vertex $V_A$ in the graph $(a)$ of Fig.\ref{VertexYM}. We begin with the case that the block $I$ in graph $(a)$ is the external gluon $k$. For this special case, two currents ${\cal J}^\mu_k=\epsilon^\mu_k$ and ${\cal J}^\mu_{A}$ enters $V_A$, and the interaction creates the the outgoing vector ${\cal V}^\mu_{A|k}$. There are three types of possible contractions at $V_A$. The first one, ${\cal J}^\mu_k=\epsilon^\mu_k$ is contracted with ${\cal J}^\mu_{A}$, this contraction contributes
\bea
P^\mu_{A;1}=\alpha_1^A\,(\epsilon_k\cdot{\cal J}_A)\,K^\mu_A\,,~~\label{piece1}
\eea
where the locality and the momentum conservation require that the combinatorial momentum $K^\mu_A$ can only depend on $k^\mu_k$ and $k^\mu_k+k^\mu_A$ which enters $V_A$ (here it is more convenience to use $k^\mu_k+k^\mu_A$ rather than $k^\mu_A$). However, the gauge invariance condition in \eref{gauge-condi} indicates
the vanishing of contribution from $k^\mu_k+k^\mu_A$, since $(k_k+k_A)\cdot{\cal J}_R=0$, where ${\cal J}^\mu_R$ is the current contributed by the remaining block $R$ in the diagram $(a)$ of Fig.\ref{VertexYM}. Therefore, the effective part of the Lorentz vector $K^\mu_A$ in \eref{piece1} only depends on $k^\mu_A$ which is purely in ${\cal S}_1$. Meanwhile, $\epsilon_k\cdot{\cal J}_A$ in \eref{piece1} is purely in ${\cal S}_1$, and the mass dimension
of the vertex forces $\a_1^A$ to be a dimensionless constant.

The second type of contractions is, $\epsilon_k^\mu$ is contracted with a momentum $K'^\mu_A$, which leads to
\bea
P^\mu_{A;2}=\alpha_2^A\,(\epsilon_k\cdot K'_{A})\,{\cal J}^\mu_A\,.~~\label{piece2}
\eea
Here $\epsilon_k\cdot K'_{A}$ is an object in ${\cal S}_1$, since $\epsilon_k$ in ${\cal S}_1$ annihilates ${\cal S}_2$ component of $K'_{A}$.
Indeed, the effective part of $K'_{A}$ always lies in ${\cal S}_1$, regardless of if $\epsilon_k$ is in ${\cal S}_1$. The reason is, the locality and the momentum conservation require $K'_{A}$ to depend on $k_k$ and $k_A$, but the on-shell condition $\epsilon_k\cdot k_k=0$ eliminates
$k_k$ (when $\epsilon_k$ is replaced by the more general current ${\cal J}_I$, the on-shell condition should be replaced by the gauge invariance condition ${\cal J}_I\cdot k_I=0$). Thus, the effective part of $K'_{A}$ only depends on $k_A$. The coefficient $\alpha_2^A$ is also a dimensionless constant.

The third type of contraction is, ${\cal J}_A^\mu$ is contracted with a momentum $K''^\mu_A$, this possibility yields
\bea
P^\mu_{A;3}=\alpha_3^A\,({\cal J}_A\cdot K''_{A})\,\epsilon^\mu_k\,.~~\label{piece3}
\eea
In the above, ${\cal J}_A\cdot K''_{A}$ is purely in ${\cal S}_1$. The locality and the momentum conservation, as well as the gauge invariance requirement ${\cal J}_A\cdot k_A=0$, guarantee that the effective part of $K''^\mu_{A}$ only depends on $k^\mu_k$. Again, the coefficient $\alpha_3^A$ is a dimensionless constant.

In summary, two vector currents $\epsilon_k^\mu$ and ${\cal J}_A^\mu$ interact at the vertex $V_A$, create a Lorentz vector ${\cal V}^\mu_{A|k}$. Each term of ${\cal V}^\mu_{A|k}$ carries a Lorentz vector purely in ${\cal S}_1$, and all Lorentz invariants formed at $V_A$ are also purely in ${\cal S}_1$.
The current ${\cal J}^\mu_{A|k}={\cal V}^\mu_{A|k}/s_{Ak}$ enters the next vertex, and serves as a component of ${\cal J}^\mu_{A\cup k}$ for the block $A\cup k$.

Then we move to the vertex in graph $(b)$ of Fig.\ref{VertexYM}, where a block $B$ from ${\cal S}_2$ is planted to $L_{k,v}$. Again, we consider the special case the block $I$ is the external gluon $k$. The effect at such $V_B$ can be directly achieved by replacing all $A$ in \eref{piece1},
\eref{piece2} and \eref{piece3} by $B$. However, since $\epsilon_k$ lies in ${\cal S}_1$, contributions in \eref{piece1} and
\eref{piece2} vanish if $A$ is replaced by $B$ (recall that the locality, the momentum conservation and the on-shell condition reduce the effective part of $K'^\mu_{A}$ in \eref{piece2} to be $k^\mu_A$, thus the effective part of corresponding $K'^\mu_{B}$ is $k^\mu_B$, this observation ensures $\epsilon_k\cdot K'_{B}=0$). On the other hand, ${\cal J}_B\cdot K''_{B}$ in $P_{B;3}^\mu$, which is parallel to \eref{piece3}, is purely in ${\cal S}_2$. Consequently, each term in the Lorentz vector ${\cal V}^\mu_{k|B}$ created at the vertex $V_B$, carries a Lorentz vector $\epsilon^\mu_k$ purely in ${\cal S}_1$, while the Lorentz invariant created at this vertex is purely in ${\cal S}_2$. The current ${\cal J}^\mu_{k|B}={\cal V}^\mu_{k|B}/s_{kB}$ is a component contributes to the entire ${\cal J}^\mu_{k\cup B}$.

Now we consider the more general case that the block $I$ also receives external legs from ${\cal S}_1$ and ${\cal S}_2$, as shown in Fig.\ref{VertexYM}. For this case, one should replace $\epsilon_k$ in \eref{piece1}, \eref{piece2} and \eref{piece3} by ${\cal J}_I$. If we assume that each term of ${\cal J}_I^\mu$ carries a Lorentz vector purely in ${\cal S}_1$, then the previous discussion for effects at $V_A$ and $V_B$ still holds,
we can conclude that the resulting ${\cal V}^\mu_{A|I}$ or ${\cal V}^\mu_{I|B}$ also carries Lorentz vectors purely in ${\cal S}_1$. Meanwhile, all Lorentz invariants created at $V_A$ are independent of blocks from ${\cal S}_2$, while all Lorentz invariants created at $V_B$ are independent of blocks from ${\cal S}_1$.

The above analysis can be directly generalized to type (a) and type (b) quadrivalent vertices in Fig.\ref{4p-vert}, along the line $L_{k,v}$. Let us again assume that ${\cal J}_I^\mu$ only carries Lorentz vectors in ${\cal S}_1$. Then, at a type (a) vertex $V_A$ with mass dimension $0$, the interaction between Berends-Giele currents ${\cal J}^\mu_{A_1}$, ${\cal J}^\mu_{A_2}$ and ${\cal J}^\mu_I$ creates ${\cal V}_{A_2|A_1|I}^\mu$, which carries Lorentz vectors purely in ${\cal S}_1$, and all Lorentz invariants created at this vertex are also in ${\cal S}_1$. Similarly, at a type (b) vertex $V_B$, the interaction between Berends-Giele currents ${\cal J}^\mu_{B_1}$, ${\cal J}^\mu_{B_2}$ and ${\cal J}^\mu_I$ creates ${\cal V}_{I|B_1|B_2}^\mu$, which carries the Lorentz vector ${\cal J}^\mu_I$ in ${\cal S}_1$, while the Lorentz invariant ${\cal J}^\mu_{B_1}\cdot{\cal J}^\mu_{B_2}$ created at this vertex is in ${\cal S}_2$.

The above statements for effects at trivalent and quadrivalent vertices, are based on the assumption that ${\cal J}_I^\mu$ only carries Lorentz vectors in ${\cal S}_1$. This assumptions is satisfied
by ${\cal J}^\mu_I=\epsilon^\mu_k$, where the block $I$ is reduced to the single external gluon $k$. On the other hand, as argued above, ${\cal V}^\mu_{A|I}$, ${\cal V}^\mu_{I|B}$, ${\cal V}_{A_2|A_1|I}^\mu$ and ${\cal V}_{I|B_1|B_2}^\mu$ solely carry Lorentz vectors in ${\cal S}_1$ if ${\cal J}_I^\mu$ does. Therefore, one can start from the external gluon $k$, move along the line $L_{k,v}$, and recursively ensure that each ${\cal J}^\mu_I$ only carries Lorentz vectors in ${\cal S}_1$.

Clearly, the similar discussion and conclusion is valid for vertices on $L_{i,v}$ and $L_{j,v}$, except $v$. For vertices on $L_{i,v}$, the situation is simpler, since they only receive blocks from ${\cal S}_2$. We emphasize that the Lorentz invariants created at vertices on $L_{i,v}$
are independent on replacing external momentum $k_i$ by $k^*_i$ given in \eref{offi-R}.
The reason is as follows. The Lorentz invariant created at a trivalent vertex on $L_{i,v}$ is ${\cal J}_B\cdot K''_{B}$, based on the argument similar as that for obtaining ${\cal J}_A\cdot K''_{A}$ in \eref{piece3}. Then, ${\cal J}_B$ annihilates the ${\cal S}_1$ component of $k_I$, eliminates all $k_a$ in \eref{offi-R}. Meanwhile, the Lorentz invariant created at a quadrivalent vertex (with mass dimension $0$) on $L_{i,v}$ also has only one candidate ${\cal J}_{B_1}\cdot {\cal J}_{B_2}$, since ${\cal J}_I$ in ${\cal S}_1$ annihilates either ${\cal J}_{B_1}$ or ${\cal J}_{B_2}$, where ${\cal J}_I$, ${\cal J}_{B_1}$ and ${\cal J}_{B_2}$ stands for three currents enter this quadrivalent vertex. This candidate is manifestly independent of shifting $k_i$ to $k_i^*$.

To finish the consideration for effects of vertices on $L_{i,v}$, $L_{j,v}$ and $L_{k,v}$, we finally discuss the Lorentz invariants created at $v$.
We first consider the case that $v$ is a trivalent vertex, and think the interaction at $v$ as three currents ${\cal J}^\mu_{I_i}$, ${\cal J}^\mu_{I_j}$ and ${\cal J}^\mu_{I_k}$ meet at $v$, where the currents ${\cal J}^\mu_{I_\ell}$ with $\ell=i,j,k$ arise from blocks $I_\ell$ in the right diagram of Fig.\ref{vertex-Y}.
The previous conclusion for effects of vertices shows that ${\cal J}^\mu_{I_i}$, ${\cal J}^\mu_{I_j}$ and ${\cal J}^\mu_{I_k}$ solely carry Lorentz vectors purely in ${\cal S}_1$. Therefore, each Lorentz invariant created at $v$ is an object in ${\cal S}_1$, independent of any block from ${\cal S}_2$, due to the interaction discussed around \eref{T-V}. If $v$ is a quadrivalent vertex, there are four currents interact at it, like in \eref{3term}. Besides ${\cal J}^\mu_{I_i}$, ${\cal J}^\mu_{I_j}$ and ${\cal J}^\mu_{I_k}$, the fourth current should be purely in ${\cal S}_1$ or ${\cal S}_2$, as indicated by the scheme of merging/splitting in Fig.\ref{splitYM}. When it is in ${\cal S}_2$, then \eref{3term} vanishes. When it is in ${\cal S}_1$, then each Lorentz invariant created at $v$ is again in ${\cal S}_1$.

As in section \ref{subsec-verification-phi3}, for an $n$-point YM amplitude, we can represent the summation over Feynman diagrams
as the summation for all possible complete sets of blocks (c.s.b),
\bea
{\cal A}_n^{\rm YM}&=&\sum_{\rm c.s.b}\,\sum_{\rm L.I}\,\Big(\sum_{\Gamma^{k,v}_{x_k,y_k}}\,\prod_{t=1}^{x_k+y_k}{1\over{\cal D}^{k,v}_t}\Big)\,\Big(\sum_{\Gamma^{j,v}_{x_j,y_j}}\,\prod_{t=1}^{x_j+y_j}{1\over{\cal D}^{j,v}_t}\Big)\,\Big(\sum_{\Gamma^{i,v}_{y_i}}\,{\prod_{s=1}^{y_i}\,{\cal L}_{V_{B^i_s}}\over\prod_{t=1}^{y_i}\,{\cal D}^{i,v}_t}\Big)\nn
& &~~~~~~~~~~~~\Big(\prod_{\ell=j,k}\,\prod_{s=1}^{x_\ell}\,{\cal L}_{V_{A^\ell_s}}\Big)\,
\Big(\prod_{\ell=j,k}\,\prod_{s=1}^{y_\ell}\,{\cal L}_{V_{B^i_s}}\Big)\,{\cal L}_v\,\Big(\prod_{\ell=j,k}\,\prod_{r=1}^{m_\ell}\,f_{A^\ell_r}\Big)\,
\Big(\prod_{\ell=i,j,k}\,\prod_{r=1}^{h_\ell}\,f_{B^\ell_r}\Big)\,.~~\label{sum-diagrams-YM}
\eea
In the above, the blocks are labeled as
\bea
& &A^\ell_r~~~~{\rm if}~{\rm it}~{\rm is}~{\rm planted}~{\rm to}~L_{\ell,v}\,,~{\rm and}~{\rm lies}~{\rm in}~{\cal S}_1\,,\nn
& &B^\ell_r~~~~{\rm if}~{\rm it}~{\rm is}~{\rm planted}~{\rm to}~L_{\ell,v}\,,~{\rm and}~{\rm lies}~{\rm in}~{\cal S}_2\,.
\eea
Furthermore, we encode vertices one three lines (except $v$) as
\bea
& &V_{A^\ell_s}~~~~{\rm if}~{\rm it}~{\rm lies}~{\rm on}~L_{\ell,v}\,,~{\rm and}~{\rm receives}~{\rm blocks}~{\rm in}~{\cal S}_1\,,\nn
& &V_{B^\ell_s}~~~~{\rm if}~{\rm it}~{\rm lies}~{\rm on}~L_{\ell,v}\,,~{\rm and}~{\rm receives}~{\rm blocks}~{\rm in}~{\cal S}_2\,.
\eea
The numbers of blocks $A^\ell_r$ and $B^\ell_r$ are denoted as $m_\ell$ and $h_\ell$, while the numbers of vertices $V_{A^\ell_s}$ and $V_{B^\ell_s}$ are labeled as $x_\ell$ and $y_\ell$. Notice that $x_\ell\neq m_\ell$, $y_\ell\neq h_\ell$, since vertices can be quadrivalent ones.
In \eref{sum-diagrams-YM}, each rational function $f$ stands for the Lorentz invariant contributed by the corresponding Berends-Giele current, while each ${\cal L}$ is understood as the Lorentz invariant created the corresponding vertex. The second summation is for all possible Lorentz invariants carried by Berends-Giele currents and created at vertices. For example, suppose there are three currents
\bea
{\cal J}_1^\mu&=&g_{11}\,v_{11}^\mu+g_{12}\,v_{12}^\mu\,,\nn
{\cal J}_2^\mu&=&g_2\,v_2^\mu\,,\nn
{\cal J}_3^\mu&=&g_3\,v_3^\mu\,,
\eea
interact at the vertex $V$ as
\bea
A_{1|2|3}=({\cal J}_1\cdot {\cal J}_3)\,({\cal J}_2\cdot K)\,.
\eea
Then we have
\bea
A_{1|2|3}&=&\sum_{\rm L.I}\,f_1\,f_2\,f_3\,{\cal L}_V\nn
&=&g_{11}\,g_2\,g_3\,(v_{11}\cdot v_3)\,(v_2\cdot K)+g_{12}\,g_2\,g_3\,(v_{12}\cdot v_3)\,(v_2\cdot K)\,.
\eea
In the above simple example, the summation contains two terms. In the first one, we have
\bea
f_1=g_{11}\,,~~f_2=g_2\,,~~f_3=g_3\,,~~{\cal L}_V=(v_{11}\cdot v_3)\,(v_2\cdot K)\,.
\eea
In the second term, we have
\bea
f_1=g_{12}\,,~~f_2=g_2\,,~~f_3=g_3\,,~~{\cal L}_V=(v_{12}\cdot v_3)\,(v_2\cdot K)\,.
\eea
It is worth to emphasize that, the previous conclusion each ${\cal L}_{V_{A^\ell_s}}$ (${\cal L}_{V_{B^\ell_s}}$) is independent of blocks from ${\cal S}_2$ (${\cal S}_1$), indicates that
changing the positions of planting blocks from ${\cal S}_2$ (${\cal S}_1$) will never alter any ${\cal L}_{V_{A^\ell_s}}$ (${\cal L}_{V_{B^\ell_s}}$). This is the reason why we need not to include $\prod_{s=1}^{x_\ell}\,{\cal L}_{V_{A^\ell_s}}$ and $\prod_{s=1}^{y_\ell}\,{\cal L}_{V_{B^\ell_s}}$ in the second line of \eref{sum-diagrams-YM}, when summing over diagrams $\Gamma^{k,v}_{x_k,y_k}$ and $\Gamma^{j,v}_{x_j,y_j}$
in the first line.

Substituting the factorization of propagators found in \eref{fac-mkb} into \eref{sum-diagrams-YM}, we get the splitting
\bea
{\cal A}_n^{\rm YM}&=&\sum_{\rm c.s.b}\,\sum_{\rm L.I}\,\Big[\,\Big(\prod_{a=1}^{x_k}\,{1\over s_{kA^k_1\cdots A^k_a}}\Big)\,\Big(\prod_{a=1}^{x_j}\,{1\over s_{jA^j_1\cdots A^j_a}}\Big)\,\Big(\prod_{\ell=j,k}\,\prod_{s=1}^{x_\ell}\,{\cal L}_{V_{A^\ell_s}}\Big)\,{\cal L}_v\,\Big(\prod_{\ell=j,k}\,\prod_{r=1}^{m_\ell}\,f_{A^\ell_r}\Big)\,\Big]\nn
& &~~~~~~~~~~~~\Big[\,\Big(\prod_{b=1}^{y_k}\,{1\over s_{kB^k_1\cdots B^k_b}}\Big)\,\Big(\prod_{b=1}^{y_j}\,{1\over s_{jB^j_1\cdots B^j_b}}\Big)\,
\Big(\sum_{\Gamma^{i,v}_{y_i}}\,{\prod_{s=1}^{y_i}\,{\cal L}_{V_{B^i_s}}\over\prod_{t=1}^{y_i}\,{\cal D}^{i,v}_t}\Big)\nn
& &~~~~~~~~~~~~\Big(\prod_{\ell=j,k}\,\prod_{s=1}^{y_\ell}\,{\cal L}_{V_{B^\ell_s}}\Big)\,
\Big(\prod_{\ell=i,j,k}\,\prod_{r=1}^{h_\ell}\,f_{B^\ell_r}\Big)\,\Big]\,.
\eea
Therefore, the resulting $A^{{\cal S}_1}_{n_1}$ and $A^{{\cal S}_2}_{n+3-n_1}$ in the $2$-split \eref{facYM-nadj-sub}, are identified as
\bea
& &A^{{\cal S}_1}_{n_1}(i,j,\cdots,k)\nn
&=&\sum_{\rm c.s.b}\,\sum_{\rm L.I}\,\Big(\prod_{a=1}^{x_k}\,{1\over s_{kA^k_1\cdots A^k_a}}\Big)\,\Big(\prod_{a=1}^{x_j}\,{1\over s_{jA^j_1\cdots A^j_a}}\Big)\,\Big(\prod_{\ell=j,k}\,\prod_{s=1}^{x_\ell}\,{\cal L}_{V_{A^\ell_s}}\Big)\,{\cal L}_v\,\Big(\prod_{\ell=j,k}\,\prod_{r=1}^{m_\ell}\,f_{A^\ell_r}\Big)\,,~~\label{JLY}
\eea
and
\bea
& &A^{{\cal S}_2}_{n+3-n_1}(k,\cdots,j)\nn
&=&\sum_{\rm c.s.b}\,\sum_{\rm L.I}\,\Big(\prod_{b=1}^{y_k}\,{1\over s_{kB^k_1\cdots B^k_b}}\Big)\,\Big(\prod_{b=1}^{y_j}\,{1\over s_{jB^j_1\cdots B^j_b}}\Big)\,
\Big(\sum_{\Gamma^{i,v}_{y_i}}\,{\prod_{s=1}^{y_i}\,{\cal L}_{V_{B^i_s}}\over\prod_{t=1}^{y_i}\,{\cal D}^{i,v}_t}\Big)\,\Big(\prod_{\ell=j,k}\,\prod_{s=1}^{y_\ell}\,{\cal L}_{V_{B^\ell_s}}\Big)\,
\Big(\prod_{\ell=i,j,k}\,\prod_{r=1}^{h_\ell}\,f_{B^\ell_r}\Big)\,.~~\label{JRY}
\eea
From the ${\cal S}$ point of view, the amplitude $A^{{\cal S}_1}_{n_1}$ in \eref{JLY} absolutely satisfies Feynman rules of pure YM theory, since any Lorentz invariant in $\{{\cal L}_{V_{A^\ell_s}},{\cal L}_v\}$
with $\ell=i,j,k$ is independent of blocks from ${\cal S}_2$. Hence, $A^{{\cal S}_1}_{n_1}$ is interpreted as the contraction between polarization
vector $\epsilon^\mu_i$, and the vector current $\W{\cal J}^\mu_{n_1}$
of pure YM theory which carries an off-shell external momentum $k_i^*$, i.e.,
\bea
A^{\rm YM}_{n_1}(i,j,\cdots,k)=\epsilon_i\cdot\W{\cal J}^{\rm YM}_{n_1}(i^*,j,\cdots,k)\,.
\eea
On the other hand, the amplitude $A^{{\cal S}_2}_{n+3-n_1}$ can not be a purely
YM one, due to the lacking of ${\cal L}_v$ which means the vertex $v$ only contributes a constant $1$, and the lacking of external polarizations $\epsilon_\ell$ (each ${\cal L}_{V_{B^\ell_s}}$ is clearly independent of $\epsilon_\ell$). To understand $A^{{\cal S}_2}_{n+3-n_1}$, we consider the special case $\{k+1,\cdots,j-1\}=i$, which forces ${\cal J}^{\rm YM}_{n_1}$ to be ${\cal A}^{\rm YM}_n$ itself, therefore reduces $A^{{\cal S}_2}_{n+3-n_1}$ to the constant $1$ which can be recognized as the $3$-point ${\rm Tr}(\phi^3)$ amplitude. The general case with un-empty $\{k+1,\cdots,j-1\}\setminus i$ can be thought as coupling blocks from ${\cal S}_2$ to this $3$-point ${\rm Tr}(\phi^3)$ skeleton, since each Feynman diagram can be understood as planting such blocks to lines $L_{\ell,v}$. Therefore,
$A^{{\cal S}_2}_{n+3-n_1}$ is interpreted as the current of YM$\oplus{\rm Tr}(\phi^3)$ theory, where $i^*$, $j$, $k$ are three ${\rm Tr}(\phi^3)$ scalars, while particles in $\{k+1,\cdots,i-1\}$ and $\{i+1,\cdots,j-1\}$ are gluons. Again, $k_i^*$ is an off-shell external momentum. Consequently, we obtain the $2$-split in \cite{Cao:2024qpp},
\bea
{\cal A}^{\rm YM}_{n}(1,\cdots,n)\,\xrightarrow[]{\eref{kinematic-condi-fac-YM}\,,\,\eref{kinematic-condi-fac-YM-add}}\,\epsilon_i\cdot{\cal J}^{\rm YM}_{n_1}(i^*_g,j_g,\dots,k_g)\,\times\,{\cal J}^{{\rm YM\oplus\phi^3}}_{n+3-n_1}(k_\phi,\cdots,i^*_\phi,\cdots,j_\phi)\,,~~~~\label{1to2-YM-fix}
\eea
as diagrammatically illustrated in Fig.\ref{splitYM}. Notice that the discussion for off-shell momenta in the $2$-split of ${\rm Tr}(\phi^3)$
amplitudes, from \eref{offi-L} to \eref{additionM},
also holds for the above $2$-split \eref{1to2-YM-fix}.

Suppose $\epsilon_\ell$ with $\ell=i,j,k$ are in ${\cal S}_2$, namely,
\bea
& &\epsilon_\ell\cdot\epsilon_a=\epsilon_\ell\cdot k_a=0\,,~~~~{\rm for}~\ell\in\{i,j,k\}\,,~~~~a\in\{j+1,\cdots,k-1\}.~~\label{kinematic-condi-fac-YM-add2}
\eea
the analogous treatment gives rise to the similar splitting
\bea
{\cal A}^{\rm YM}_{n}(1,\cdots,n)\,\xrightarrow[]{\eref{kinematic-condi-fac-YM}\,,\,\eref{kinematic-condi-fac-YM-add2}}\,{\cal J}^{\rm YM\oplus\phi^3}_{n_1}(i^*_\phi,j_\phi,\dots,k_\phi)\,\times\,\epsilon_i\cdot{\cal J}^{{\rm YM}}_{n+3-n_1}(k_g,\cdots,i^*_g,\cdots,j_g)\,,~~~~\label{1to2-YM-fix2}
\eea

The similar argument also fix resulting amplitudes in \eref{facYM-adj-sub} and \eref{1to3-YM} as
\bea
{\cal A}^{\rm YM}_{n}(1,\cdots,2n)\,\xrightarrow[]{\eref{kinematic-condi-fac-YM-adj}\,,\,\eref{kinematic-condi-fac-YM-add-adj}}\,\epsilon_i\cdot{\cal J}^{\rm YM}_{n_1}(i^*_g,i+1_g,\dots,k_g)\,\times\,{\cal J}^{\rm YM\oplus\phi^3}_{n+3-n_1}(k_\phi,\cdots,i_\phi,i+1^*_\phi)\,,~~~~\label{1to2express-YM}
\eea
and
\bea
& &{\cal A}^{\rm YM}_{n}(1,\cdots,2n)\,\xrightarrow[]{\eref{condi-1to3-YM}\,,\,\eref{condi-1to3-YM-2}}\nn
& &~~~~~~~~~~~~~~~~~~{\cal J}^{\rm YM\oplus\phi^3}_{n_1}(i^*_\phi,j_\phi,\dots,k_\phi)\,\times\,\epsilon_j\cdot{\cal J}^{{\rm YM}}_{n_2}(j^*_g,k_g,\cdots,i_g)\,\times\,{\cal J}^{{\rm YM}\oplus\phi^3}_{n+6-n_1-n_2}(k^*_\phi,i_\phi,\cdots,j_\phi)\,,~~\label{1to3express-YM}
\eea
The above $3$-split in \eref{1to3express-YM}, is the generalization of smooth splitting of scalar amplitudes in \cite{Cachazo:2021wsz},
to YM amplitudes.
Similar as in section \ref{subsec-fac-zero}, this $3$-split \eref{1to3express-YM} can be verified by further splitting ${\cal J}^{\rm YM}_{n+3-n_1}$ in \eref{1to2-YM-fix2}, in the manner of \eref{1to2express-YM}. On the other hand, further splitting
also leads to factorizations near zeros discovered in \cite{Arkani-Hamed:2023swr}.

Finally, the technic in this subsection also yields another interpretation for zeros of YM amplitudes, which serves as the generalization of the method at the end of section \ref{subsec-verification-phi3}. Based on the merging in Fig.\ref{share2}, we can formulate the YM amplitude as
\bea
{\cal A}^{\rm YM}_n(1,\cdots,n)=\sum_{\rm c.s.b}\,\Big[\sum_{\Gamma^{i,j}_{x,y}}\,{\big(\prod_{s'=1}^{x}\,{\cal L}_{V_{A_{s'}}}\big)\,\big(\prod_{s''=1}^{y}\,{\cal L}_{V_{B_{s''}}}\big)\over\prod_{t=1}^{x+y-1}\,{\cal D}^{i,j}_t}\Big]\,\Big(\prod_{r=1}^m\,f_{A_r}\Big)
\,\Big(\prod_{r=1}^{h}\,f_{B_r}\Big)\,,~~\label{np-YM-for0}
\eea
where the set of blocks planted onto $L_{i,j}$ is $\{A_1,\cdots,A_m\}\cup\{B_1,\cdots,B_h\}$, while the set of vertices on $L_{i,j}$ is
$\{V_{A_1},\cdots V_{A_x}\}\cup\{V_{B_1},\cdots,V_{B_y}\}$. In the above, the Lorentz invariants created at vertices $V_{A_s}$ and $V_{B_s}$ are included when summing over diagrams $\Gamma^{i,j}_{x,y}$, since the polarization vectors $\epsilon_i$ and $\epsilon_j$ are kept to be general. However, one can always decompose $\epsilon_i$ into components in orthogonal subspaces, namely, $\epsilon_i=\epsilon^{{\cal S}_1}_i+\epsilon^{{\cal S}_2}_i$. This decomposition separates ${\cal A}^{\rm YM}_n$ in \eref{np-YM-for0} into two parts with $a=1$ and $a=2$,
\bea
{\cal A}^{\rm YM}_n(1,\cdots,n)=\sum_{a=1,2}\,\sum_{\rm c.s.b}\,\Big(\sum_{\Gamma^{i,j}_{x,y}}\,{1\over\prod_{t=1}^{x+y-1}\,{\cal D}^{i,j}_t}\Big)\,\Big(\prod_{s=1}^{x}\,{\cal L}^a_{V_{A_{s}}}\Big)\,\Big(\prod_{s=1}^{y}\,{\cal L}^a_{V_{B_{s}}}\Big)\,\Big(\prod_{r=1}^m\,f_{A_r}\Big)
\,\Big(\prod_{r=1}^{h}\,f_{B_r}\Big)\,,~~\label{np-YM-for0-2}
\eea
where ${\cal L}^a_{V_{A_{s}}}$ and ${\cal L}^a_{V_{B_{s}}}$ correspond to the external polarization vector $\epsilon_i^{{\cal S}_a}$. When the polarization carried by $i$ is $\epsilon_i^{{\cal S}_a}$, localized in ${\cal S}_a$, we can directly apply the previous conclusion: each ${\cal L}_{V_{A_s}}$ (${\cal L}_{V_{B_s}}$) is independent of blocks from ${\cal S}_2$ (${\cal S}_1$), therefore is invariant for all diagrams $\Gamma^{i,j}_{x,y}$. Thus, the summation over $\Gamma^{i,j}_{x,y}$ in \eref{np-YM-for0-2} does not include ${\cal L}^a_{V_{A_s}}$ and ${\cal L}^a_{V_{B_s}}$. Then we can use the method at the end of section \ref{subsec-verification-phi3}, as shown in \eref{fro0-step0}, \eref{for0-step1} and \eref{for0-step2}, to prove that \eref{np-YM-for0-2} vanishes when $s_{iA_1\cdots A_m B_1\cdots B_h}=k_j^2=0$.

\section{NLSM amplitudes}
\label{sec-NLSM}

In this section, we study zeros and splittings of ordered tree level NLSM amplitudes.
To specify the theory under consideration, here we give the Lagrangian of NLSM in the Cayley parametrization,
\bea
{\cal L}_{\rm NLSM}={1\over 8\lambda^2}\,{\rm Tr}(\partial_\mu {\rm U}^\dag\partial^\mu {\rm U})\,,~~\label{Lag-N}
\eea
with
\bea
{\rm U}=(\mathbb{I}+\lambda\Phi)\,(\mathbb{I}-\lambda\Phi)^{-1}\,,
\eea
where $\mathbb{I}$ is the identity matrix, and $\Phi=\phi_IT^I$, with $T^I$ the generators of $U(N)$.
The Lagrangian \eref{Lag-N} indicates that the mass dimension of the kinematic part of any $n$-point NLSM amplitude, with coupling constants stripped
off, is always ${\cal D}=2$.

The most convenient way for getting zeros and splittings of NLSM amplitudes,
is to use the elegant observation that a tree NLSM amplitude ${\cal A}_{2n}^{\rm NLSM}$ can be generated from the ${\rm Tr}(\phi^3)$ one ${\cal A}_{2n}^{\phi^3}$, via a simple shift of kinematic variables \cite{Arkani-Hamed:2023swr,Arkani-Hamed:2024nhp}. More explicitly,
\bea
{\cal A}_{2n}^{\rm NLSM}(1,\cdots,2n)\,&=&\,\lim_{\delta\to\infty}\,\delta^{2n-2}\,{\cal A}^{\phi^3}_{2n}(s_{e\cdots o}\to s_{e\cdots o}+\delta,s_{o\cdots e}\to s_{o\cdots e}-\delta)\,.~~\label{phi3-NLSM}
\eea
In the above, the subscript $e$ stands for an even number, while $o$ denotes an odd one. For example, $s_{12}$ is shifted as $s_{12}+\delta$,
while $s_{2345}$ is shifted as $s_{2345}-\delta$.
As can be directly verified, the shift in \eref{phi3-NLSM} has no influence on conditions \eref{kinematic-condi-zero}, \eref{kinematic-condi-1to2}, \eref{kinematic-condi2-1to2} and \eref{condi-1to3split}. It means the manipulation in \eref{phi3-NLSM} is commutable with imposing these conditions, therefore the loci in kinematic space for zeros and splittings also hold for the NLSM case. Such commutativity allows us to apply \eref{phi3-NLSM} to factorized ${\rm Tr}(\phi^3)$ amplitudes in \eref{1to2express-fix}, \eref{1to2express-2-fix} and \eref{1to3express-fix}, to generate the resulting currents in splittings of NLSM amplitudes.

Along the line in previous sections, we want to understand zeros and splittings of tree NLSM amplitudes via Feynman diagrams. In subsection \ref{subsec-zero-NLSM}, we extend the arguments in sections \ref{subsec-zero-phi3}, \ref{subsec-1to2-phi3} and \ref{subsec-fac-zero}
to the current NLSM case, to determine the loci in kinematic space, for zeros and splittings. Then, subsection \ref{subsec-flow-NLSM}
aims to figure out certain interpretation for resulting amplitudes in splittings. The NLSM theory involves an infinite tower of vertices, thus it
is not wise to derive resulting amplitudes by using explicit Feynman rules. An attractive idea is to extend the method in sections 
\ref{subsec-verification-phi3} and \ref{subsec-flow-YM} by utilizing locality and the specific soft behavior called Adler zero,
since it is well known any tree NLSM amplitude can be fully fixed by locality and Adler zero \cite{Cheung:2014dqa,Cheung:2015ota}.
Unfortunately, for general $2n$-point NLSM amplitudes, it is hard to realize the above goal, due to various technical difficulties. Thus we leave
this idea to the future work. Instead, in this section we determine
the resulting amplitudes via another alternative method: besides performing the shift \eref{phi3-NLSM} to factorized ${\rm Tr}(\phi^3)$ amplitudes, the zeros and splittings of NLSM amplitudes can also be generated by acting transmutation operators
which turn YM and YM$\oplus$BAS amplitudes to NLSM and NLSM$\oplus$BAS ones, to zeros and splittings of YM amplitudes discussed in section \ref{subsec-flow-YM}.

\subsection{Kinematic conditions for zeros and splittings}
\label{subsec-zero-NLSM}

%
\begin{figure}
  \centering
   \includegraphics[width=8cm]{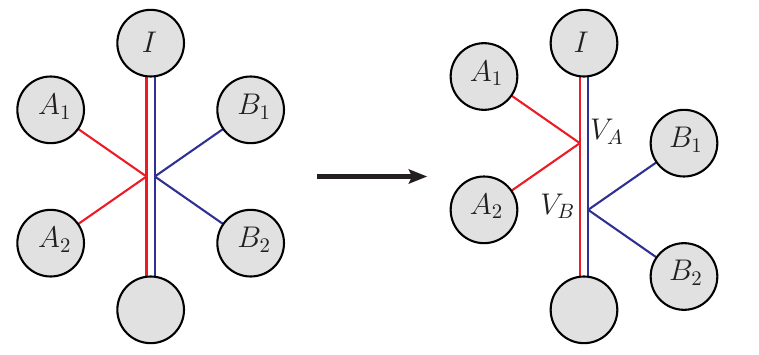}
   ~~~~
   \includegraphics[width=8cm]{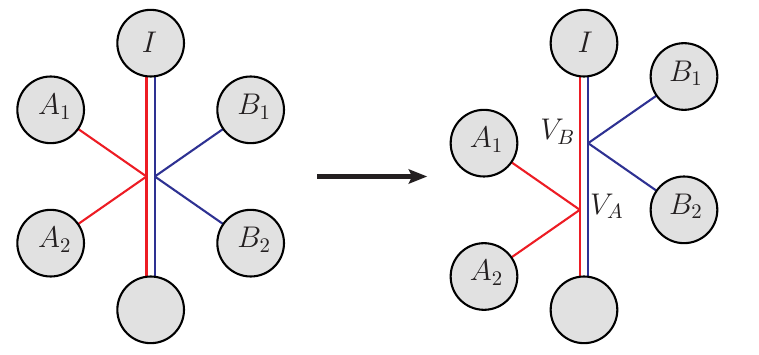} \\
  \caption{Split the $6$-point NLSM vertex along the glued line, into lower-point ones.}\label{split6pvNL}
\end{figure}

In this subsection, we use the idea of orthogonal spaces to determine the certain loci in kinematic space for zeros and splittings of tree NLSM amplitudes. Again, the argument can not determine the corresponding theory (theories) of resulting amplitudes.

As well known, all vertices in NLSM theory are even valence ones, with mass dimension $2$. The even valence vertices along the glued line can receive blocks from ${\cal S}_1$ and ${\cal S}_2$ simultaneously, thus prevent us from regarding the merging as the spurious one without introducing any interaction between pieces in ${\cal S}_1$ and ${\cal S}_2$. Similar as in the YM case, to extend the idea in section \ref{subsec-zero-phi3} to the NLSM case, we can use the technic of inserting ${\cal D}/{\cal D}$, to separate vertices along the glued line which receive blocks from both ${\cal S}_1$ and ${\cal S}_2$, into lower-point vertices which receive blocks solely from one of ${\cal S}_1$ and ${\cal S}_2$. Let us explain that the lower-point vertices obtained in the above separation satisfy the locality. Since NLSM is a theory of scalars, and each vertex has mass dimension $2$, thus the effect at each vertex is creating various Mandelstam variables at it. To analyze such Mandelstam variables, we take the $6$-point vertex in Fig.\ref{split6pvNL} as the example. The locality and momentum conservation indicates that the potential independent Lorentz invariants for this vertex
\bea
& &k_I^2\,,\nn
& &k_{A_1}^2\,,~~~~k_{A_2}^2\,,~~~~k_I\cdot k_{A_1}\,,~~~~k_I\cdot k_{A_2}\,,~~~~k_{A_1}\cdot k_{A_2}\,,\nn
& &k_{B_1}^2\,,~~~~k_{B_2}^2\,,~~~~k_I\cdot k_{B_1}\,,~~~~k_I\cdot k_{B_2}\,,~~~~k_{B_1}\cdot k_{B_2}\,,\nn
& &k_{A_1}\cdot k_{B_1}\,,~~~~k_{A_1}\cdot k_{B_2}\,,~~~~k_{A_2}\cdot k_{B_1}\,,~~~~k_{A_2}\cdot k_{B_2}\,.~~\label{Mandel}
\eea
All allowed Mandelstam variables which can be formed at this $6$-point vertex, can be expressed as the combination of the above Lorentz invariants in \eref{Mandel}. If two blocks $A_1$ and $A_2$ lie in ${\cal S}_1$, while $B_1$ and $B_2$ lie in ${\cal S}_2$, then terms in the last line of \eref{Mandel} automatically vanish. Then, the contribution from this $6$-point vertex can involve terms only in first three lines of \eref{Mandel}. For any term in the second line of \eref{Mandel}, for instance $k_I\cdot k_{A_2}$, we split the $6$-point vertex as in the left graph of Fig.\ref{split6pvNL}, namely, ${\cal L}_{V_A}=k_I\cdot k_{A_2}$, and ${\cal L}_{V_B}=k_{A_2A_1I}^2$ since we inserted $k_{A_2A_1I}^2/k_{A_2A_1I}^2$, where ${\cal L}_V$ denotes the Lorentz invariant created at $V$. For any term in the third line of \eref{Mandel}, we split the $6$-point vertex as in the right graph of Fig.\ref{split6pvNL}. For $k_I^2$ in the first line of of \eref{Mandel}, we can split the $6$-point vertex in either two manners in Fig.\ref{split6pvNL}. The above manipulation for the $6$-point vertex in Fig.\ref{split6pvNL},
can be directly generalized to arbitrary even valence vertices. In this scheme of separation, all resulting new vertices satisfy the locality. This feature ensures that any vertex which receives blocks from ${\cal S}_1$ (${\cal S}_2$) can never sense the particles in ${\cal S}_2$ (${\cal S}_1$).
After performing the above separation, the argument in section \ref{subsec-zero-phi3} also makes sense for new diagrams of NLSM amplitudes.
Again, suppose one merge two amplitudes in ${\cal S}_1$ and ${\cal S}_2$ by gluing only a pair of lines
as in Fig.\ref{share2}, then the bigger amplitude in ${\cal S}$ has not been created, due to the lacking of interaction between two pieces in ${\cal S}_1$ and ${\cal S}_2$. Thus, zeros of a NLSM amplitude also occur when the condition \eref{kinematic-condi-zero} is satisfied.

Then we consider the splitting without taking residue at any pole. The same as in sections \ref{subsec-1to2-phi3} and \ref{subsec-zero-YM},
for any Feynman diagram of NLSM amplitude, one can also start from three external legs $i$, $j$ and $k$ with $i<j<k$, find three lines $L_{\ell,v}$ with $\ell=i,j,k$ meet at the vertex $v$, and cut the diagram along these $L_{\ell,v}$. Again, we set particles $a\in\{j+1,\cdots,k-1\}$ to be in ${\cal S}_1$, while those $b\in\{k+1,\cdots,j-1\}\setminus i$ to be in ${\cal S}_2$. Furthermore, we separate vertices along $L_{\ell,v}$, which receive blocks from both of ${\cal S}_1$ and ${\cal S}_2$, into lower-point vertices which solely receive blocks from ${\cal S}_1$ or ${\cal S}_2$. The resulting new vertices satisfy the locality, as discussed before. Then,
the independence of scattering processes in orthogonal spaces, forces the full amplitude to factorize into two pieces as in \eref{fac-nadj-sub}. The first piece is the off-shell amplitude $A^{{\cal S}_1}_{n_1}(i,j,\cdots,k)$, where external momenta $k_a$ with $a\in\{j+1,\cdots,k-1\}$ are the same as those of the full amplitude, while particles $\{i,j,k\}$ only carry components in ${\cal S}_1$. Analogously, the second piece is the off-shell amplitude $A^{{\cal S}_2}_{2n+3-n_1}(k,\cdots,j)$, where external momenta $k_b$ with $b\in\{k+1,\cdots,j-1\}\setminus i$ are inherited from the full amplitude, while particles $\{i,j,k\}$ only carry ${\cal S}_2$ components. Notice that in general one should expect the splitting to take the generalized form \eref{fac-nadj-sub-gen} instead of \eref{fac-nadj-sub}, since the NLSM theory involves complicated vertices. However, the splitting of tree NLSM amplitudes does take the simpler form \eref{fac-nadj-sub}, since when the number of external legs of $A^{{\cal S}_1}_{n_1}$ or $A^{{\cal S}_2}_{2n+3-n_1}$ is odd, the external particles $\{i,j,k\}$ should be understood as ${\rm Tr}(\phi^3)$ scalars rather than pions, as will be shown in the next subsection. Then, the vertex $v$ which connects three ${\rm Tr}(\phi^3)$ scalars behaves as the vertex in the ${\rm Tr}(\phi^3)$ theory, thus contributes only a constant. The above phenomenon ensures that the splitting is in the simple form \eref{fac-nadj-sub}.

One can also decompose the full spacetime into three orthogonal subspaces as ${\cal S}={\cal S}_1\oplus{\cal S}_2\oplus{\cal S}_3$,
then use the similar argument to conclude that the NLSM amplitude factorizes into three amplitudes like in \eref{1to3express},
under the constraint \eref{condi-1to3split} for external kinematics.

\subsection{Generating zeros and splittings via transmutation operators}
\label{subsec-flow-NLSM}

In this subsection, we derive zeros and splittings of tree NLSM amplitudes, including the explicit interpretations for resulting amplitudes in the splittings,
by applying transmutation operators which turn YM and YM$\oplus$BAS amplitudes to NLSM and NLSM$\oplus$BAS ones \cite{Cheung:2017ems,Zhou:2018wvn,Bollmann:2018edb}. The derivation in this subsection is independent of the assumption that scattering processes in orthogonal spaces are independent of each other.

We first give a brief introduction to such operators.
For a $2n$-point YM amplitude, define an operator ${\cal L}$ as
\bea
{\cal L}\,\equiv\,\Big(\prod_{h\in\{1,\cdots,2n\}\setminus\{a,b\}}\,{\cal L}_h\Big)\,\partial_{\epsilon_a\cdot\epsilon_b}\,,~~\label{defin-L}
\eea
where
\bea
{\cal L}_h\,\equiv\,\sum_{g\neq h}\,k_h\cdot k_g\,\partial_{\epsilon_h\cdot k_g}\,,~~\label{define-Li}
\eea
and $\epsilon_a$, $\epsilon_b$ are polarization vectors of two external gluons, such two gluons can be chosen arbitrary. This operator transmutes the $2n$-point
YM amplitude ${\cal A}^{\rm YM}_{2n}(1,\cdots,2n)$ as
\bea
{\cal A}^{\rm NLSM}_{2n}(1,\cdots,2n)={\cal L}\,{\cal A}^{\rm YM}_{2n}(1,\cdots,2n)\,.~~\label{trans-YM}
\eea
For a $(2n+3)$-point YMS amplitude which contains $2n$ external gluons $\{1,\cdots,2n\}$ and $3$ external BAS scalars $\{a,b,c\}$,
the similar operator
\bea
\W{\cal L}\,\equiv\,\prod_{h\in\{1,\cdots,2n\}}\,{\cal L}_h\,,
\eea
with ${\cal L}_h$ defined in \eref{define-Li}, transmutes the YM$\oplus$BAS amplitude as
\bea
& &{\cal A}^{\rm NLSM\oplus BAS}_{2n+3}(\sigma_a,\sigma_b,\sigma_c|1,\cdots,a,\cdots,b,\cdots,c,\cdots,2n)\nn
&=&\W{\cal L}\,{\cal A}^{\rm YM\oplus BAS}_{2n+3}(\sigma_a,\sigma_b,\sigma_c|1,\cdots,a,\cdots,b,\cdots,c,\cdots,2n)\,.~~\label{trans-YMS}
\eea
In the above, the YM$\oplus$BAS amplitude carries two orderings, one is $(1,\cdots,a,\cdots,b,\cdots,c,\cdots,2n)$ among all external legs,
while another one is $(\sigma_a,\sigma_b,\sigma_c)$ among only external BAS scalars. Here $\sigma$ denotes the permutation for three elements in
$\{a,b,c\}$. The NLSM$\oplus$BAS amplitude has $2n$ external pions $\{1,\cdots,2n\}$ and $3$ external BAS scalars $\{a,b,c\}$, with the same interpretation for two orderings. The YM$\oplus$BAS and NLSM$\oplus$BAS amplitudes in \eref{trans-YMS} can be reinterpreted as YM$\oplus{\rm Tr}(\phi^3)$
and NLSM$\oplus{\rm Tr}(\phi^3)$ ones, due to the following reason. A general BAS amplitude ${\cal A}^{\rm BAS}_n(1,\cdots,n|\sigma_1,\cdots,\sigma_n)$ carries two different orderings $(1,\cdots,n)$ and $(\sigma_1,\cdots,\sigma_n)$, while the ${\rm Tr}(\phi^3)$ amplitude ${\cal A}^{\phi^3}_n(1,\cdots,n)$ serves as the special case of such BAS amplitude with $\sigma_i=i$, namely, two orderings are the same. Meanwhile, the cyclic invariance of orderings indicates that there are only two inequivalent orderings among elements in $\{a,b,c\}$,
which are $(a,b,c)$ and $(c,b,a)$. The YM$\oplus$BAS and NLSM$\oplus$BAS amplitudes with above two inequivalent orderings are related by the order-reversing relation
\bea
& &{\cal A}^{\rm YM\oplus BAS}_{2n+3}(a,b,c|1,\cdots,a,\cdots,b,\cdots,c,\cdots,2n)\nn
&=&-{\cal A}^{\rm YM\oplus BAS}_{2n+3}(c,b,a|1,\cdots,a,\cdots,b,\cdots,c,\cdots,2n)\,,\nn
& &{\cal A}^{\rm NLSM\oplus BAS}_{2n+3}(a,b,c|1,\cdots,a,\cdots,b,\cdots,c,\cdots,2n)\nn
&=&-{\cal A}^{\rm NLSM\oplus BAS}_{2n+3}(c,b,a|1,\cdots,a,\cdots,b,\cdots,c,\cdots,2n)\,.
\eea
Therefore, when the number of external BAS scalars is $3$, there are only one independent YM$\oplus$BAS amplitude, and only one independent NLSM$\oplus$BAS amplitude,
which can be identifies as YM$\oplus{\rm Tr}(\phi^3)$ and NLSM$\oplus{\rm Tr}(\phi^3)$ ones,
\bea
& &{\cal A}^{\rm YMS}_{2n+3}(a,b,c|1,\cdots,a,\cdots,b,\cdots,c,\cdots,2n)={\cal A}^{\rm YM\oplus{\rm Tr}(\phi^3)}_{2n+3}(1,\cdots,a,\cdots,b,\cdots,c,\cdots,2n)\,,\nn
& &{\cal A}^{\rm NLSM\oplus BAS}_{2n+3}(a,b,c|1,\cdots,a,\cdots,b,\cdots,c,\cdots,2n)={\cal A}^{\rm NLSM\oplus{\rm Tr}(\phi^3)}_{2n+3}(1,\cdots,a,\cdots,b,\cdots,c,\cdots,2n)\,,
\eea
since orderings among legs in $\{a,b,c\}$ are the same in $(a,b,c)$ and $(1,\cdots,a,\cdots,b,\cdots,c,\cdots,2n)$.
The above observation allows us to reformulate the transmutation \eref{trans-YMS} as
\bea
{\cal A}^{\rm NLSM\oplus{\rm Tr}(\phi^3)}_{2n+3}(1,\cdots,a,\cdots,b,\cdots,c,\cdots,2n)=\W{\cal L}\,{\cal A}^{\rm YM\oplus{\rm Tr}(\phi^3)}_{2n+3}(1,\cdots,a,\cdots,b,\cdots,c,\cdots,2n)\,.~~\label{trans-YMphi}
\eea

Now we apply the operator ${\cal L}$ in \eref{defin-L} to the factorized YM amplitude in \eref{1to2-YM-fix} (with $n$ replaced by $2n$).
Two resulting currents in \eref{1to2-YM-fix} carry $n_1$ and $(2n+3-n_1)$ external legs. Let us assume $n_1$ to be an even number, then $(2n+3-n_1)$
is forced to be an odd one. Furthermore, we can choose $a$ and $b$ in \eref{defin-L} as $a=j$, $b=k$. Under the constraints in \eref{kinematic-condi-fac-YM} and \eref{kinematic-condi-fac-YM-add}, the operator ${\cal L}$ automatically factorizes as
\bea
{\cal L}\,\xrightarrow[]{\eref{kinematic-condi-fac-YM}\,\eref{kinematic-condi-fac-YM-add}}\,{\cal L}^{{\cal S}_1}\,\times\,\W{\cal L}^{{\cal S}_2}\,,~~\label{fac-L}
\eea
where
\bea
{\cal L}^{{\cal S}_1}&=&\Big(\prod_{h\in \{i,j+1\cdots,k-1\}}\,{\cal L}_h\Big)\,\partial_{\epsilon_j\cdot\epsilon_k}\,,\nn
\W{\cal L}^{{\cal S}_2}&=&\prod_{h'\in\{k+1\cdots,j-1\}\setminus i}\,{\cal L}_{h'}\,.~~\label{L1L2}
\eea
In the operator ${\cal L}^{{\cal S}_1}$, each ${\cal L}_h$ only acts on $\epsilon_h\cdot k_g$ with $g\in \{i,j,\cdots,k\}$,
since $\epsilon_h$ lies in ${\cal S}_1$. Similarly, in the operator $\W{\cal L}^{{\cal S}_2}$, each ${\cal L}_{h'}$ only acts on
$\epsilon_{h'}\cdot k_{g'}$ with $g'\in\{k,\cdots,j\}$. The operator ${\cal L}^{{\cal S}_1}$ transmutes ${\cal J}^{\rm YM}_{n_1}$
in \eref{1to2-YM-fix}
to ${\cal J}^{\rm NLSM}_{n_1}$, due to the relation \eref{trans-YM}, and annihilates ${\cal J}^{\rm YM\oplus\phi^3}_{2n+3-n_1}$.
Meanwhile, the operator ${\cal L}^{{\cal S}_2}$ transmutes ${\cal J}^{\rm YM\oplus\phi^3}_{2n+3-n_1}$ to ${\cal J}^{\rm NLSM\oplus\phi^3}_{2n+3-n_1}$, due to the relation \eref{trans-YMphi}, and annihilates ${\cal J}^{\rm YM}_{n_1}$. Here a subtle point is,
we have directly extended the transmutations \eref{trans-YM} and \eref{trans-YMphi} for on-shell amplitudes to off-shell currents. The reason is,
in the off-shell momentum $k_i^*$ in \eref{offi-L},
carried by ${\cal J}^{\rm YM}_{n_1}$, only the on-shell component $k_i$ contributes to $\epsilon_h\cdot k_i^*$. It means the operator ${\cal L}^{{\cal S}_1}$ can not distinguish $k_i$ and $k_i^*$. Similarly, the effect of the operator $\W{\cal L}^{{\cal S}_2}$ is insensitive for shifting
the on-shell momentum $k_i$ as in \eref{offi-R}.
Consequently, the operator ${\cal L}$ acts on \eref{1to2-YM-fix} as
\bea
{\cal L}\,{\cal A}^{\rm YM}_{2n}(1,\cdots,2n)\,\xrightarrow[]{\eref{kinematic-condi-fac-YM}\,\eref{kinematic-condi-fac-YM-add}}\,\Big[{\cal L}^{{\cal S}_1}\,{\cal J}^{\rm YM}_{n_1}(i^*_g,j_g,\dots,k_g)\Big]\,\times\,\Big[\W{\cal L}^{{\cal S}_2}\,{\cal J}^{{\rm YM\oplus\phi^3}}_{n+3-n_1}(k_\phi,\cdots,i^*_\phi,\cdots,j_\phi)\Big]\,,
\eea
gives rise to the $2$-split of NLSM amplitudes in \cite{Cao:2024qpp},
\bea
{\cal A}^{\rm NLSM}_{2n}(1,\cdots,2n)\,\xrightarrow[]{\eref{kinematic-condi2-1to2}}\,{\cal J}^{\rm NLSM}_{n_1}(i^*_\pi,j_\pi,\dots,k_\pi)\,\times\,{\cal J}^{{\rm NLSM\oplus\phi^3}}_{2n+3-n_1}(k_\phi,\cdots,i^*_\phi,\cdots,j_\phi)\,.~~~~\label{1to2-NLSM-fix}
\eea
In the above, the kinematic conditions \eref{kinematic-condi-fac-YM} and \eref{kinematic-condi-fac-YM-add} are reduced to \eref{kinematic-condi2-1to2}, since scalar particles do not carry any polarization. The resulting
${\cal J}^{\rm NLSM}_{n_1}$ is understood as the current of pure NLSM theory, with $n_1$ external pions in $i\cup\{j,\cdots,k\}$, and the external momentum
$k^*_i$ is off-shell. ${\cal J}^{{\rm NLSM\oplus\phi^3}}_{2n+3-n_1}$ is the current of NLSM$\oplus$BAS theory, with $(2n+3-n_1)$ external pions $\{k+1,\cdots,j-1\}\setminus i$, and $3$ external ${\Tr}(\phi^3)$ scalars $\{i,j,k\}$, while the external momentum
$k^*_i$ is again off-shell.

The above splitting \eref{1to2-NLSM-fix} makes sense when the number of legs in $\{j+1,\cdots,k-1\}$ is odd. If $\{j+1,\cdots,k-1\}$
includes an even number of elements, then the analogous manipulation leads to the similar splitting
\bea
{\cal A}^{\rm NLSM}_{2n}(1,\cdots,2n)= {\cal J}^{\rm NLSM\oplus\phi^3}_{n_1}(i^*_\phi,j_\phi,\dots,k_\phi)\,\times\,{\cal J}^{{\rm NLSM}}_{2n+3-n_1}(k_\pi,\cdots,i^*_\pi,\cdots,j_\pi)\,.~~~~\label{1to2-NLSM-fix2}
\eea

Applying the same operators, one can also transmutes the $2$- and $3$- splits in \eref{1to2express-YM} and \eref{1to3express-YM} to
\bea
{\cal A}^{\rm NLSM}_{2n}(1,\cdots,2n)\,&\xrightarrow[]{\eref{kinematic-condi-1to2}}&\, {\cal J}^{\rm NLSM}_{n_1}(i^*_\pi,i+1_\pi,\dots,k_\pi)\,\times\,{\cal J}^{\rm NLSM\oplus\phi^3}_{2n+3-n_1}(k_\phi,\cdots,i_\phi,i+1^*_\phi)\,,~~~~\label{1to2express-N}
\eea
and
\bea
& &{\cal A}^{\rm NLSM}_{2n}(1,\cdots,2n)\,\xrightarrow[]{\eref{condi-1to3split}}\nn
& &~~~~~~~~~~~~{\cal J}^{{\rm NLSM}\oplus\phi^3}_{n_1}(i^*_\phi,j_\phi,\cdots,k_\phi)\,\times\,{\cal J}^{\rm NLSM}_{n_2}(j^*_\pi,k_\pi,\cdots,i_\pi)\,\times\,{\cal J}^{{\rm NLSM}\oplus\phi^3}_{2n+6-n_1-n_2}(k^*_\phi,i_\phi,\cdots,j_\phi)\,.~~\label{1to3express-N}
\eea
The above \eref{1to3express-N} is the same as the smooth splitting found in \cite{Cachazo:2021wsz}. In \eref{1to3express-N}, either ${\cal J}^{\rm NLSM\oplus\phi^3}_{n_1}$ or ${\cal J}^{{\rm NLSM}\oplus\phi^3}_{n+6-n_1-n_2}$ carries the odd number of external legs, since the NLSM interactions require the number of external pions to be even. On the other hand, applying the $2$-split in \eref{1to2express-N} to
$ {\cal J}^{\rm NLSM}_{2n+3-n_1}$ in \eref{1to2-NLSM-fix2} also leads to \eref{1to3express-N}.

Clearly, zeros of NLSM amplitudes, as well as factorizations near zeros, can also be generated from those for YM amplitudes, by acting the transmutation operators.

\section{Discussion}
\label{sec-conclusion}

In this paper, we proposed the novel interpretations of zeros and splittings of ordered tree level amplitudes discovered in \cite{Cachazo:2021wsz,Arkani-Hamed:2023swr,Cao:2024gln,Arkani-Hamed:2024fyd,Cao:2024qpp}, by utilizing Feynman diagrams.
We have considered three theories, which are ${\rm Tr}(\phi^3)$, YM and NLSM. For ${\rm Tr}(\phi^3)$ amplitudes, zeros and splittings are understood via two alternative ways, one is based on the assumption that scattering processes in orthogonal spaces are independent of each other, while another one exploits the property in \eref{fac-mkb} which can be called "factorization of propagators". The first method can not answer the theory (theories) of resulting amplitudes in splittings, while the second one completely fixes the interpretations for resulting amplitudes.
The above first method also determines the certain loci in kinematic space for zeros and splittings of YM and NLSM amplitudes.
On the other hand, the second method was extended to YM amplitudes, by utilizing locality and gauge invariance.
We also provided the method of applying transmutation operators,
to generate zeros and splittings of NLSM amplitudes, from those of YM amplitudes. It is with great regret that, in this work we have not extended the second method to the NLSM case, by exploiting locality and Adler zero. We leave this problem to the feature work.

Besides considering Feynman diagrams, in this work we have frequently used the assumption that scattering processes in orthogonal
spaces are independent of each other. It is worth to emphasize again
one can also prove zeros and splittings by utilizing Feynman diagrams, but without this assumption.
As repeatedly emphasized in sections \ref{subsec-verification-phi3} and \ref{subsec-flow-YM}, our second method is valid for ${\rm Tr}(\phi^3)$ and YM amplitudes even if this assumption does not hold. Meanwhile, the approach of performing transmutation operators, which generates zeros and splittings of NLSM amplitudes from zeros and splittings of YM ones, is absolutely independent of this assumption. On the other hand,
as can be seen through discussions in this paper, this assumption is helpful for thinking zeros and splittings, and leads to the correct loci in kinematic space for zeros and splittings. It is natural to ask, if this assumption is satisfied by a wider range of theories beyond ${\rm Tr}(\phi^3)$, YM and NLSM. We will study this issue else where.

The amplitudes considered in this paper are color/flavor ordered ones. However, the splitting is an universal behavior also holds for unordered amplitudes such as gravitational and SG amplitudes, as uncovered in \cite{Cao:2024qpp}. Therefore, it is important to generalize the method in this paper to unordered cases. The obstacle is, the propagators such as $1/s_{ab}$, with $k_a$ in ${\cal S}_1$ and $k_b$ in ${\cal S}_2$, will occur in unordered amplitudes, thus cause the divergence.
For ordered amplitudes, this situation will not happen, since legs $a$ and $b$ are separated by the glued lines. One may expect that, such divergences in an unordered amplitude are eliminated by Lorentz invariants such like $k_a\cdot k_b$, $\epsilon_a\cdot k_b$ and $\epsilon_a\cdot \epsilon_b$ in numerators, which also vanish on the special loci in kinematic space. However, the explicit mechanism of such cancelations is unclear. Interpreting zeros and splittings of unordered amplitudes, especially the above cancelations, by generalizing the approach in this paper, is an interesting future direction. Another potential generalization of our method, is to the splittings at loop orders discussed in \cite{Arkani-Hamed:2024fyd}.

Finally, a fascinating question is, if the universal splittings of amplitudes can be used to construct amplitudes effectively. As revealed in \cite{Cao:2024gln,Arkani-Hamed:2024fyd,Cao:2024qpp}, the splittings have deep connections to soft theorems. It is well known, amplitudes of a large variety of theories can be constructed by exploiting the universal soft behaviors \cite{Nguyen:2009jk,Boucher-Veronneau:2011rwd,Rodina:2018pcb,Ma:2022qja,Cheung:2014dqa,Cheung:2015ota,Luo:2015tat,Elvang:2018dco,Zhou:2022orv,
Du:2024dwm,Zhou:2024qwm,Zhou:2024qjh}. One can expect that the methods of inverse soft theorems can be generalized by utilizing splittings, due to the relations between splittings and soft behaviors.

\section*{Acknowledgments}

The author would thank Prof. Yijian Du, Prof. Song He, Congsi Xie and Canxin Shi for useful discussions.
We also thank Prof. Yijian Du and Congsi Xie for very valuable comments and suggestions on the draft. This work is supported by NSFC under Grant No. 11805163.


\bibliographystyle{JHEP}

\bibliography{reference}

\providecommand{\href}[2]{#2}\begingroup\raggedright\begin{thebibliography}{10}

\bibitem{Britto:2004ap}
R.~Britto, F.~Cachazo, and B.~Feng, {\it {New recursion relations for tree
  amplitudes of gluons}},  {\em Nucl. Phys. B} {\bf 715} (2005) 499--522,
  [\href{http://arxiv.org/abs/hep-th/0412308}{{\tt hep-th/0412308}}].

\bibitem{Britto:2005fq}
R.~Britto, F.~Cachazo, B.~Feng, and E.~Witten, {\it {Direct proof of tree-level
  recursion relation in Yang-Mills theory}},  {\em Phys. Rev. Lett.} {\bf 94}
  (2005) 181602, [\href{http://arxiv.org/abs/hep-th/0501052}{{\tt
  hep-th/0501052}}].

\bibitem{Bern:1994zx}
Z.~Bern, L.~J. Dixon, D.~C. Dunbar, and D.~A. Kosower, {\it {One loop n point
  gauge theory amplitudes, unitarity and collinear limits}},  {\em Nucl. Phys.
  B} {\bf 425} (1994) 217--260,
  [\href{http://arxiv.org/abs/hep-ph/9403226}{{\tt hep-ph/9403226}}].

\bibitem{Bern:1994cg}
Z.~Bern, L.~J. Dixon, D.~C. Dunbar, and D.~A. Kosower, {\it {Fusing gauge
  theory tree amplitudes into loop amplitudes}},  {\em Nucl. Phys. B} {\bf 435}
  (1995) 59--101, [\href{http://arxiv.org/abs/hep-ph/9409265}{{\tt
  hep-ph/9409265}}].

\bibitem{Britto:2004nc}
R.~Britto, F.~Cachazo, and B.~Feng, {\it {Generalized unitarity and one-loop
  amplitudes in N=4 super-Yang-Mills}},  {\em Nucl. Phys. B} {\bf 725} (2005)
  275--305, [\href{http://arxiv.org/abs/hep-th/0412103}{{\tt hep-th/0412103}}].

\bibitem{Cachazo:2021wsz}
F.~Cachazo, N.~Early, and B.~Gim\'enez~Umbert, {\it {Smoothly splitting
  amplitudes and semi-locality}},  {\em JHEP} {\bf 08} (2022) 252,
  [\href{http://arxiv.org/abs/2112.14191}{{\tt arXiv:2112.14191}}].

\bibitem{GimenezUmbert:2024jjn}
B.~Gim\'enez~Umbert and K.~Yeats, {\it {$\Phi^p$ Amplitudes from the Positive
  Tropical Grassmannian: Triangulations of Extended Diagrams}},
  \href{http://arxiv.org/abs/2403.17051}{{\tt arXiv:2403.17051}}.

\bibitem{Arkani-Hamed:2023swr}
N.~Arkani-Hamed, Q.~Cao, J.~Dong, C.~Figueiredo, and S.~He, {\it {Hidden zeros
  for particle/string amplitudes and the unity of colored scalars, pions and
  gluons}},  {\em JHEP} {\bf 10} (2024) 231,
  [\href{http://arxiv.org/abs/2312.16282}{{\tt arXiv:2312.16282}}].

\bibitem{Arkani-Hamed:2023lbd}
N.~Arkani-Hamed, H.~Frost, G.~Salvatori, P.-G. Plamondon, and H.~Thomas, {\it
  {All Loop Scattering As A Counting Problem}},
  \href{http://arxiv.org/abs/2309.15913}{{\tt arXiv:2309.15913}}.

\bibitem{Arkani-Hamed:2023mvg}
N.~Arkani-Hamed, H.~Frost, G.~Salvatori, P.-G. Plamondon, and H.~Thomas, {\it
  {All Loop Scattering For All Multiplicity}},
  \href{http://arxiv.org/abs/2311.09284}{{\tt arXiv:2311.09284}}.

\bibitem{Arkani-Hamed:2023jry}
N.~Arkani-Hamed, Q.~Cao, J.~Dong, C.~Figueiredo, and S.~He, {\it
  {Scalar-Scaffolded Gluons and the Combinatorial Origins of Yang-Mills
  Theory}},  \href{http://arxiv.org/abs/2401.00041}{{\tt arXiv:2401.00041}}.

\bibitem{Arkani-Hamed:2024nhp}
N.~Arkani-Hamed, Q.~Cao, J.~Dong, C.~Figueiredo, and S.~He, {\it {Nonlinear
  Sigma model amplitudes to all loop orders are contained in the
  Tr(\ensuremath{\Phi}3) theory}},  {\em Phys. Rev. D} {\bf 110} (2024), no.~6
  065018, [\href{http://arxiv.org/abs/2401.05483}{{\tt arXiv:2401.05483}}].

\bibitem{Arkani-Hamed:2024vna}
N.~Arkani-Hamed, C.~Figueiredo, H.~Frost, and G.~Salvatori, {\it {Tropical
  Amplitudes For Colored Lagrangians}},
  \href{http://arxiv.org/abs/2402.06719}{{\tt arXiv:2402.06719}}.

\bibitem{Arkani-Hamed:2024yvu}
N.~Arkani-Hamed and C.~Figueiredo, {\it {Circles and Triangles, the NLSM and
  Tr($\Phi^3$)}},  \href{http://arxiv.org/abs/2403.04826}{{\tt
  arXiv:2403.04826}}.

\bibitem{Arkani-Hamed:2024tzl}
N.~Arkani-Hamed, Q.~Cao, J.~Dong, C.~Figueiredo, and S.~He, {\it {Surface
  Kinematics and ''The'' Yang-Mills Integrand}},
  \href{http://arxiv.org/abs/2408.11891}{{\tt arXiv:2408.11891}}.

\bibitem{Cao:2024gln}
Q.~Cao, J.~Dong, S.~He, and C.~Shi, {\it {A universal splitting of tree-level
  string and particle scattering amplitudes}},  {\em Phys. Lett. B} {\bf 856}
  (2024) 138934, [\href{http://arxiv.org/abs/2403.08855}{{\tt
  arXiv:2403.08855}}].

\bibitem{Arkani-Hamed:2024fyd}
N.~Arkani-Hamed and C.~Figueiredo, {\it {All-order splits and multi-soft limits
  for particle and string amplitudes}},
  \href{http://arxiv.org/abs/2405.09608}{{\tt arXiv:2405.09608}}.

\bibitem{Cao:2024qpp}
Q.~Cao, J.~Dong, S.~He, C.~Shi, and F.~Zhu, {\it {On universal splittings of
  tree-level particle and string scattering amplitudes}},  {\em JHEP} {\bf 09}
  (2024) 049, [\href{http://arxiv.org/abs/2406.03838}{{\tt arXiv:2406.03838}}].

\bibitem{Cachazo:2013hca}
F.~Cachazo, S.~He, and E.~Y. Yuan, {\it {Scattering of Massless Particles in
  Arbitrary Dimensions}},  {\em Phys. Rev. Lett.} {\bf 113} (2014), no.~17
  171601, [\href{http://arxiv.org/abs/1307.2199}{{\tt arXiv:1307.2199}}].

\bibitem{Cachazo:2013iea}
F.~Cachazo, S.~He, and E.~Y. Yuan, {\it {Scattering of Massless Particles:
  Scalars, Gluons and Gravitons}},  {\em JHEP} {\bf 07} (2014) 033,
  [\href{http://arxiv.org/abs/1309.0885}{{\tt arXiv:1309.0885}}].

\bibitem{Cachazo:2014nsa}
F.~Cachazo, S.~He, and E.~Y. Yuan, {\it {Einstein-Yang-Mills Scattering
  Amplitudes From Scattering Equations}},  {\em JHEP} {\bf 01} (2015) 121,
  [\href{http://arxiv.org/abs/1409.8256}{{\tt arXiv:1409.8256}}].

\bibitem{Cachazo:2014xea}
F.~Cachazo, S.~He, and E.~Y. Yuan, {\it {Scattering Equations and Matrices:
  From Einstein To Yang-Mills, DBI and NLSM}},  {\em JHEP} {\bf 07} (2015) 149,
  [\href{http://arxiv.org/abs/1412.3479}{{\tt arXiv:1412.3479}}].

\bibitem{Berends:1987me}
F.~A. Berends and W.~T. Giele, {\it {Recursive Calculations for Processes with
  n Gluons}},  {\em Nucl. Phys. B} {\bf 306} (1988) 759--808.

\bibitem{Wu:2021exa}
K.~Wu and Y.-J. Du, {\it {Off-shell extended graphic rule and the expansion of
  Berends-Giele currents in Yang-Mills theory}},  {\em JHEP} {\bf 01} (2022)
  162, [\href{http://arxiv.org/abs/2109.14462}{{\tt arXiv:2109.14462}}].

\bibitem{Cheung:2014dqa}
C.~Cheung, K.~Kampf, J.~Novotny, and J.~Trnka, {\it {Effective Field Theories
  from Soft Limits of Scattering Amplitudes}},  {\em Phys. Rev. Lett.} {\bf
  114} (2015), no.~22 221602, [\href{http://arxiv.org/abs/1412.4095}{{\tt
  arXiv:1412.4095}}].

\bibitem{Cheung:2015ota}
C.~Cheung, K.~Kampf, J.~Novotny, C.-H. Shen, and J.~Trnka, {\it {On-Shell
  Recursion Relations for Effective Field Theories}},  {\em Phys. Rev. Lett.}
  {\bf 116} (2016), no.~4 041601, [\href{http://arxiv.org/abs/1509.03309}{{\tt
  arXiv:1509.03309}}].

\bibitem{Cheung:2017ems}
C.~Cheung, C.-H. Shen, and C.~Wen, {\it {Unifying Relations for Scattering
  Amplitudes}},  {\em JHEP} {\bf 02} (2018) 095,
  [\href{http://arxiv.org/abs/1705.03025}{{\tt arXiv:1705.03025}}].

\bibitem{Zhou:2018wvn}
K.~Zhou and B.~Feng, {\it {Note on differential operators, CHY integrands, and
  unifying relations for amplitudes}},  {\em JHEP} {\bf 09} (2018) 160,
  [\href{http://arxiv.org/abs/1808.06835}{{\tt arXiv:1808.06835}}].

\bibitem{Bollmann:2018edb}
M.~Bollmann and L.~Ferro, {\it {Transmuting CHY formulae}},  {\em JHEP} {\bf
  01} (2019) 180, [\href{http://arxiv.org/abs/1808.07451}{{\tt
  arXiv:1808.07451}}].

\bibitem{Nguyen:2009jk}
D.~Nguyen, M.~Spradlin, A.~Volovich, and C.~Wen, {\it {The Tree Formula for MHV
  Graviton Amplitudes}},  {\em JHEP} {\bf 07} (2010) 045,
  [\href{http://arxiv.org/abs/0907.2276}{{\tt arXiv:0907.2276}}].

\bibitem{Boucher-Veronneau:2011rwd}
C.~Boucher-Veronneau and A.~J. Larkoski, {\it {Constructing Amplitudes from
  Their Soft Limits}},  {\em JHEP} {\bf 09} (2011) 130,
  [\href{http://arxiv.org/abs/1108.5385}{{\tt arXiv:1108.5385}}].

\bibitem{Rodina:2018pcb}
L.~Rodina, {\it {Scattering Amplitudes from Soft Theorems and Infrared
  Behavior}},  {\em Phys. Rev. Lett.} {\bf 122} (2019), no.~7 071601,
  [\href{http://arxiv.org/abs/1807.09738}{{\tt arXiv:1807.09738}}].

\bibitem{Ma:2022qja}
S.~Ma, R.~Dong, and Y.-J. Du, {\it {Constructing EYM amplitudes by inverse soft
  limit}},  {\em JHEP} {\bf 05} (2023) 196,
  [\href{http://arxiv.org/abs/2211.10047}{{\tt arXiv:2211.10047}}].

\bibitem{Luo:2015tat}
H.~Luo and C.~Wen, {\it {Recursion relations from soft theorems}},  {\em JHEP}
  {\bf 03} (2016) 088, [\href{http://arxiv.org/abs/1512.06801}{{\tt
  arXiv:1512.06801}}].

\bibitem{Elvang:2018dco}
H.~Elvang, M.~Hadjiantonis, C.~R.~T. Jones, and S.~Paranjape, {\it {Soft
  Bootstrap and Supersymmetry}},  {\em JHEP} {\bf 01} (2019) 195,
  [\href{http://arxiv.org/abs/1806.06079}{{\tt arXiv:1806.06079}}].

\bibitem{Zhou:2022orv}
K.~Zhou, {\it {Tree level amplitudes from soft theorems}},  {\em JHEP} {\bf 03}
  (2023) 021, [\href{http://arxiv.org/abs/2212.12892}{{\tt arXiv:2212.12892}}].

\bibitem{Du:2024dwm}
Y.-J. Du and K.~Zhou, {\it {Multi-trace YMS amplitudes from soft behavior}},
  {\em JHEP} {\bf 03} (2024) 081, [\href{http://arxiv.org/abs/2401.03879}{{\tt
  arXiv:2401.03879}}].

\bibitem{Zhou:2024qwm}
K.~Zhou and C.~Hu, {\it {Towards tree Yang-Mills and Yang-Mills-scalar
  amplitudes with higher-derivative interactions}},
  \href{http://arxiv.org/abs/2406.03034}{{\tt arXiv:2406.03034}}.

\bibitem{Zhou:2024qjh}
K.~Zhou, {\it {Constructing tree amplitudes of scalar EFT from double soft
  theorem}},  \href{http://arxiv.org/abs/2406.03784}{{\tt arXiv:2406.03784}}.

\end{thebibliography}\endgroup

\end{document}